\PassOptionsToPackage{
  colorlinks=true,
  linkcolor=Maroon,
  citecolor=Blue,
  urlcolor=Teal
}{hyperref}

\documentclass[11pt]{article}
\usepackage{fullpage}

\usepackage[letterpaper,
left=1in,right=1in,top=1in,bottom=1in]{geometry}
\usepackage{amsmath}
\usepackage{amssymb}
\usepackage{mathtools}
\usepackage{amsthm}
\usepackage{multicol}
\usepackage{tabularx}
\usepackage{longtable}
\usepackage{array}
\usepackage{booktabs}
\usepackage{ragged2e}
\usepackage{graphicx}
\usepackage{enumitem}
\usepackage[dvipsnames]{xcolor}
\definecolor{Teal}{rgb}{0.0, 0.5, 0.5}

\usepackage[
  backend=biber,
  backref=false,
  backrefstyle=none,
  date=year,
  doi=false,
  giveninits=true,
  hyperref=true,
  maxbibnames=10,
  sortcites=false,
  style=alphabetic,
  url=false, 
]{biblatex}
\usepackage{hyperref}

\theoremstyle{plain}
\newtheorem{theorem}{Theorem}[section]

\theoremstyle{definition}

\theoremstyle{remark}

\def\4o{GPT-4o}
\def\o1{o1-preview}

\begin{document}

\title{Assessing GPT Performance in a Proof-Based University-Level Course Under Blind Grading}

\author{
  Ming Ding\\ 
	\texttt{ming.ding@inf.ethz.ch}\\
	Department of Computer Science\\
	ETH Zurich
\and
Rasmus Kyng\\ 
\texttt{kyng@inf.ethz.ch}\\
Department of Computer Science\\
ETH Zurich
  \and 
Federico Soldà\\ 
\texttt{federico.solda@inf.ethz.ch}\\
Department of Computer Science\\
ETH Zurich
\and
Weixuan Yuan\\ 
\texttt{weyuan@inf.ethz.ch}\\
Department of Computer Science\\
ETH Zurich
}

\date{} 

\clearpage\maketitle
\thispagestyle{empty}

\begin{abstract}
As large language models (LLMs) advance, their role in higher education, particularly in free-response problem-solving, requires careful examination. This study assesses the performance of GPT-4o and o1-preview under realistic educational conditions in an undergraduate algorithms course. Anonymous GPT-generated solutions to take-home exams were graded by teaching assistants unaware of their origin. Our analysis examines both coarse-grained performance (scores) and fine-grained reasoning quality (error patterns). Results show that GPT-4o consistently struggles, failing to reach the passing threshold, while o1-preview performs significantly better, surpassing the passing score and even exceeding the student median in certain exercises. However, both models exhibit issues with unjustified claims and misleading arguments. These findings highlight the need for robust assessment strategies and AI-aware grading policies in education.
\end{abstract}

\newpage
\pagenumbering{gobble}
\sloppy

\tableofcontents

\newpage

\pagenumbering{arabic}

\section{Introduction}
\label{sect:intro}

The emergence of large language models (LLMs) has revolutionized how we interact with artificial intelligence (AI), and garnered significant attention from researchers and the public. While LLMs 
have demonstrated impressive capabilities, their true level of intelligence and reasoning remains a subject of debate.

The classical Turing Test proposes that a machine demonstrating human-like responses in conversation could be considered intelligent. Over the past few years, substantial efforts have been devoted to evaluating LLMs from various angles \cite{CWWWYZCYWWE24}.  
For example, LLMs can generate essays with their quality rated higher than those produced by humans \cite{herbold2023large}; 
pass questions involving communication skills, ethics, empathy, and professionalism in a United States Medical Licensing Examination (USMLE) \cite{brin2023comparing}; 
achieve passing scores on the reading comprehension test of the Program for International Student Assessment (PISA), a global standardized student assessment \cite{vazquez2023chatgpt};
and demonstrate strong performance in solving middle school-level math word problems, with multiple LLMs achieving passing scores and some exceeding 90\% accuracy 
\cite{vidal2024evaluation}.
However, existing evaluation protocols may fall short of comprehensively assessing their reasoning and problem-solving capabilities. Most benchmarks involve structured datasets, such as multiple-choice or numeric questions, which may not fully capture the complexities of real-world tasks. While prior research has demonstrated their potential to assist humans or improve responses through prompt engineering \cite{YYZSGCN24, stribling2024model}, there remains a need to evaluate LLMs in scenarios that require independent reasoning and creativity.

The study presented in this article addresses the gap by evaluating GPT models’ performance in solving complex free-response questions, specifically proof-based problems from an undergraduate algorithms course. The tasks require not only creative algorithm design but also rigorous justification of correctness—a skill fundamental to human cognitive problem-solving.
To ensure unbiased evaluation, we anonymously submitted GPT-generated responses to exam graders, who assessed them without knowing whether they were generated by a student or AI. Unlike prior work that focuses on coarse-grained metrics such as pass/fail rates or overall accuracy, we conduct a fine-grained analysis by examining grader feedback and error patterns. This provides deeper insights into LLMs’ limitations in proof-based and free-response problem-solving. Beyond assessing performance, our study contributes to the broader discussion on AI’s role in education by highlighting both its potential and its constraints. By exploring these aspects, we aim to clarify how AI can reshape learning and assessment practices, as well as the thresholds it must surpass to genuinely emulate undergraduate students.

Recent advancements in LLMs have introduced several competing products, such as OpenAI’s ChatGPT \cite{Ochatgpt2024}, Google’s Gemini \cite{gemini2024}, GitHub’s Copilot \cite{copilot2024}, and SciSpace \cite{scispace2024}. 
As these models continue to improve, their progress necessitates a reassessment of the role and effectiveness of academic assessments in the era of AI-assisted learning.
In this work, we focus on ChatGPT-4o and {\o1}, the latest models from OpenAI available during our experiment period.
GPT-4o (``o'' for ``omni'') is a multilingual, multimodal generative pre-trained transformer developed by OpenAI and released in May 2024. In September 2024, OpenAI released {\o1}, the first of a new series of products, with the full version released later. Despite their rapid development, the performance of OpenAI’s latest models in realistic educational settings remains largely unexamined. Our study evaluates their effectiveness in solving complex academic problems.

With a particular focus on {\4o} and {\o1}, our study makes the following key contributions:

\begin{enumerate}
    \item Our study is conducted in real educational settings with authentic grading conditions, ensuring that the evaluation closely reflects actual academic environments. The experimental design serves as a reference for measuring AI performance in an unbiased and reliable manner.
    Specifically, we implement \textit{blind grading}, ensuring that the teaching team remains unaware of the experiment and the origin of the submissions. This prevents any potential bias in evaluation. Additionally, we generate solutions using \textit{naive prompts}—without providing course materials or additional guidance—ensuring that GPT does not receive any advantages from instructor input. 
    Different from other benchmarks that are designed to explore the upper limits of LLM problem-solving capabilities, our approach aims to establish a realistic \textit{lower bound} on model effectiveness in educational environments, reflecting how LLMs might perform when used by students in typical, unstructured settings.

    \item 
    The dataset we use is original and highly difficult.
    We leverage take-home exams in an undergraduate-level algorithms theory course, where exercises are carefully designed to be challenging even with access to all course materials. This ensures that the evaluation assesses genuine problem-solving skills rather than mere retrieval-based responses. 
    Our selected exercises are original, which sets us apart from existing popular evaluations on standardized examinations, which often have extensive learning material available online for trainees (including large sets of example questions and answers).
    
    Moreover, our dataset consists of free-response, proof-based questions, which require a combination of background knowledge, creativity, and multi-stage reasoning. These characteristics make our benchmark particularly suitable for evaluating GPT’s ability to generate correct and logically coherent responses to expert-level queries. 
    Unlike most existing evaluation settings, which often use closed-book exam questions, our approach presents a stronger test of AI reasoning. 
    We release the full dataset in the Appendix, including problem sets, master solutions, and AI-generated solutions. It not only complements existing benchmarks focused on closed-ended or structured problems but also enriches the landscape of high-difficulty evaluations, such as BIG-Bench Hard \cite{suzgun2022challenging}, by providing open-ended, proof-based tasks.
    
    \item 
    We employ expert human grading to ensure a precise and trustworthy evaluation of both student and AI-generated solutions. Although this approach requires substantial human effort, it significantly enhances the reliability of comparisons across students, GPT-4o, and o1-preview.  Additionally, our graders provide detailed feedback and comments, allowing for a \textit{fine-grained analysis} of AI errors. By categorizing error patterns into unjustified claims, misleading statements, and mathematical inaccuracies, 
    this qualitative assessment is beyond numerical scoring, and provides insights into how the models function at a deeper level and highlights specific areas where improvements are needed. 
    
\end{enumerate}

\section{Related Work}
\label{sect:related_work}

Large Language Models (LLMs), such as GPT-4, have been extensively evaluated in mathematics and computer science to assess their reasoning and problem-solving capabilities. The evaluation of LLMs has been extensively reviewed in \cite{CWWWYZCYWWE24}, we focus here on the studies most relevant to our work.

\paragraph{Tests against standard datasets.}
A comprehensive description of how LLMs are traditionally evaluated can be found, for instance, in GPT4 Technical Report \cite{OAAAAAAAAAE24}.
In mathematics, the GSM8K dataset \cite{CKBCJKPTHNE21} contains 8.5K high-quality, linguistically diverse grade-school math word problems designed to evaluate and enhance the mathematical reasoning abilities of language models. 
Hendrycks et al. proposed the more challenging MATH dataset \cite{HBKABTSS21}, which includes 12.5K problems spanning a wide range of topics and difficulty levels. 
Recent LLMs models have also been evaluated on a variety of benchmarks: the AIME 2024 benchmark, which features advanced competition-level mathematics problems \cite{LDJZZZZD25}; the GPQA Diamond dataset \cite{rein2023gpqa}, which focuses on numerical and multiple-choice questions; and the MMLU benchmark \cite{HBBZMSS}, a suite of multiple-choice questions covering 57 subjects. 
Other evaluations include ARC, the AI2 Reasoning Challenge \cite{CCEKSST18}; WinoGrande, an adversarial Winograd schema challenge at scale \cite{SBBC21}; HumanEval, which tests functional correctness in code generation from docstrings \cite{CTJYPKEBJBE21}; DROP, a reading comprehension benchmark requiring discrete reasoning over paragraphs \cite{DWDSSG19}; and HellaSwag \cite{ZHBFC19}; and others.

Most recently, Latif et al. \cite{LZGSGNBYZ24} evaluated the ChatGPT o1-preview model on established benchmarks. They observed that while o1-preview excelled in structured tasks, it displayed limitations in problem-solving and adaptive reasoning. Notably, it outperformed undergraduate and postgraduate participants in critical thinking, systematic thinking, data literacy, and creative thinking tasks, and surpassed the highest human scores in scientific reasoning.

\paragraph{The use of LLMs in higher education.}
The application of AI in higher education has also been widely explored in recent years. For example, Nikolić et al. \cite{nikolic2024chatgpt} conducted a multi-institutional, multi-disciplinary evaluation of various Generative AI tools, including ChatGPT-3.5, ChatGPT-4, Copilot, Gemini, SciSpace, and Wolfram, across ten engineering subjects in seven Australian universities. The study evaluated AI performance on diverse assessment types. 
The tasks were rated as Pass, Fail, Component Pass, or Possible Pass, providing a coarse-grained evaluation of AI's educational capabilities.
In addition, Susnjak et al. \cite{susnjak2024chatgpt} explored how LLMs challenge the integrity of online exams. Their work emphasized multi-step reasoning strategies, guiding LLMs through iterative self-review and refinement to improve their responses.
Yeadon et al. \cite{yeadon2024comparison} evaluated the performance of GPT-3.5 and GPT-4 in university-level physics coding assignments using the Python language, concluding that AI-generated work closely approaches the quality of university students' work while remaining detectable by human evaluators.

\paragraph{Tests with human comparison.}
Several studies have directly compared AI and human performance in academic settings. 
Stribling et al. \cite{stribling2024model} assess the capability of GPT-4 to answer questions from nine graduate-level final examinations for scientific trainees in the biomedical sciences. They also examine the impact of multiple styles of GPT-4 query (prompt patterns) and compare results to student performance. Grading has been performed blinded for most examinations.
Kipp \cite{kipp2024gpt} tested GPT-4o on 30 unique anatomical multiple-choice questions from German medical licensing exams, where it consistently outperformed medical students across six state exams.
In the study of \cite{bayer2024chatgpt} ChatGPT-4, GPT-4o, and Google Gemini were tested on a physiology exam for veterinary students, with their results compared to those of the students.
Finally, Richards et al. \cite{RWSPWG24} conducted a dual-anonymous ``quality assurance'' marking exercise on four end-of-module assessments within a distance-learning computer science curriculum. A total of 90 scripts, either student-submitted or ChatGPT-generated, were blindly graded. All ChatGPT-generated undergraduate scripts received at least a passing grade, whereas none of the ChatGPT-generated postgraduate scripts achieved a passing grade.

\paragraph{High-difficulty datasets.}
Several recent studies have explored the limitations and capabilities of large language models (LLMs) in proof-based and symbolic reasoning tasks using relatively small, high-difficulty datasets.
BIG-Bench Hard \cite{suzgun2022challenging} focuses on a suite of 23 challenging tasks from BIG-Bench that were found to be beyond the capabilities of current language models. These tasks are ones where prior language model evaluations did not outperform the average human-rater, making them valuable for stress-testing reasoning abilities. 
Building on this, Program-Aided Language models (PAL) \cite{gao2023pal} were evaluated on 13 tasks spanning mathematical, symbolic, and algorithmic reasoning drawn from BIG-Bench Hard and other sources. 
DeepMind's AlphaProof system\footnote{\url{https://deepmind.google/discover/blog/ai-solves-imo-problems-at-silver-medal-level/}} evaluated model performance on six problems from the International Mathematical Olympiad (IMO), four of which were successfully solved. 
Despite the small dataset, this evaluation demonstrated that carefully selected high-difficulty problems can yield meaningful insights into model capabilities—enough to support claims of ``silver medal'' performance.
Similarly, the miniF2F benchmark \cite{yousefzadeh2024lean} for the Lean theorem prover contains a test set of 20 formalized IMO problems, with only six problems having fully verified Lean proofs. This benchmark highlights the difficulty of formal reasoning tasks and has become a standard for evaluating autoformalization and proof synthesis systems in low-data, high-complexity regimes.

\section{Methodology}
\label{sect:methodology}

\subsection{Experiment Design}

The study presented in this article aims to assess the effectiveness of LLMs in performing algorithm exercises assigned to university students, using a blinded grading approach to evaluate solutions written by both students and AI.

\paragraph{Course and examination information.}

The study was conducted in a third-year undergraduate-level course at ETH Zurich in the Computer Science Department.\footnote{\url{https://inf.ethz.ch}} As a core course in the Bachelor’s major ``Theoretical Computer Science'', the class tends to attract a more specialized group of students, which introduces some selection bias compared to a typical undergraduate cohort.
This introductory course on algorithm theory carries 8 ECTS credits and involves an estimated workload of 200 hours over a 14-week semester. Topics covered include Randomized Search Trees, Point Location, Bootstrapping Algorithms, Linear Programming, Randomized Algebraic Algorithms, and Parallel Algorithms.

The course evaluation comprises two take-home exams featuring free-response proof-based questions.\footnote{The course also includes written in-person exams, which we excluded from evaluation.} 
For each take-home exam, students have two weeks to submit solutions, with unrestricted access to all course materials.
They are required to propose algorithmic solutions and provide formal proof of correctness and complexity analysis of proposed algorithms. All solutions must be typeset in {\LaTeX} and submitted as PDF.

Since this is a take-home task, students have access to the Internet and potentially AI tools. To ensure academic integrity as much as possible, besides requiring students to cite any external sources for results not covered in the lectures, the course instructors take additional measures in exercise design. Before release, each exercise undergoes a rigorous design process that includes: (1) Non-expert validation – Ensuring that no off-the-shelf solutions are readily available online or in past problem sets.
(2) Expert validation – Conducting test-solving sessions among the teaching team to verify the problem's clarity, difficulty, and originality.
These measures help maintain exercises' effectiveness while minimizing the risk of students relying on existing solutions.

\paragraph{Data collection and grading process.}
Each take-home exam consists of four exercises with multiple subquestions. 
For the first exam, we gathered 134 student submissions and two AI-generated responses, while for the second, we collected 122 student submissions along with two AI-generated responses.

Each exercise was graded independently, and we collected the data per exercise.
Grading was conducted by a teaching team member unaware of the submission source (student or AI), ensuring a blind evaluation process. 
This approach ensures an unbiased comparison between AI-generated and student solutions, making the study reflective of a realistic educational setting.
Our grading team consists of PhD students and top-performing bachelor’s and master’s students (top 10\%) from previous years of the course.
To ensure transparency, consistency, and minimize subjectivity in grading, we implemented a rigorous multi-stage process: graders first drafted a grading scheme, which was reviewed by course lecturers; this was followed by a first round of grading and a second-round review conducted by PhD-level graders.
After grading was completed, we revealed our experiment to all graders and asked for their permission to use the results. All graders gave permission.

\paragraph{AI-solution generation.}
For each exercise, a member in the teaching team generated solutions using both GPT-4o and {\o1}.
This person did not participate in any grading.
Since the problem statements are well-structured and clearly worded, no prompt engineering was applied.
We used only text-based prompts, without incorporating vision inputs or attachments. Additionally, as the course lecture notes were not provided, the LLMs had to rely solely on prior knowledge from their pre-training to solve the problems.
However, an exception was made for exercises that explicitly required the use of certain results from the lecture notes. 
In such cases, we explicitly supplied the relevant content to ensure the model had access to the necessary information and asked the model to summarize the results to make sure it understood correctly.
The detailed prompts used in this experiment are provided in Table \ref{tbl:prompt}.

\begin{table*}[t]
\caption{Prompt patterns.}
\label{tbl:prompt}
\vskip 0.15in
\begin{center}
\begin{small}
\begin{tabular}{>{\centering\arraybackslash}p{2.5cm} p{13cm}}
\toprule
\textbf{Types} & \textbf{Prompts} \\
\midrule
Problem-Solving Request  & 
``Please try to solve the following problem. Note that there are three questions in total. Generate your answer in LaTeX codes. Your solution should be logically consistent and contain necessary steps and explanations toward the final answers.\\
& \textsc{$\langle$LaTeX source code of the problem statement$\rangle$}''\\
&``Now consider the following problem. There are 4 subquestions in total. Please take the hints into consideration and do not forget the instructions about generating your solution in LaTeX codes.'' \\
\midrule
Reference to Lecture Notes & 
''Please summarize the following Theorem 4.6 to me.'' 

\textsc{$\langle$LaTeX source code of the theorem statement$\rangle$}\\
\midrule
Formatting Request    & 
``Please directly output LaTeX codes so that I can copy.'' 

``Can you clearly state your algorithm and analyze it (still in latex codes)?''\\
\bottomrule
\end{tabular}
\end{small}
\end{center}
\vskip -0.1in
\end{table*}

For each exercise and for each GPT model, we repeated the query process three times and selected the best response for submission.
We observed that AI-generated solutions exhibit distinct patterns and formatting, making them easily distinguishable from student submissions. To address this,  we made \textit{minimal edits} to remove obvious AI-specific artifacts while preserving the content and reasoning of the solutions. These adjustments helped prevent graders from unintentionally identifying AI-generated responses based on formatting alone. More specifically, our  edits include:
\begin{enumerate}
    \item Removing repeated problem statements, as GPT often restates the question before providing an answer;
    \item Eliminating excessive headings and subheadings, since GPT frequently structures responses with multiple hierarchical levels;
    \item Omitting unnecessary thought processes, as GPT sometimes explicitly outlines its reasoning before presenting the actual solution;
    \item Reformatting bullet points into standard paragraphs to make the writing style more natural and human-like.\footnote{Exceptions are made when bullet points are a natural way to present the content.}
\end{enumerate}

In our experiment, we found that only minimal edits to AI-generated responses were needed to support effective blind grading. 
 In proof-based mathematical problems, student submissions naturally vary in structure and formatting due to differences in reasoning style and experience. 
 This natural diversity in student submissions reduces the likelihood that AI-generated solutions stand out, even if with slight presentation deviation from typical answers.

To provide a point of comparison for typical formatting, in the Appendix, we include master solutions alongside the AI-generated responses.  Master solutions are drafted by course lecturers and refined by graders during the grading process.

\subsection{Highlights of the Experiment Settings}

We emphasize several unique aspects of our experimental setup that set it apart from other studies.

\paragraph{High-difficulty benchmark.}
Correctly solving these exercises is highly challenging. Students are given two weeks to complete four exercises, with full access to all course materials (e.g., lecture notes, blackboard notes, past exercises and solutions). Additionally, though we encouraged students to rely on course content and their own reasoning, we cannot prevent students from using the Internet and possibly AI tools.

Successfully solving these problems requires more than just background knowledge of existing theorems, algorithms, and analytical tools. It demands creativity, the ability to generalize known methods to novel problems, deep mathematical thinking, and rigorous mathematical proof construction.

While extensive research has explored AI's capabilities in structured problem-solving, such as coding tasks \cite{yeadon2024comparison} and multiple-choice questions \cite{rein2023gpqa}, there is limited investigation into AI’s performance on complex and proof-based problems. These take-home examinations thus serve as a strong benchmark for evaluating GPT’s ability to generate correct, logically consistent, and well-structured responses to expert-level questions.

\paragraph{Fine-grained expert solution evaluation.}
Our solution evaluation is pure human-based and in a fine-grained pattern.
Each exercise resembles a simplified research problem, and their solutions share similarities with mini research papers. 
While we provide master solutions after student attempts have been handed in, there is a broad range of acceptable answers, as students may use different notations, methodologies, and proof techniques. Consequently, grading requires a case-by-case analysis to differentiate valid alternative approaches from incorrect ones.

This meticulous grading process evaluates logical consistency and mathematical rigor, providing a robust measure of AI’s reliability in generating high-quality solutions.
Moreover, graders provided detailed comments and feedback on each solution, enabling a fine-grained analysis of the error types in AI-generated responses. This level of detail sets our study apart from others, allowing for deeper insights into the specific limitations and strengths of AI in academic problem-solving.

\paragraph{Realistic educational condition.}
Our experiment is conducted in a realistic educational setting, closely mirroring how students might use AI models for academic problem-solving.
GPT receives the exact problem statements given to students, without prompting using additional context or explanations.
Besides, AI-generated solutions are minimally edited to match a student-like writing style, reducing potential bias from graders recognizing AI-generated responses.
By simulating real-world student interactions with AI, this experimental setup provides a more accurate and practical proxy for assessing AI’s capabilities in a higher education context and its potential impact on academic learning.

\section{Results}
\label{sect:results}

In this section, we start with a coarse-grained analysis of score distributions, followed by a fine-grained examination of error patterns based on graders' comments. Additionally, we conduct supplementary experiments using exercises from previous years to further validate our findings.

Appendices \ref{sect:appendix_1} and \ref{sect:appendix_2} contain our full dataset, including all 8 exercises, master solutions, responses generated by GPT-4o and {\o1}, as well as the corresponding grader feedback and comments.

\subsection{Score Comparison}

\begin{figure}\label{fig:scores_pre_exercise}
    \centering
    \includegraphics[width=0.6\linewidth]{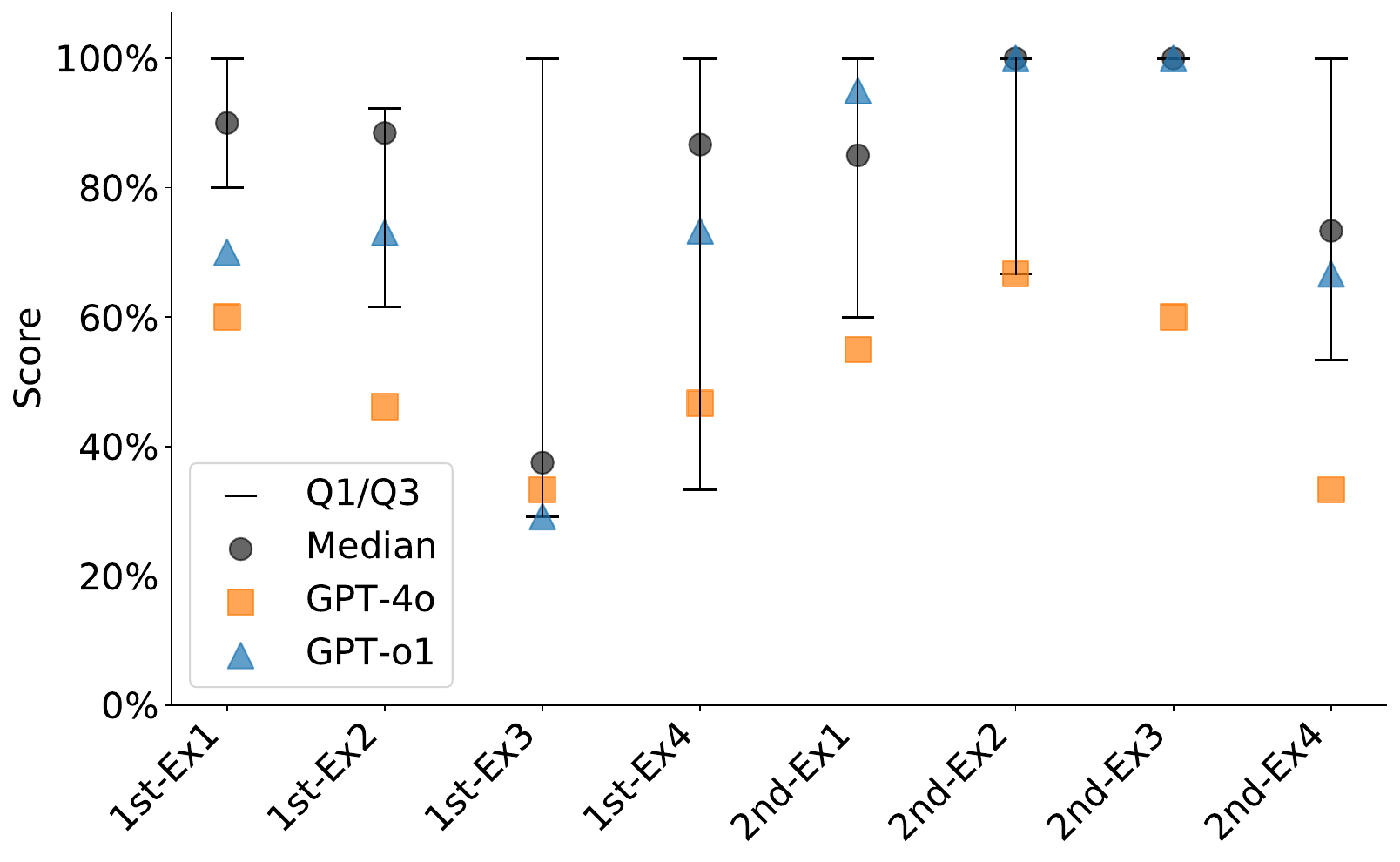}
    \caption{Score per exercise.}
\end{figure}

Figure \ref{fig:scores_pre_exercise} illustrates the score information for eight exercises across two exams, highlighting variability in performance among students and AI models (GPT-4o and {\o1}). The horizontal axis represents the exercises, while the vertical axis indicates the percentage of full scores achieved. 
Each exercise’s score distribution is visualized with three key quantile markers: the .25 quantile/Q1 (lower marker), the .5 quantile/median (middle marker), and the .75 quantile/Q3 (upper marker).

The data reveals significant differences in score distributions across exercises, reflecting differences in difficulty levels. For example, in 1st-Ex3, the range between the .25 and .75 quantile is wide, indicating a high level of difficulty. Students' scores are widely dispersed, with some struggling while others excelled. In contrast, for 2nd-Ex3, over 75\% of students achieved full points, suggesting that this exercise was relatively easy for most students.

GPT-4o consistently performs below or slightly above the .25 quantile. This indicates significant challenges for GPT-4o in tackling the exercises effectively.
Except for 1st-Ex3 where both {\o1} and GPT-4o performed badly, {\o1} constantly outperforms GPT-4o, even by a substantial margin. Notably, in the second exam, {\o1} demonstrates particularly strong performance, exceeding the median and achieving scores close to full marks for 2nd-Ex1, 2nd-Ex2, and 2nd-Ex3.

\paragraph{Exam passing score.}
Aggregating the scores across all exercises in each exam provides insights into the AI's ability to meet the passing criteria. Table \ref{tbl:exam_pass} summarizes the results.

\begin{table}[t]
\caption{Exam Passing Information}
\label{tbl:exam_pass}
\vskip 0.15in
\begin{center}
\begin{small}
\begin{tabular}{lccccc}
\toprule
\textbf{Exam} & \textbf{Passing} & \multicolumn{2}{c}{\textbf{GPT-4o}} & \multicolumn{2}{c}{\textbf{{\o1}}} \\
 \cmidrule(lr){3-4} \cmidrule(lr){5-6} 
& \textbf{score}& \textbf{score} & \textbf{quantile} & \textbf{score} & \textbf{quantile}  \\
\midrule
 First &  50  & 46 & 17.65\% & 62 & 36.03\% \\
\midrule
 Second &  60 & 56 & 6.45\% & 92 & 58.07\%\\
\bottomrule
\end{tabular}
\end{small}
\end{center}
\vskip -0.1in
\end{table}

For the first exam, where the passing score is set at 50\% of the total score across the four exercises.
GPT-4o obtained 46\%, falling short of the passing threshold and outperforming only 17.65\% of students. In contrast, {\o1} achieved 62\%, exceeding the passing threshold and outperforming 36.03\% of students.

For the second exam, given a relatively easier problem set, the passing score is higher, set at 60\% of the total score across the four exercises. GPT-4o obtained 56\%, falling short of the passing threshold and outperforming only 6.45\% of students. In contrast, {\o1} achieved an impressive 92\%, exceeding the passing threshold by a large margin and outperforming 58.07\% of students.

This analysis highlights the clear superiority of {\o1} over GPT-4o in both of the two exams and demonstrates the variability in their relative performances depending on the difficulty of the exercises and the context of the exam.

\subsection{Error Types}
Through the above score comparison, we conclude {\o1} outperforms GPT-4o. In particular, {\o1} has the capability to pass the university-level course with a relatively large margin, while GPT-4o slightly falls below the passing score. 
In this part, we conduct a fine-grained analysis of each exercise. Based on the graders' feedback and comments, we draw some interesting error patterns and findings, and deep dive into the problem solving ability of AI tools.

We categorize the errors identified by graders into three major types:
\begin{enumerate}
\item \textbf{Unjustified Claims}: Arguments that appear plausible but lack sufficient rigor or proper justification.
\item \textbf{Misleading Claims}: Incorrect or invalid arguments presented as valid reasoning.
\item \textbf{Mathematical Errors}: These include basic arithmetic or algebraic mistakes, incorrect application of theorems, failure to address edge cases, misuse of technical terms, and related issues.
\end{enumerate}

The error types identified in solutions generated by GPT-4o and {\o1} are summarized in Table \ref{tbl:error_type}.
In Appendix \ref{sect:appendix_1}, \ref{sect:appendix_2}, we highlight the error types and graders' feedback for each AI-generated solution in red.
\begin{table*}[t]
\caption{Error types per exercise.}
\label{tbl:error_type}
\vskip 0.15in
\begin{center}
\begin{small}
\resizebox{\textwidth}{!}{%
\begin{tabular}{lcccccc}
\toprule
\textbf{Exercise} & \multicolumn{2}{c}{\textbf{Unjustified Claims}} & \multicolumn{2}{c}{\textbf{Misleading Claims}} & \multicolumn{2}{c}{\textbf{Mathematical Errors}}  \\
\cmidrule(lr){2-3} \cmidrule(lr){4-5} \cmidrule(lr){6-7}
& \textbf{GPT-4o} & \textbf{{\o1}} & \textbf{GPT-4o} & \textbf{{\o1}} & \textbf{GPT-4o} & \textbf{{\o1}} \\
\midrule
1st-Ex1: Random Binary Search Trees (\ref{sect:appendix_1_ex1}) & $\times$ &  & &  & $\times$ & $\times$ \\
\midrule
1st-Ex2: Number of Spanning Trees (\ref{sect:appendix_1_ex2}) &  $\times$ &  & $\times$ &  & $\times$ & $\times$ \\
\midrule
1st-Ex3: Point Location (\ref{sect:appendix_1_ex3}) & $\times$  &  & $\times$ & $\times$ &  &  \\
\midrule
1st-Ex4: Dynamic Counting (\ref{sect:appendix_1_ex4}) &  $\times$ &  $\times$ & $\times$ &  & $\times$ & $\times$ \\
\midrule
2nd-Ex1: Page Ranking (\ref{sect:appendix_2_ex1}) & $\times$ & $\times$ & $\times$ &  & $\times$ & \\
\midrule
2nd-Ex2: NOT-OR Circuits (\ref{sect:appendix_2_ex2}) &  $\times$ &  &  $\times$ &  &  & \\
\midrule
2nd-Ex3: Pizza Slicing (\ref{sect:appendix_2_ex3}) &  &  & $\times$ &  &  & \\
\midrule
2nd-Ex4: Shortest Cycle (\ref{sect:appendix_2_ex4}) & $\times$ & $\times$ &  &  & $\times$ & \\
\bottomrule
\end{tabular}
}  
\end{small}
\end{center}
\vskip -0.1in
\end{table*}

When analyzing the error patterns of GPT-4o and {\o1}, we observe clear differences in their strengths and weaknesses across multiple categories of errors.

\paragraph{Unjustified Claims.}
One of the most prominent issues with GPT-4o is its frequent tendency to make unjustified claims, which occurs in seven out of eight exercises. This suggests that GPT-4o often generates plausible-sounding but ultimately unsupported reasoning. In contrast, {\o1} exhibits this error significantly less frequently, appearing in only three exercises.

A key reason for this discrepancy is that {\o1} produces longer, more structured, and more detailed responses compared to GPT-4o. While verbosity does not necessarily equate to correctness, it appears to help {\o1} provide better justification and avoid making claims without supporting arguments.

This pattern is particularly evident in 1st-Ex1 (Appendix \ref{sect:appendix_1_ex1}), where GPT-4o derives a complex recursive formula but fails to explicitly show the intermediate steps needed to reach the final result. In contrast, {\o1} not only arrives at the correct solution but also presents a clear, step-by-step derivation, demonstrating better logical structuring and justification.

\paragraph{Misleading Claims.}
Another major issue with GPT-4o is its tendency to make misleading claims, which appear significantly more frequently compared to {\o1}. These errors typically involve incorrect arguments presented in a confident manner, making them particularly problematic for grading, as they may appear superficially correct but lack validity upon closer inspection.
For instance, GPT-4o generates large portions of text that are logically incoherent or irrelevant. For example, in subquestion (c) of 1st-Ex2 (Appendix \ref{sect:appendix_1_ex2}), subquestion (c) of 1st-Ex4 (Appendix \ref{sect:appendix_1_ex4}), and 2nd-Ex3 (Appendix \ref{sect:appendix_2_ex3}), GPT-4o produces lengthy parts that fail to meaningfully contribute to solving the problem.

However, {\o1} is not entirely free from this issue. In 1st-Ex3 (Appendix \ref{sect:appendix_1_ex3}), both GPT-4o and {\o1} adopt the same incorrect approach. The significant underperformance compared to students might be AI models' lack of familiarity with the course content compared to the students.
Moreover, the inability to reason effectively about geometric concepts may also explain this outcome. According to the master solution of 1st-Ex3 (Point Location), it requires spatial intuition and geometric reasoning—areas where current LLMs tend to struggle. As a result, both models exhibit a tendency to make false claims in their responses when attempting to reason through geometry-based problems.

\paragraph{Mathematical Errors.}
Both GPT-4o and {\o1} are susceptible to mathematical errors, but GPT-4o commits these errors more frequently and with greater severity. This suggests that GPT-4o may be less precise in its numerical and algebraic reasoning, leading to more fundamental mistakes in problem-solving.

In contrast, {\o1}'s mathematical errors tend to be more localized and less impactful on the overall correctness of its solutions. For instance, in question (b) of 1st-Ex2 (Appendix \ref{sect:appendix_1_ex2}), {\o1} misinterprets notation, leading to minor computational errors. However, despite these small mistakes, the overall reasoning and logical approach remain intact, meaning the solution still demonstrates a good understanding of the underlying concepts. This is in stark contrast to GPT-4o, which sometimes fails in a more fundamental way by misapplying mathematical principles altogether.

\subsection{Supplementary Experiments} 
Through score comparison and analysis on the error types of GPT-4o and {\o1}, we find that {\o1} outperforms GPT-4o in overall performance, and significantly makes fewer unjustified claims and misleading claims. 
We have obtained some additional insights.

Course instructors often use the strategy of camouflaging existing problems to prevent students from directly finding solutions online when designing new questions.

For example, in 2nd-Ex2 (Appendix \ref{sect:appendix_2_ex2}), students were tasked with analyzing a linear programming technique used in designing an approximation algorithm for the Boolean satisfiability problem (CNF-SAT). To obscure its connection to well-documented materials, the instructor reframed the problem in the context of circuit analysis, substituting key terms with alternative terminology. Specifically, ``CNF formula'' became ``NOT-OR circuit'', ``literal'' was replaced with ``input wire'', ``clause'' became ``intermediate wire'', and ``number of satisfied clauses'' was termed ``output current''.

Despite these deliberate modifications, {\o1} successfully identified the underlying mathematical structure and provided a strong solution, whereas {\4o} struggled. 
One possible interpretation of this is that {\o1} is more adept at leveraging pre-existing knowledge from publicly available sources, potentially explaining its superior performance on problems with well-documented solutions.

\paragraph{Experiment design.}
Given the apparent correlation between {\o1}'s stronger performance and the availability of similar solutions on the open web, we conducted an additional experiment to gather further evidence or refute this hypothesis. We applied the same ``choose best of three'' naive-prompting procedure to solve an exercise from the previous year’s exams. This exercise was sourced from a book \cite{LP09}, where it was incorrectly posed both in the main text and in the book's errata, and no solution was provided. The authors also attempted to find a solution online without success. This strongly suggests that the solution to this exercise was not included in OpenAI's training data, making it effectively ``new'' for both {\4o} and {\o1}. 

\paragraph{Results and hypothesis.}
The text of the additional question, master solution, and AI-generated solutions can be found in Appendix \ref{sect:extended_data}. When confronted with subquestion (b), {\o1} generates text that, at first glance, appears well-reasoned and coherent but ultimately lacks logical consistency. The response consists of fragments of seemingly correct arguments that do not form a valid explanation, giving the impression of an attempt to justify an answer without actually understanding the solution.

We hypothesize that {\o1}’s strong problem-solving performance may partly arise from exposure to similar solutions available online. While we do not have conclusive empirical evidence to support this hypothesis, our results represent a small but meaningful step toward understanding the true nature of LLMs’ problem-solving abilities: whether they stem from genuine reasoning or from leveraging correlations and latent patterns acquired during training.
This remains a fundamental and open research question in the study of language models.

\section{Discussion}
\label{sect:discussion}
In this section, we share grader reflections after revealing the GPT experiment to them and discuss potential implications for university educators.
We also discuss several limitations of our study.

\subsection{Grader Reflections}
After completing the experiment, we informed the graders about the study, and several shared their experiences evaluating AI-generated solutions. Surprisingly, except for the grader of 1st-Ex4 (Appendix \ref{sect:appendix_1_ex4}), who initially expressed suspicion while grading, all other graders stated they could not distinguish AI-generated responses from student submissions, even after being told which ones were AI-generated. This underscores the effectiveness of our minimal editing approach in making AI-written solutions to resemble student writing styles.

More broadly, graders consistently identified ``misleading claims'' as a major challenge in evaluating AI-generated free-response answers, particularly for complex reasoning tasks. Unlike clearly incorrect responses, misleading claims often appear well-structured and confident, making them deceptively plausible. This forced graders to actively disprove incorrect but superficially logical arguments, which was significantly more time-consuming and cognitively demanding than simply verifying a correct solution.

\paragraph{Implications for university educators.}

Given the strong performance of LLMs on university-level assignments, we recommend gradually reducing the weight of take-home exams and supplementing them with in-class assessments. In the future, take-home assignments may be more effective when used as \textit{formative} rather than \textit{summative} assessments.

Additionally, it is important to adopt AI-aware evaluation strategies that emphasize transparent and well-justified reasoning. For instance, our analysis of AI-generated responses reveals that LLMs frequently make unjustified or misleading claims. Grading rubrics can be adapted to penalize these types of errors more heavily, helping to distinguish genuine human reasoning from AI-generated artifacts.

The growing capabilities of LLMs also pose new challenges for question design. Simple ``camouflaging'' of existing problems is no longer sufficient to prevent high-performing LLMs from generating plausible answers. Instructors may need to craft questions that target areas where LLMs still struggle, such as geometric or spatial reasoning and multi-step logical analysis.

\subsection{Limitations}
\paragraph{Dataset size.}
Our dataset is relatively small, consisting of 8 exercises with 20 subquestions. Given the limited sample size, the variance of AI-generated solutions may be high, which could introduce some uncertainty in the observed patterns.

However, we emphasize that our dataset is of exceptionally high quality, requiring substantial effort in both exercise design and solution evaluation. The careful construction of this dataset makes it a valuable resource for evaluating LLMs in an academic setting, and we believe sharing it for further experimentation will contribute meaningfully to the field.

\paragraph{Non-penalized errors.}
The grading schemes used in this study may not capture all errors made by GPT models. In particular, for ``unjustified claims'', graders did not always deduct points when such reasoning issues arose, as they were not explicitly penalized under the grading scheme. Instead, graders provided written comments highlighting these issues. As a result, there may be a slight discrepancy between the numerical scores and the fine-grained error analysis.

Despite this, the grading scheme was applied consistently across all student and AI-generated submissions, ensuring that the comparative evaluation between students, GPT-4o, and {\o1} remains reliable. The overall findings still provide meaningful insights into AI performance in solving complex academic problems.

\paragraph{Potential underestimation of LLMs' capabilities.}

When generating AI solutions, we deliberately use naive prompts—simple, zero-shot inputs without advanced prompt engineering—to avoid human bias. 
While this approach may not fully activate the reasoning capabilities of LLMs, it aligns with the intended purpose of our benchmark: not to probe the upper limits of LLM performance, but to establish a realistic lower bound of their effectiveness in educational contexts under minimal prompting.

Additionally, since the evaluated tasks were take-home assignments, students may have had access to AI tools and potentially interacted with them using more advanced strategies, such as chain-of-thought (CoT) prompting, which can enhance performance. While this may have inflated student scores, it does not undermine the validity of our assessment of LLM performance. On the contrary, it strengthens our conclusions about {\o1}’s capabilities: its ability to achieve comparable or even superior scores in such a setting highlights its problem-solving ability and underscores the need for reassessing how student performance is evaluated in the presence of powerful LLMs.

\section{Conclusion}
\label{sect:conclusion}

This study assessed the performance of {\4o} and {\o1} in university-level theoretical computer science assignments. AI-generated solutions were produced using naive prompting and evaluated under blind grading conditions alongside over 100 student submissions from two separate take-home exams in algorithm theory.

Our analysis reveals that {\o1} consistently surpasses {\4o} in accuracy and coherence. A detailed examination of their respective strengths and weaknesses suggests that {\o1} may be leveraging memorized or pattern-matched content from existing sources, enabling it to provide more structured and correct answers in specific cases. Conversely, {\4o} exhibits a tendency to make unjustified claims, highlighting a key limitation in its reasoning and verification processes.

An intriguing question remains regarding {\o1}’s ability to replicate its success in scenarios beyond coursework, for instance, in research settings where problems may be novel, unsolved, and potentially unsolvable. While the model achieved a passing grade on both exams, it is unclear whether it would maintain this level of performance when faced with entirely new challenges requiring original problem-solving.

Another critical consideration is the evaluation process itself—specifically, the implications of having a passing threshold that an automated tool can meet. This raises important questions about the standards used to assess student performance and the evolving role of AI in education.

These findings contribute to a deeper understanding of LLMs capabilities in academic settings and underscore the importance of designing evaluation methods that account for both their strengths and limitations. Future research should explore the extent to which these models rely on memorization versus true generalization and develop strategies to enhance their ability to construct well-reasoned and verifiable arguments.

\printbibliography

@article{nikolic2024chatgpt,
  title={ChatGPT, Copilot, Gemini, SciSpace and Wolfram versus higher education assessments: an updated multi-institutional study of the academic integrity impacts of Generative Artificial Intelligence (GenAI) on assessment, teaching and learning in engineering},
  author={Nikolic, Sasha and Sandison, Carolyn and Haque, Rezwanul and Daniel, Scott and Grundy, Sarah and Belkina, Marina and Lyden, Sarah and Hassan, Ghulam M and Neal, Peter},
  journal={Australasian journal of engineering education},
  volume={29},
  number={2},
  pages={126--153},
  year={2024},
  publisher={Taylor \& Francis}
}

@article{stribling2024model,
  title={The model student: GPT-4 performance on graduate biomedical science exams},
  author={Stribling, Daniel and Xia, Yuxing and Amer, Maha K and Graim, Kiley S and Mulligan, Connie J and Renne, Rolf},
  journal={Scientific Reports},
  volume={14},
  number={1},
  pages={5670},
  year={2024},
  publisher={Nature Publishing Group UK London}
}

@article{susnjak2024chatgpt,
  title={ChatGPT: The end of online exam integrity?},
  author={Susnjak, Teo and McIntosh, Timothy R},
  journal={Education Sciences},
  volume={14},
  number={6},
  pages={656},
  year={2024},
  publisher={MDPI}
}

@online{CKBCJKPTHNE21,
  title = {Training {{Verifiers}} to {{Solve Math Word Problems}}},
  author = {Cobbe, Karl and Kosaraju, Vineet and Bavarian, Mohammad and Chen, Mark and Jun, Heewoo and Kaiser, Lukasz and Plappert, Matthias and Tworek, Jerry and Hilton, Jacob and Nakano, Reiichiro and Hesse, Christopher and Schulman, John},
  date = {2021-11-18},
  eprint = {2110.14168},
  eprinttype = {arXiv},
  eprintclass = {cs},
  doi = {10.48550/arXiv.2110.14168},
  url = {http://arxiv.org/abs/2110.14168},
  urldate = {2025-01-27},
  pubstate = {prepublished}
}

@online{LZGSGNBYZ24,
  title = {Can {{OpenAI}} O1 Outperform Humans in Higher-Order Cognitive Thinking?},
  author = {Latif, Ehsan and Zhou, Yifan and Guo, Shuchen and Shi, Lehong and Gao, Yizhu and Nyaaba, Matthew and Bewerdorff, Arne and Yang, Xiantong and Zhai, Xiaoming},
  date = {2024-12-07},
  eprint = {2412.05753},
  eprinttype = {arXiv},
  eprintclass = {cs},
  doi = {10.48550/arXiv.2412.05753},
  url = {http://arxiv.org/abs/2412.05753},
  urldate = {2025-01-27},
  pubstate = {prepublished}
}

@article{RWSPWG24,
  title = {Bob or {{Bot}}: {{Exploring ChatGPT}}'s {{Answers}} to {{University Computer Science Assessment}}},
  shorttitle = {Bob or {{Bot}}},
  author = {Richards, Mike and Waugh, Kevin and Slaymaker, Mark and Petre, Marian and Woodthorpe, John and Gooch, Daniel},
  date = {2024-03-31},
  journaltitle = {ACM Transactions on Computing Education},
  shortjournal = {ACM Trans. Comput. Educ.},
  volume = {24},
  number = {1},
  pages = {1--32},
  issn = {1946-6226},
  doi = {10.1145/3633287},
  url = {https://dl.acm.org/doi/10.1145/3633287},
  urldate = {2025-01-27},
  langid = {english}
}

@article{yeadon2024comparison,
  title={A comparison of human, GPT-3.5, and GPT-4 performance in a university-level coding course},
  author={Yeadon, Will and Peach, Alex and Testrow, Craig},
  journal={Scientific Reports},
  volume={14},
  number={1},
  pages={23285},
  year={2024},
  publisher={Nature Publishing Group UK London}
}

@article{bayer2024chatgpt,
  title={ChatGPT-4o, ChatGPT-4 and Google Gemini are Compared with Students: A Study in Higher Education.},
  author={Bayer, Harun and Araci, Fazilet G{\"u}l Ince and G{\"u}rkan, G{\"u}lsah},
  journal={International Journal of Technology in Education and Science},
  volume={8},
  number={4},
  pages={627--644},
  year={2024},
  publisher={ERIC}
}

@article{kipp2024gpt,
  title={From GPT-3.5 to GPT-4. o: A Leap in AI’s Medical Exam Performance},
  author={Kipp, Markus},
  journal={Information},
  volume={15},
  number={9},
  pages={543},
  year={2024},
  publisher={MDPI}
}

@article{rein2023gpqa,
  title={Gpqa: A graduate-level google-proof q\&a benchmark},
  author={Rein, David and Hou, Betty Li and Stickland, Asa Cooper and Petty, Jackson and Pang, Richard Yuanzhe and Dirani, Julien and Michael, Julian and Bowman, Samuel R},
  journal={arXiv preprint arXiv:2311.12022},
  year={2023}
}

@article{YYZSGCN24,
  title = {Tree of Thoughts: {{Deliberate}} Problem Solving with Large Language Models},
  shorttitle = {Tree of Thoughts},
  author = {Yao, Shunyu and Yu, Dian and Zhao, Jeffrey and Shafran, Izhak and Griffiths, Tom and Cao, Yuan and Narasimhan, Karthik},
  date = {2024},
  journaltitle = {Advances in Neural Information Processing Systems},
  volume = {36},
  url = {https://proceedings.neurips.cc/paper_files/paper/2023/hash/271db9922b8d1f4dd7aaef84ed5ac703-Abstract-Conference.html},
  urldate = {2025-01-28}
}

@online{OAAAAAAAAAE24,
  title = {{{GPT-4 Technical Report}}},
  author = {OpenAI}, 
  date = {2024-03-04},
  eprint = {2303.08774},
  eprinttype = {arXiv},
  eprintclass = {cs},
  doi = {10.48550/arXiv.2303.08774},
  url = {http://arxiv.org/abs/2303.08774},
  urldate = {2025-01-29},
  pubstate = {prepublished}
}

@article{herbold2023large,
  title={A large-scale comparison of human-written versus ChatGPT-generated essays},
  author={Herbold, Steffen and Hautli-Janisz, Annette and Heuer, Ute and Kikteva, Zlata and Trautsch, Alexander},
  journal={Scientific reports},
  volume={13},
  number={1},
  pages={18617},
  year={2023},
  publisher={Nature Publishing Group UK London}
}

@article{brin2023comparing,
  title={Comparing ChatGPT and GPT-4 performance in USMLE soft skill assessments},
  author={Brin, Dana and Sorin, Vera and Vaid, Akhil and Soroush, Ali and Glicksberg, Benjamin S and Charney, Alexander W and Nadkarni, Girish and Klang, Eyal},
  journal={Scientific Reports},
  volume={13},
  number={1},
  pages={16492},
  year={2023},
  publisher={Nature Publishing Group UK London}
}

@article{vazquez2023chatgpt,
  title={ChatGPT: The brightest student in the class},
  author={V{\'a}zquez-Cano, Esteban and Ramirez-Hurtado, Jose M and Saez-Lopez, Jose M and Lopez-Meneses, Eloy},
  journal={Thinking Skills and Creativity},
  volume={49},
  pages={101380},
  year={2023},
  publisher={Elsevier}
}

@article{vidal2024evaluation,
  title={Evaluation of the Performance of State-of-the-Art Large Language Models (LLMs) in Solving Math Word Problems},
  author={Vidal, Jonathan},
  journal={Available at SSRN 4902960},
  year={2024}
}

@online{CCEKSST18,
  title = {Think You Have {{Solved Question Answering}}? {{Try ARC}}, the {{AI2 Reasoning Challenge}}},
  shorttitle = {Think You Have {{Solved Question Answering}}?},
  author = {Clark, Peter and Cowhey, Isaac and Etzioni, Oren and Khot, Tushar and Sabharwal, Ashish and Schoenick, Carissa and Tafjord, Oyvind},
  date = {2018-03-14},
  eprint = {1803.05457},
  eprinttype = {arXiv},
  eprintclass = {cs},
  doi = {10.48550/arXiv.1803.05457},
  url = {http://arxiv.org/abs/1803.05457},
  urldate = {2025-01-29},
  pubstate = {prepublished}
}

@online{CTJYPKEBJBE21,
  title = {Evaluating {{Large Language Models Trained}} on {{Code}}},
  author = {Chen, Mark and Tworek, Jerry and Jun, Heewoo and Yuan, Qiming and Pinto, Henrique Ponde de Oliveira and Kaplan, Jared and Edwards, Harri and Burda, Yuri and Joseph, Nicholas and Brockman, Greg and Ray, Alex and Puri, Raul and Krueger, Gretchen and Petrov, Michael and Khlaaf, Heidy and Sastry, Girish and Mishkin, Pamela and Chan, Brooke and Gray, Scott and Ryder, Nick and Pavlov, Mikhail and Power, Alethea and Kaiser, Lukasz and Bavarian, Mohammad and Winter, Clemens and Tillet, Philippe and Such, Felipe Petroski and Cummings, Dave and Plappert, Matthias and Chantzis, Fotios and Barnes, Elizabeth and Herbert-Voss, Ariel and Guss, William Hebgen and Nichol, Alex and Paino, Alex and Tezak, Nikolas and Tang, Jie and Babuschkin, Igor and Balaji, Suchir and Jain, Shantanu and Saunders, William and Hesse, Christopher and Carr, Andrew N. and Leike, Jan and Achiam, Josh and Misra, Vedant and Morikawa, Evan and Radford, Alec and Knight, Matthew and Brundage, Miles and Murati, Mira and Mayer, Katie and Welinder, Peter and McGrew, Bob and Amodei, Dario and McCandlish, Sam and Sutskever, Ilya and Zaremba, Wojciech},
  date = {2021-07-14},
  eprint = {2107.03374},
  eprinttype = {arXiv},
  eprintclass = {cs},
  doi = {10.48550/arXiv.2107.03374},
  url = {http://arxiv.org/abs/2107.03374},
  urldate = {2025-01-29},
  pubstate = {prepublished}
}

@inproceedings{DWDSSG19,
  title = {{{DROP}}: {{A Reading Comprehension Benchmark Requiring Discrete Reasoning Over Paragraphs}}},
  shorttitle = {{{DROP}}},
  booktitle = {Proceedings of the 2019 {{Conference}} of the {{North American Chapter}} of the {{Association}} for {{Computational Linguistics}}: {{Human Language Technologies}}, {{Volume}} 1 ({{Long}} and {{Short Papers}})},
  author = {Dua, Dheeru and Wang, Yizhong and Dasigi, Pradeep and Stanovsky, Gabriel and Singh, Sameer and Gardner, Matt},
  date = {2019},
  pages = {2368--2378},
  url = {https://aclanthology.org/N19-1246/},
  urldate = {2025-01-29}
}

@inproceedings{HBBZMSS,
  title = {Measuring {{Massive Multitask Language Understanding}}},
  booktitle = {International {{Conference}} on {{Learning Representations}}},
  author = {Hendrycks, Dan and Burns, Collin and Basart, Steven and Zou, Andy and Mazeika, Mantas and Song, Dawn and Steinhardt, Jacob},
  url = {https://openreview.net/forum?id=d7KBjmI3GmQ},
  urldate = {2025-01-29}
}

@online{LDJZZZZD25,
  title = {Search-O1: {{Agentic Search-Enhanced Large Reasoning Models}}},
  shorttitle = {Search-O1},
  author = {Li, Xiaoxi and Dong, Guanting and Jin, Jiajie and Zhang, Yuyao and Zhou, Yujia and Zhu, Yutao and Zhang, Peitian and Dou, Zhicheng},
  date = {2025-01-09},
  eprint = {2501.05366},
  eprinttype = {arXiv},
  eprintclass = {cs},
  doi = {10.48550/arXiv.2501.05366},
  url = {http://arxiv.org/abs/2501.05366},
  urldate = {2025-01-29},
  pubstate = {prepublished}
}

@article{SBBC21,
  title = {{{WinoGrande}}: An Adversarial Winograd Schema Challenge at Scale},
  shorttitle = {{{WinoGrande}}},
  author = {Sakaguchi, Keisuke and Bras, Ronan Le and Bhagavatula, Chandra and Choi, Yejin},
  date = {2021-09},
  journaltitle = {Communications of the ACM},
  shortjournal = {Commun. ACM},
  volume = {64},
  number = {9},
  pages = {99--106},
  issn = {0001-0782, 1557-7317},
  doi = {10.1145/3474381},
  url = {https://dl.acm.org/doi/10.1145/3474381},
  urldate = {2025-01-29},
  langid = {english}
}

@inproceedings{ZHBFC19,
  title = {{{HellaSwag}}: {{Can}} a {{Machine Really Finish Your Sentence}}?},
  shorttitle = {{{HellaSwag}}},
  booktitle = {Proceedings of the 57th {{Annual Meeting}} of the {{Association}} for {{Computational Linguistics}}},
  author = {Zellers, Rowan and Holtzman, Ari and Bisk, Yonatan and Farhadi, Ali and Choi, Yejin},
  date = {2019},
  pages = {4791--4800},
  url = {https://aclanthology.org/P19-1472/},
  urldate = {2025-01-29}
}

@book{LP09,
  title = {Matching Theory},
  author = {Lovász, László and Plummer, Michael D.},
  date = {2009},
  volume = {367},
  publisher = {American Mathematical Soc.},
  url = {https://books.google.com/books?hl=en&lr=&id=OaoJBAAAQBAJ&oi=fnd&pg=PR7&dq=Matching+Theory+-+L%C3%A1szl%C3%B3+Lov%C3%A1sz,+M.+D.+Plumme&ots=45UOn0ptAc&sig=7lpaP9C92Nb3uQFV3vZebVKKzho},
  urldate = {2025-01-29}
}

@software{Ochatgpt2024,
  author = "{OpenAI}",
  title = "{ChatGPT: Optimizing Language Models for Dialogue}",
  year = {2024},
  url = {https://openai.com/chatgpt},
}

@software{gemini2024,
  author = "{Google DeepMind}",
  title = "{Gemini AI}",
  year = {2024},
  url = {https://deepmind.google/technologies/gemini/}
}

@software{copilot2024,
  author = "{GitHub}",
  title = "{GitHub Copilot: Your AI Pair Programmer}",
  year = {2024},
  url = {https://github.com/features/copilot}
}

@software{scispace2024,
  author = "{SciSpace}",
  title = "{SciSpace: AI-Powered Research Platform}",
  year = {2024},
  url = {https://www.scispace.com/}
}

@article{CWWWYZCYWWE24,
  title = {A {{Survey}} on {{Evaluation}} of {{Large Language Models}}},
  author = {Chang, Yupeng and Wang, Xu and Wang, Jindong and Wu, Yuan and Yang, Linyi and Zhu, Kaijie and Chen, Hao and Yi, Xiaoyuan and Wang, Cunxiang and Wang, Yidong and Ye, Wei and Zhang, Yue and Chang, Yi and Yu, Philip S. and Yang, Qiang and Xie, Xing},
  date = {2024-06-30},
  journaltitle = {ACM Transactions on Intelligent Systems and Technology},
  shortjournal = {ACM Trans. Intell. Syst. Technol.},
  volume = {15},
  number = {3},
  pages = {1--45},
  issn = {2157-6904, 2157-6912},
  doi = {10.1145/3641289},
  url = {https://dl.acm.org/doi/10.1145/3641289},
  urldate = {2025-01-27},
  langid = {english}
}

@online{HBKABTSS21,
  title = {Measuring {{Mathematical Problem Solving With}} the {{MATH Dataset}}},
  author = {Hendrycks, Dan and Burns, Collin and Kadavath, Saurav and Arora, Akul and Basart, Steven and Tang, Eric and Song, Dawn and Steinhardt, Jacob},
  date = {2021-11-08},
  eprint = {2103.03874},
  eprinttype = {arXiv},
  eprintclass = {cs},
  doi = {10.48550/arXiv.2103.03874},
  url = {http://arxiv.org/abs/2103.03874},
  urldate = {2025-01-27},
  pubstate = {prepublished}
}

@article{yousefzadeh2024lean,
  title={A Lean Dataset for International Math Olympiad: Small Steps towards Writing Math Proofs for Hard Problems},
  author={Yousefzadeh, Roozbeh and Cao, Xuenan and Ospanov, Azim},
  journal={arXiv preprint arXiv:2411.18872},
  year={2024}
}

@article{suzgun2022challenging,
  title={Challenging big-bench tasks and whether chain-of-thought can solve them},
  author={Suzgun, Mirac and Scales, Nathan and Sch{\"a}rli, Nathanael and Gehrmann, Sebastian and Tay, Yi and Chung, Hyung Won and Chowdhery, Aakanksha and Le, Quoc V and Chi, Ed H and Zhou, Denny and others},
  journal={arXiv preprint arXiv:2210.09261},
  year={2022}
}

@inproceedings{gao2023pal,
  title={Pal: Program-aided language models},
  author={Gao, Luyu and Madaan, Aman and Zhou, Shuyan and Alon, Uri and Liu, Pengfei and Yang, Yiming and Callan, Jamie and Neubig, Graham},
  booktitle={International Conference on Machine Learning},
  pages={10764--10799},
  year={2023},
  organization={PMLR}
}

\newpage
\appendix

\newcommand{\func}{\textsc{SampleAndFactor}}
\newcommand{\funcs}{\func\xspace}
\newcommand{\fA}{\mathcal{A}}
\newcommand{\expec}[2]  {{\,{\mathbf{ E}}_{#1}\!\left[#2\right]}}
\newcommand{\N}{\mathbb{N}}
\newcommand{\setof}[1]{\{ #1 \}}
\newcommand{\var}[1]    {{\,{\mathrm{var}}\!\left[#1\right]}}
\newcommand{\prob}[2]  {{\,{\mathbf{Pr}}_{#1}\!\left[#2\right]}}

\section{First Exam: Exercises and AI-generated Solutions}
\label{sect:appendix_1}

\subsection{Exercise 1: Random Binary Search Tree}
\label{sect:appendix_1_ex1}
Let $n\in \N$, the keys $2,4,8,\dots,2^n$ are inserted in a uniformly random order in a binary search tree.
Let $k_1,k_2,\dots,k_\ell$ denote the value of the keys on the right spine of the tree, i.e. on the path from the root to the biggest key. In this exercise, we want to compute $X_n = \expec{}{k_1+k_2,\dots+k_\ell}$. Note that $\ell$ is a random number, $k_1$ is the value in the root node, and $k_\ell =2^n$.

\begin{longtable}{|p{0.44\textwidth}|p{0.5\textwidth}|}
\multicolumn{2}{p{0.96\textwidth}}{\textbf{(a)} Write $X_n$ as a function of $n$ and $X_{i}$ for $i=1,2,\dots,n-1$.} \\ 

\multicolumn{2}{p{0.96\textwidth}}{
\vspace{.1em}
\textbf{Master Solution} 

Working out the two cases for $n=1$ yields $X_1 = 2$. As usual, we condition on the rank of the root node. Here, the key observation is that if the root has rank $i$, the right subtree of the root is a random tree with the keys $2,4,\dots,2^{n-i}$ where all the keys are multiplied by $2^{i}$. Further notice that if the root has rank $n$, it is the only node on the right spine of the tree and we get $k_1 = 2^n$ for our value. Hence, for $n\geq 2$, we can write the following recurrence.
        \begin{eqnarray*}
            X_n & = & \frac{2^n}{n} + \frac{1}{n} \sum_{i=1}^{n-1} (2^i + 2^i X_{n-i}) \\
            & = & \frac{1}{n}\sum_{i=1}^{n} 2^i + \frac{1}{n}  \sum_{i=1}^{n-1}  2^{n-i} X_{i} \\
            & = & \frac{2^{n+1} - 2}{n} + \frac{1}{n}  \sum_{i=1}^{n-1}  2^{n-i} X_{i}
        \end{eqnarray*}\ } \\ 
\hline
\textbf{GPT-4o} & \textbf{{\o1}} \\ \hline
\endfirsthead
\hline
\textbf{GPT-4o} & \textbf{{\o1}} \\ \hline
\endhead
\hline \multicolumn{2}{r}{\textit{Continued on the next page...}} \\ \hline
\endfoot
\hline
\endlastfoot
We start by considering how the tree is constructed when the keys \( 2, 4, 8, \dots, 2^n \) are inserted in random order. The first step is choosing the root key \( k_1 \), which is randomly selected from the set \( \{2, 4, 8, \dots, 2^n\} \). 
    Let \( k_1 = 2^i \), meaning the root key is \( 2^i \). The keys on the right spine of the tree will be those greater than \( 2^i \), which are \( \{ 2^{i+1}, 2^{i+2}, \dots, 2^n \} \). The right spine thus contains the keys from this subset, and its expected sum is \( X_{n-i} \), the expected sum of the keys on the right spine of a tree with \( n-i \) keys. \textcolor{red}{[\textbf{Unjustified Claims}: Too vague on conditional probability]}

    &
    
     To express \(X_n\) as a function of \(n\) and \(X_i\) for \(i = 1, 2, \dots, n-1\), we consider the process of building the BST by inserting the keys in random order. Each key \(2^i\) has an equal probability \(\frac{1}{n}\) of being the root of the tree.
    If the root is \(2^n\) (the largest key), then the right spine consists only of the root, so \(X_n = 2^n\) in this case. If the root is \(2^i\) for some \(i < n\), then the expected sum of the keys on the right spine is \(2^i + X_{n - i}\), where \(X_{n - i}\) represents the expected sum of the keys on the right spine in the subtree containing the keys \(\{2^{i+1}, 2^{i+2}, \dots, 2^n\}\).
     Therefore, we can write \(X_n\) as:
    \[
    X_n = \frac{1}{n} \cdot 2^n + \sum_{i=1}^{n-1} \frac{1}{n} \left( 2^i + X_{n - i} \right).
    \]
   
    \\
    The probability of selecting \( 2^i \) as the root is \( \frac{1}{n} \). Hence, the expected value \( X_n \) is the sum of the root value \( 2^i \) and the expected sum of the right spine, weighted by the probability of selecting \( 2^i \) as the root.
    
    Thus, the recurrence relation for \( X_n \) is:
    \[
    X_n = \frac{1}{n} \sum_{i=1}^{n} \left( 2^i + X_{n-i} \right)
    \]
    \textcolor{red}{[\textbf{Mathematical Errors}: Missing a factor $2^i$, which should be $X_n = \frac{1}{n} \sum_{i=1}^{n} \left( 2^i + 2^i X_{n-i} \right)$]}
    
    This expression relates \( X_n \) to \( X_{n-i} \) for all \( i = 1, 2, \dots, n-1 \).  
    
    &

    Simplifying the expression, we combine the terms:
    \[
    X_n = \frac{1}{n} \left( 2^n + \sum_{i=1}^{n-1} \left( 2^i + X_{n - i} \right) \right).
    \]
    The sum of the powers of 2 is:
    \[
    \sum_{i=1}^{n-1} 2^i = 2^n - 2,
    \]
    
    so the expression becomes:

    \[
    X_n 
    = \frac{1}{n} \left( 2^{n+1} - 2 + \sum_{i=1}^{n-1} X_{n - i} \right).
    \]
    Noting that \(n - i\) ranges from \(1\) to \(n - 1\) as \(i\) ranges from \(n - 1\) to \(1\), we can rewrite the sum as:
    \[
    \sum_{i=1}^{n-1} X_{n - i} = \sum_{j=1}^{n-1} X_j.
    \]
    Thus, the expression simplifies to:
    \[
    X_n = \frac{1}{n} \left( 2^{n+1} - 2 + \sum_{j=1}^{n-1} X_j \right).
    \]
    \textcolor{red}{[\textbf{Mathematical Errors}: Missing $2^{n-j}$ inside summation, which should be $\sum_{j=1}^{n-1} 2^{n-j} X_j$.]}
    \end{longtable}

\begin{longtable}{|p{0.4\textwidth}|p{0.54\textwidth}|}
\multicolumn{2}{p{0.96\textwidth}}{\textbf{(b)} Write $X_n$ as a function of only $n$ and $X_{n-1}$.} \\ 
\multicolumn{2}{p{0.96\textwidth}}{
\vspace{.1em}
\textbf{Master Solution} 

For $n \geq 3$, we get that
        \begin{eqnarray*}
        n X_n  - 2(n-1) X_{n-1}= & 2 +   2 X_{n-1}
        \end{eqnarray*}
        Or equivalently,
        \[
        X_n = \begin{cases}
        2 X_{n-1} + \frac{2}{n} & \mathrm{if}~n\geq 3 \\
        5 & \mathrm{if}~n=2 \\
        2 & \mathrm{if}~n=1
        \end{cases}
        \]
}\\
\hline
\textbf{GPT-4o} & \textbf{{\o1}} \\ \hline
\endfirsthead
\hline
\textbf{GPT-4o} & \textbf{{\o1}} \\ \hline
\endhead
\hline \multicolumn{2}{r}{\textit{Continued on the next page...}} \\ \hline
\endfoot
\hline
\endlastfoot
We now express \( X_n \) only in terms of \( n \) and \( X_{n-1} \). To do this, notice that when the root key is chosen to be \( 2^n \), the right spine contains no further keys, and the contribution to the sum is just \( 2^n \).
    For any other root key \( 2^i \) with \( i < n \), the tree still has a right spine, and the expected sum of the right spine is \( X_{n-1} \), since the structure of the right subtree is essentially the same as in the case with \( n-1 \) keys.
    The recurrence relation can thus be simplified to:
    \[
    X_n = \frac{1}{n} \left( 2^n + (n-1) X_{n-1} \right)
    \]
    This relation uses only the previous value \( X_{n-1} \) and the current root value \( 2^n \), simplifying the recursion.

    &

    We aim to express \(X_n\) in terms of only \(n\) and \(X_{n-1}\). From the result of Part (a), we have:
\[
X_n = \frac{1}{n} \left( 2^{n+1} - 2 + \sum_{i=1}^{n-1} X_i \right).
\]
Multiplying both sides by \(n\), we get:
\[
n X_n = 2^{n+1} - 2 + \sum_{i=1}^{n-1} X_i.
\]
Similarly, for \(X_{n-1}\):
\[
(n - 1) X_{n-1} = 2^{n} - 2 + \sum_{i=1}^{n-2} X_i.
\]
Subtracting the second equation from the first, we obtain:
\[
n X_n - (n - 1) X_{n-1} = 2^{n+1} - 2^{n} + X_{n-1}.
\]
Simplifying, we get:
\[
n X_n - n X_{n-1} + X_{n-1} = 2^{n} + X_{n-1},
\]
which further simplifies to:
\[
n (X_n - X_{n-1}) = 2^{n}.
\]
Therefore, we can express \(X_n\) as:
\[
X_n = X_{n-1} + \frac{2^{n}}{n}.
\]
\end{longtable}

\begin{longtable}{|p{0.42\textwidth}|p{0.52\textwidth}|}
\multicolumn{2}{p{0.96\textwidth}}{\textbf{(c)} Write $X_n$ in a closed form. The result can contain sums and products but should not be written as a function of some $X_i$ for $i\in \N$.} \\ 
\multicolumn{2}{p{0.96\textwidth}}{
\vspace{.1em}
\textbf{Master Solution} 

The solution is $$X_n = \sum_{i=1}^{n} \frac{2^{n-i+1}}{i}, $$ for all $n\geq 1.$ Which can be shown by simple induction over $n$. The base case $X_1$ is easily checked. Let $m\geq 2$ be arbitrary but fixed, we get:\[
    X_m = 2X_{m-1} + \frac{2}{m} = 2\sum_{i=1}^{m-1}\frac{2^{(m-1)-i+1}}{i} + \frac{2}{m} = \sum_{i=1}^{m-1}\frac{2^{m-i+1}}{i} + \frac{2}{m} = \sum_{i=1}^{m}\frac{2^{m-i+1}}{i},
    \] where we used the induction hypothesis in the second step.\
}\\
\hline
\textbf{GPT-4o} & \textbf{{\o1}} \\ \hline
\endfirsthead
\hline
\textbf{GPT-4o} & \textbf{{\o1}} \\ \hline
\endhead
\hline \multicolumn{2}{r}{\textit{Continued on the next page...}} \\ \hline
\endfoot
\hline
\endlastfoot
To obtain a closed form for \( X_n \), we solve the recurrence relation from part (b). First, let's unroll the recurrence:
    \[
    X_n = \frac{1}{n} \left( 2^n + (n-1) X_{n-1} \right)
    \]
    \[
    X_{n-1} = \frac{1}{n-1} \left( 2^{n-1} + (n-2) X_{n-2} \right)
    \]
    Substituting \( X_{n-1} \) into the equation for \( X_n \), we get:
    \begin{align*}
    X_n = \frac{2^n + (n-1) \frac{1}{n-1} \left( 2^{n-1} + (n-2) X_{n-2} \right)}{n}
    \end{align*}
    Continuing this process, we eventually find:
    \[
    X_n = \sum_{i=1}^{n} \frac{2^i}{i}
    \]
    This expression gives the expected sum of the keys on the right spine of the binary search tree as a sum of terms involving powers of 2 divided by their index.
    & 
    To find a closed-form expression for \(X_n\) without involving \(X_i\), we use the recursive formula from Part (b):
    \[
    X_n = X_{n-1} + \frac{2^{n}}{n}.
    \]
    By repeatedly applying this recurrence, we get:
    
    \parbox{0.4\textwidth}{
    \begin{align*}
    X_n &= X_{n-1} + \frac{2^{n}}{n} \\
    &= \left( X_{n-2} + \frac{2^{n-1}}{n-1} \right) + \frac{2^{n}}{n} \\
    &= X_{n-2} + \frac{2^{n-1}}{n-1} + \frac{2^{n}}{n} \\
    &\vdots \\
    &= X_1 + \sum_{k=2}^{n} \frac{2^{k}}{k}.
    \end{align*}
    }
    
    Since \(X_1 = 2\) (as there is only one key, \(2\), and it forms the root), we have:
    \[
    X_n = 2 + \sum_{k=2}^{n} \frac{2^{k}}{k} = \sum_{k=1}^{n} \frac{2^{k}}{k}.
    \]
    Thus, the closed-form expression for \(X_n\) is:
    \[
    X_n = \sum_{k=1}^{n} \frac{2^{k}}{k}.
    \]
\end{longtable}

\subsection{Exercise 2: Number of Spanning Trees in a Complete Graph}
\label{sect:appendix_1_ex2}
The objective of the exercise is to find the number of distinct spanning trees in a complete graph with $n$ vertices labeled with $n$ distinct labels. An edge between vertices $i$ and $j$ is labeled $\{i,j\}$, and two spanning trees are considered different if they contain any differently labeled edges. For example, on the graph with $n = 3$ labeled vertices, we have 3 distinct spanning trees.
\\
Let $K_n = (V, E)$ be the complete graph with $n$ vertices and suppose that the vertices are labeled $1,2,\dots, n$.
Let $R \subseteq V : |R| = k$ be a set of $k$ vertices that are fixed (for simplicity, you can think of $R = \{1,2,\dots,k\}$).
Denote with $T_{n,k}$ the number of (labeled) forests on $\setof{1, \dots,n}$ consisting of $k$ trees whose roots are the vertices in $R$. 
Note that $T_{n,k}$ does not depend on $R$ but only on its size and by fixing the roots, the number of possible forests decreases, eg. $T_{3,2} = 2$ as the vertices $1$ and $2$ cannot be in the same tree (using $R=\{1,2\}$). For coherence, define also $T_{0,0} =1 $ and $ T_{n,0} = 0$ for $n>0$.

\begin{longtable}{|p{0.4\textwidth}|p{0.55\textwidth}|}
\multicolumn{2}{p{0.96\textwidth}}{\textbf{(a)} Compute $T_{n,n}$ for $n\geq 1$.} \\ 
\multicolumn{2}{p{0.96\textwidth}}{
\vspace{.1em}
\textbf{Master Solution}

If we have a forest with $n$ vertices and $n$ trees, the only possibility is that each tree consists of only one vertex. This proves that $T_{n,n} = 1$ for all $n\geq 1$.
}\\
\hline
\textbf{GPT-4o} & \textbf{{\o1}} \\ \hline
\endfirsthead
\hline
\textbf{GPT-4o} & \textbf{{\o1}} \\ \hline
\endhead
\hline \multicolumn{2}{r}{\textit{Continued on the next page...}} \\ \hline
\endfoot
\hline
\endlastfoot
First, we observe that \( T_{n,n} \) represents the number of forests in a graph with \( n \) vertices, where each vertex is its own tree (i.e., there are no edges between the vertices). Since each vertex is isolated, the number of such forests is trivially 1, as the only possible configuration is having \( n \) isolated vertices with no edges.
    Thus:
    \[
    T_{n,n} = 1 \quad \text{for all } n \geq 1.
    \]
    & 
When \( k = n \), the set \( R \) includes all vertices (\( R = V \)). Each vertex is a root of its own tree, and since the forest must consist of \( n \) trees rooted at these vertices, there are no edges connecting any vertices. Therefore, there is only one such forest, which consists of \( n \) isolated vertices.
Thus,
\[
T_{n,n} = 1, \quad \text{for all } n \geq 1.
\]
\end{longtable}

\begin{longtable}{|p{0.4\textwidth}|p{0.55\textwidth}|}
\multicolumn{2}{p{0.96\textwidth}}{\textbf{(b)} Show that for all $1\leq k\leq n$,
$$ T_{n,k} = \sum_{i=0}^{n-k} \binom{n-k}{i} T_{n-1,k-1+i} .$$ \emph{Hint: It might be helpful to use different sets $R$ in recursive cases.}} \\ 
\multicolumn{2}{p{0.96\textwidth}}{
\vspace{.1em}
\textbf{Master Solution}

Suppose that we have a forest that consists of $k$ trees. Let $v$ be the root of one of the trees and suppose we remove $v$. $v$ can be directly connected to $i$ vertices for $i=0, 1,\dots, n-k$ (at least $k-1$ are not connected to $v$ because they are the roots of the other trees). If $v$ is directly connected to $i$ vertices, after $v$ has been removed, the tree with root $v$ splits into $i$ trees and there are $\binom{n-k}{i}$ possible combinations to connect $i$ of the $n-k$ non-root nodes to $v$.
Furthermore, we have obtained a forest with $(n-1)$ vertices and $(k-1+i)$ connected components.
Note that we are not double-counting or not-counting any of the combinations here. 
{\emph{Showing this can be made explicit, but we do not expect the argument below to be stated in this much detail to get a full mark.}}

{Below follows an explicit exposition of this not double-counting and not-counting.
Towards this goal, let's define the following sets of graphs
\begin{align*}
    \mathcal T_{n,k} &:= \big\{\big([n], E(F) \cup \{ \{r,n\} : r \in R^*\}\big) : T \in \mathcal T_{n-1,k-1+i}, R^* \in \binom {[n-k]}i, 0 \le i \le n-k \big\} \\
    \mathcal F_{n,k} &:= \big\{\text{ all rooted labelled forests on } n \text{ nodes and } k \text{ roots with } R=\{n-k+1,\dots,n\}\big\}
\end{align*}
In the following, we compare $\mathcal T_{n,k}$ (the forests that we construct, i.e. $T_{n,k} = |\mathcal T_{n,k}|$) with $\mathcal F_{n,k}$ (the actual forests that we should obtain) and want to show that they are equal.
We use $[n] = \{1,2,\dots,n\}$ and $\binom {[n]}i :=$ are all $i$-element subsets of $[n]$ as in A\&W.
\vspace{.1em}
1. No double-counting: Consider $T_1 \not= T_2 \in \mathcal T_{n,k}$. Note that the tree which contains the node $n$ (it is the root of this tree) is different for every element as the edges that go out of $n$ are different, i.e. the $R^*$ (the $i$ direct neighbours of $n$) are different.
}
}\\

\multicolumn{2}{p{0.96\textwidth}}{\vspace{.1em}
2. No not-counting, i.e. $F \in \mathcal F_{n,k} \implies F \in \mathcal T_{n,k}$: Assume towards a contradiction that $\exists F: F \in \mathcal F_{n,k} \wedge F \notin \mathcal T_{n,k}$. 
Consider vertex $n$ in $F$ ($n$ is a root as $n \in R$). It has to have $0 \le \deg(n) \le n-k$ ($n$ can only be connected to non-root nodes, namely to elements in $[n-k]$). But we consider all of these possibilities in the definition of $\mathcal T_{n,k}$, namely all subsets of $[n-k]$ are in $\bigcup_{0\le i \le n-k} \binom {[n-k]}i$. 
Now consider $F\backslash n$. It has to have $k-1+\deg(n)$ many trees. Assuming as our induction hypothesis that the above holds for $n'<n$, we know inductively that $F\backslash n \in \mathcal F_{n-1,k-1+\deg(n)} \implies F\backslash n \in \mathcal T_{n-1,k-1+\deg(n)}$.} \\ 

\multicolumn{2}{p{0.96\textwidth}}{\vspace{.1em}
3. All are valid, i.e. $F \in \mathcal T_{n,k} \implies F \in \mathcal F_{n,k}$: We can argue inductively that all our $F \in \mathcal T_{n,k}$ are valid. We can assume that this is the case inductively for $E(T)$, the edges of $T \in \mathcal T_{n-1,k-1+i}$. There, our $R' = R\backslash\{n\} \cup R^*$. The only edges that we now add are from root $n$ to root nodes $r \in R^*$; because root nodes are not connected with each other, we don't form any cycles. Furthermore, we add $1$ node and $i=|R'|$ edges, thus going from $k-1+i$ to $k-1+i+1-i=k$ connected components. Our $R=\{n-k+1,\dots,n\}$, so because we connected $n$ only to nodes in $[n-k]$, we indeed have rooted trees with roots in $R$. Note that in the induction step, we actually use a different $R$; being precise, we should actually define $\mathcal T_{n,k,R}$ and $\mathcal F_{n,k,R}$ and prove the properties for all $R: |R|=k$, but this is just a permutation of the indices and thus a trivial generalization of the above.} \\ 
\end{longtable}

\begin{longtable}{|p{0.4\textwidth}|p{0.55\textwidth}|}
\hline
\textbf{GPT-4o} & \textbf{{\o1}} \\ \hline
\endfirsthead
\hline
\textbf{GPT-4o} & \textbf{{\o1}} \\ \hline
\endhead
\hline \multicolumn{2}{r}{\textit{Continued on the next page...}} \\ \hline
\endfoot
\hline
\endlastfoot
We fix one vertex (say, vertex 1), and consider different ways in which this vertex can be part of a forest: Vertex 1 could either belong to one of the \( k \) roots or a subtree of any other root.
    If vertex 1 belongs to a tree rooted in \( R \) (with \( |R| = k \)), we recursively consider the remaining \( n-1 \) vertices and form \( k-1+i \) disjoint trees. \textcolor{red}{[\textbf{Unjustified Claims}: Why \( k-1+i \)? Not explained]} The number of ways to choose \( i \) vertices from the remaining \( n-k \) vertices to add to the forest is given by \( \binom{n-k}{i} \). Thus, the recurrence relation is:
    \[
    T_{n,k} = \sum_{i=0}^{n-k} \binom{n-k}{i} T_{n-1,k-1+i}
    \]
    \textcolor{red}{[\textbf{Unjustified Claims}: Why \( n-k \)? Not explained.]}
    & 
We aim to establish a recursive relationship for \( T_{n,k} \) by considering the role of vertex \( n \) in the forests.
Let \( S = V \setminus R \) be the set of non-root vertices, with \( |S| = n - k \). We consider how vertex \( n \) can be connected in the forest:
Since \( R \) is fixed and \( n \) may or may not be in \( R \), but for consistency, we can assume \( R \) does not contain \( n \) (if \( n \in R \), the argument is similar). \textcolor{red}{[\textbf{Mathematical Errors}: No, but you can assume without loss of generality.]}
Vertex \( n \) is not a root (since \( n \notin R \)), so it must be connected to some other vertex in the forest.
We consider all possible ways to connect vertex \( n \):

\begin{enumerate}[leftmargin=1.2em]
    \item Connecting to a Root in \( R \):
   Vertex \( n \) connects directly to one of the \( k \) roots in \( R \). This does not change the number of trees in the forest. However, since \( R \) is fixed, we cannot choose which roots are in \( R \), so we proceed differently.
   \item Connecting to Non-root Vertices in \( S \):
   Vertex \( n \) connects to one or more \textcolor{red}{[\textbf{Mathematical Errors}: Should be ``zero or more''.]} vertices in \( S \). By connecting vertex \( n \) to some vertices in \( S \), we might merge their subtrees, possibly reducing the number of trees in the forest.
\end{enumerate}

We proceed by considering the number of ways vertex \( n \) can be connected to vertices in \( S \):
\begin{enumerate}[leftmargin=1.2em]
    \item Choose \( i \) vertices from \( S \) to connect to \( n \): There are \( \binom{n - k}{i} \) \textcolor{red}{[\textbf{Mathematical Errors}: Should be \( \binom{n - k-1}{i} \).]} ways to choose \( i \) vertices from \( S \).
    \item Resulting Number of Trees:
  By connecting \( n \) to these \( i \) vertices, we potentially reduce the number of trees by \( i \). However, since \( n \) is not a root, the total number of roots remains \( k \).
  So after removing vertex \( n \), the forest consists of \( n - 1 \) vertices and \( k - 1 + i \) trees (since \( i \) subtrees may have been merged) \textcolor{red}{[\textbf{Mathematical Errors}: $i$ in our formula \( k-1+i \) is $\text{deg}(n)$ but $i+1$ is $\text{deg}(n)$ in your solution.]}.
\end{enumerate}

\\\hline
    
    &
Therefore, for each \( i \) from \( 0 \) to \( n - k \), the number of forests is:
\[
\binom{n - k}{i} \, T_{n - 1, k - 1 + i}.
\]

Summing over all possible \( i \), we obtain:
\[
T_{n,k} = \sum_{i=0}^{n - k} \binom{n - k}{i} \, T_{n - 1, k - 1 + i}.
\]
\end{longtable}

\begin{longtable}{|p{0.38\textwidth}|p{0.53\textwidth}|}
\multicolumn{2}{p{0.96\textwidth}}{\textbf{(c)} Using the recursive expression above, prove that 
\[
T_{n,k} = kn^{n-k-1} .
\]
} \\ 
\multicolumn{2}{p{0.96\textwidth}}{
\textbf{Master Solution}

We prove the result via induction.
The \textbf{Base Case} $T_{1,k}$ is trivial with $T_{1,0} = 0$ by definition and $T_{1,1} = 1$ by (a) and the Base Case $T_{n,0}=0$ by definition also satisfies our hypothesis. 

The \textbf{Induction Hypothesis} is given as in the exercise as $T_{a,b} = a b^{a-b-1}\  \forall\ b \le a<n$. 

So let's focus on the \textbf{Induction Step}:

\parbox{0.96\textwidth}{
\begin{eqnarray*}
T_{n,k} & = &  \sum_{i=0}^{n-k} \binom{n-k}{i} T_{n-1,k-1+i} \\
& \overset{I.H.}= &  \sum_{i=0}^{n-k} \binom{n-k}{i} (k-1+i) (n-1)^{n-1 - k -i} \\
 & = & \sum_{i'=0}^{n-k} \binom{n-k}{i'} (n-1-i') (n-1)^{i'-1} \\
& = & \sum_{i'=0}^{n-k} \binom{n-k}{i'} (n-1)^{i'} - \sum_{i'=1}^{n-k} \binom{n-k}{i'} i' (n-1)^{i'-1} \\
 & = & (1+(n-1))^{n-k} - (n-k) \sum_{i'=1}^{n-k} \binom{n-k-1}{i'-1} (n-1)^{i'-1} \\
 & = & n^{n-k} - (n-k) \sum_{i''=0}^{n-k-1} \binom{n-k-1}{i''} (n-1)^{i''} \\
 & = & n^{n-k} - (n-k)(1+(n-1))^{n-1-k} \\
& = & n^{n-k} - (n-k)n^{n-1-k} \\
& = & k n^{n-1-k}
\end{eqnarray*}
}

Where we used the substitutions $i'=n-k-i$, $i''=i'-1$ and the fact that $\sum_{i=0}^{q} \binom q i (n-1)^i = n^q$ is a special case of the binomial theorem $(a+b)^q$ where $a=1$ and $b=n-1$.
}
\end{longtable}

\begin{longtable}{|p{0.38\textwidth}|p{0.53\textwidth}|}
\hline
\textbf{GPT-4o} & \textbf{{\o1}} \\ \hline
\endfirsthead
\hline
\textbf{GPT-4o} & \textbf{{\o1}} \\ \hline
\endhead
\hline \multicolumn{2}{r}{\textit{Continued on the next page...}} \\ \hline
\endfoot
\hline
\endlastfoot

We will prove this by induction.
    
\textbf{Base Case:} 

Consider the case where \( n = 1 \) and \( k = 1 \): \( T_{1,1} \) represents the number of forests consisting of one tree with one vertex, which is trivially 1. Thus, \( T_{1,1} = 1 \).
    Now, let's check if the formula holds for this case:
    \[
    T_{1,1} = 1 \cdot 1^{1-2} = 1 \cdot 1^{-1} = 1,
    \]
    which is true. Therefore, the base case holds.
    \textcolor{red}{[\textbf{Mathematical Errors}: Missing another base case.]}

\vspace{1em}

\textbf{Inductive Hypothesis:}  

    Assume that for some \( n \geq 1 \), the formula holds for all \( m < n \). That is, for all \( m < n \) and for all \( k \leq m \), we assume:
    \[
    T_{m,k} = k m^{m-k-1}.
    \]
  \textcolor{red}{[\textbf{Misleading Claims}: The rest part is wrong.]}
    This is a weighted sum over the binomial expansion of \( (n-1)^{n-k-1} \). The sum of the binomial coefficients \( \binom{n-k}{i} \) is \( n-k \). Thus, the recurrence simplifies to:
    \[
    T_{n,k} = k (n-1)^{n-k-1} + (n-k)(n-1)^{n-k-1}.
    \]
    Factor out \( (n-1)^{n-k-1} \):
    \begin{align*}
    T_{n,k} = (k + (n-k))(n-1)^{n-k-1} \\= n(n-1)^{n-k-1}.
    \end{align*}
    This matches the desired formula:
    \[
    T_{n,k} = k n^{n-k-1}.
    \]
&

We will prove by induction on \( n \) that:
\[
T_{n,k} = k \, n^{n - k - 1}.
\]
\textbf{Base Case:} For \( n = k \), from part (a), we have:
\[
T_{n,n} = 1 = n \cdot n^{n - n - 1} = n \cdot n^{-1} = 1.
\]
\textcolor{red}{[\textbf{Mathematical Errors}: Missing another base case.]}

\textbf{Inductive Step:} Assume that for all \( m < n \) and all \( 1 \leq k \leq m \), we have:
\[
T_{m,k} = k \, m^{m - k - 1}.
\]
We will show that this holds for \( n \).
Using the recursive formula from part (b):
\[
T_{n,k} = \sum_{i=0}^{n - k} \binom{n - k}{i} \, T_{n - 1, k - 1 + i}.
\]

By the induction hypothesis:
\begin{align*}
T_{n - 1, k - 1 + i} = (k - 1 + i) (n - 1)^{(n - 1) - (k - 1 + i) - 1} \\= (k - 1 + i) (n - 1)^{n - k - i - 1}.
\end{align*}
Substituting back:
\[
T_{n,k} = \sum_{i=0}^{n - k} \binom{n - k}{i} (k - 1 + i) (n - 1)^{n - k - i - 1}.
\]
We can simplify this sum by recognizing that it resembles the binomial theorem. Consider the expression:
\[
S = \sum_{i=0}^{n - k} \binom{n - k}{i} (k - 1 + i) (n - 1)^{n - k - i - 1}.
\]

\\
        
& 
Let's factor out \( (n - 1)^{n - k - 1} \):
\[
S = (n - 1)^{n - k - 1} \sum_{i=0}^{n - k} \binom{n - k}{i} (k - 1 + i) (n - 1)^{-i}.
\]
Now, observe that:
\[
(n - 1)^{-i} = \left( \frac{1}{n - 1} \right)^{i}.
\]
Also, note that:
\[
k - 1 + i = (k - 1) + i.
\]
Therefore, we have:
\[
S = (n - 1)^{n - k - 1} \sum_{i=0}^{n - k} \binom{n - k}{i} \left( (k - 1) + i \right) \left( \frac{1}{n - 1} \right)^{i}.
\]
Split the sum into two parts:
\begin{align*}
S = (n - 1)^{n - k - 1} \left[ (k - 1) \sum_{i=0}^{n - k} \binom{n - k}{i} \left( \frac{1}{n - 1} \right)^{i} \right.\\\left. + \sum_{i=0}^{n - k} \binom{n - k}{i} i \left( \frac{1}{n - 1} \right)^{i} \right].
\end{align*}
The first sum is:
\begin{align*}
    \sum_{i=0}^{n - k} \binom{n - k}{i} \left( \frac{1}{n - 1} \right)^{i} = \left( 1 + \frac{1}{n - 1} \right)^{n - k} \\= \left( \frac{n}{n - 1} \right)^{n - k}.
\end{align*}
The second sum involves \( i \binom{n - k}{i} \left( \frac{1}{n - 1} \right)^{i} \). This sum can be computed by differentiating the generating function. \textcolor{red}{[\textbf{Mathematical Errors}: Should be binomial function.]}
Compute the second sum:

\\

&

\parbox{0.4\textwidth}{
\begin{align*}
&\sum_{i=0}^{n - k} i \binom{n - k}{i} \left( \frac{1}{n - 1} \right)^{i}\\ =& (n - k) \left( \frac{1}{n - 1} \right) \left( 1 + \frac{1}{n - 1} \right)^{n - k - 1} \\=& \frac{n - k}{n - 1} \left( \frac{n}{n - 1} \right)^{n - k - 1}.
\end{align*}
}

Now, combine the two sums:

\parbox{0.4\textwidth}{
\begin{align*}
S =& (n - 1)^{n - k - 1} \left[ (k - 1) \left( \frac{n}{n - 1} \right)^{n - k} \right.
\\ &+ \left.\frac{n - k}{n - 1} \left( \frac{n}{n - 1} \right)^{n - k - 1} \right] \\
=& (n - 1)^{n - k - 1} \left( \frac{n}{n - 1} \right)^{n - k - 1} \cdot\\&\left[ (k - 1) \left( \frac{n}{n - 1} \right) + \frac{n - k}{n - 1} \right] \\
=& n^{n - k - 1} \left[ (k - 1) \left( \frac{n}{n - 1} \right) + \frac{n - k}{n - 1} \right].
\end{align*}
}

Simplify the expression inside the brackets:
\parbox{0.4\textwidth}{
\begin{align*}
(k - 1) \left( \frac{n}{n - 1} \right) &+ \frac{n - k}{n - 1} = \frac{(k - 1) n + n - k}{n - 1} \\
&= \frac{(k - 1) n + n - k}{n - 1} = \frac{n k - n}{n - 1}.
\end{align*}
}

Simplify numerator:
\[
n k - n = n (k - 1).
\]
Thus,
\[
S = n^{n - k - 1} \cdot \frac{n (k - 1)}{n - 1} = n^{n - k - 1} \cdot \left( \frac{n}{n - 1} \cdot (k - 1) \right).
\]
As \( n \to n \), the factor \( \frac{n}{n - 1} \) approaches 1, but we can leave it as is for exact calculation.
However, this suggests that \( S = k \, n^{n - k - 1} \).
Therefore, we conclude:
\[
T_{n,k} = k \, n^{n - k - 1}.
\]
\end{longtable}

\begin{longtable}{|p{0.4\textwidth}|p{0.55\textwidth}|}
\multicolumn{2}{p{0.96\textwidth}}{\textbf{(d)} Use the previous result to deduce the number of different spanning trees in a complete graph with $n$ labelled vertices.} \\ 
\multicolumn{2}{p{0.96\textwidth}}{
\vspace{.1em}
\textbf{Master Solution}

Substitute $k=1$ in the formula above and obtain $T_n = n^{n-2}$. This is known as Cayley's formula.
}\\
\hline
\textbf{GPT-4o} & \textbf{{\o1}} \\ \hline
\endfirsthead
\hline
\textbf{GPT-4o} & \textbf{{\o1}} \\ \hline
\endhead
\hline \multicolumn{2}{r}{\textit{Continued on the next page...}} \\ \hline
\endfoot
\hline
\endlastfoot
We know from (c) that the number of spanning trees in a complete graph with \( n \) labelled vertices is given by:
    \[
    T_{n,1} = n^{n-2}.
    \]

    & 
A spanning tree is a connected acyclic subgraph that includes all \( n \) vertices and has \( n - 1 \) edges. This corresponds to the case where the forest consists of a single tree (\( k = 1 \)).
Using the formula from part (c) with \( k = 1 \):
\[
T_{n,1} = 1 \times n^{n - 1 - 1} = n^{n - 2}.
\]
Therefore, the number of different spanning trees in a complete graph with \( n \) labelled vertices is:
\[
T = n^{n - 2}.
\]
\end{longtable}

\subsection{Exercise 3: Point Location}
\label{sect:appendix_1_ex3}
Given a collection $S$ of $n$ points in the plane and a constant $d>0$, consider the problem of, given a query point $q$, finding all the points in $S$ that are at most at distance $d$ from $q$. Devise a data structure for this problem. In order to get the maximum score, the preprocessing time should be polynomial in the number of points $n$ and if $k$ is the number of points to report, the query should take $O(\log(n)+k)$ operations in expectation.

\vspace{.5em}
\noindent\textbf{Master Solution}

\begin{figure}[h]
    \centering
    \includegraphics[width=0.5\linewidth]{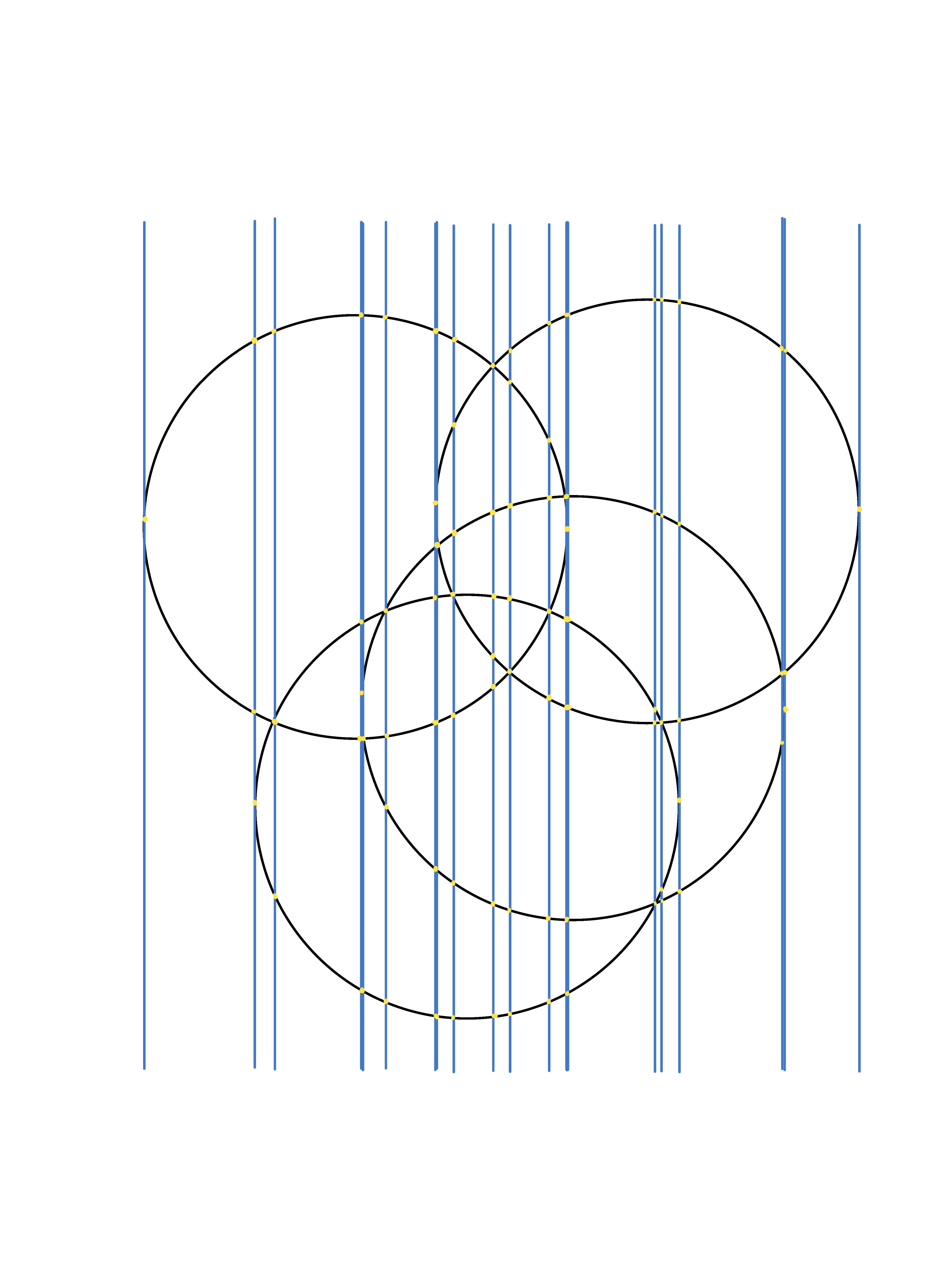}
    \caption{Vertical lines are added at the extreme points of circles and at their intersections with each other. The vertices of the final subdivision are in yellow.}
    \label{fig:enter-label}
\end{figure}

Draw $n$ circles of radius $d$, centered on the input points. The arcs of these circles and the intersection points define a planar graph that tiles the plane. Furthermore, for each tile, all points in the same tile share the same points at a distance of at most $d$. 

At this point, we want to use a strategy somewhat similar to trapezoidal decomposition, where, upon receiving a query point $q$, we find the tile that contains $q$ and report the precomputed answer. To make the subdivision more amenable to strategies like binary search, we’d like to add some structure. It is always possible to add vertical lines to divide the plane further, such that we can determine in $O(\log n)$ time which “vertical strip” contains our query point. Once we have found the correct strip, we still need to locate our point within one of the tiles in that strip. Therefore, we would like these strips to be, in some sense, “nice”.

If we trace vertical lines such that they pass through each leftmost and rightmost point of a circle (i.e., its vertical tangents) and through each intersection point of a circle with another circle, we obtain a decomposition such that each strip (the region between two successive vertical lines) contains different arcs that have one endpoint on the left side of the strip and one endpoint on the right side. Furthermore, the interiors of the arcs do not intersect. In particular, for any $x$-coordinate in an interval associated with a strip, we encounter the arcs in the same order when moving along the $y$-axis. Thus, for each such interval/strip, we can maintain a sorted array of arcs, through which we can perform binary search to locate our query point (we assume that checking whether a point is below an arc can be done in constant time).

It remains to prove an upper bound on the number of vertices in the graph. Each vertex of the graph is either the intersection of a circle with a line or the point where a tangent line touches a circle. Each vertical line intersects at most $2n$ times with the circles. The number of vertical lines is $2n + 2\binom{n}{2}$ (the tangent ones and the ones corresponding to circle-circle intersections), giving an upper bound of $2n(2n + \binom{n}{2}) = O(n^3)$ on the number of vertices. 

\begin{longtable}{|p{0.48\textwidth}|p{0.46\textwidth}|}
\hline
\textbf{GPT-4o} & \textbf{{\o1}} \\ \hline
\endfirsthead
\hline
\textbf{GPT-4o} & \textbf{{\o1}} \\ \hline
\endhead
\hline \multicolumn{2}{r}{\textit{Continued on the next page...}} \\ \hline
\endfoot
\hline
\endlastfoot
\textcolor{red}{[\textbf{Misleading Claims}: Wrong approach.]} 

We propose using a $k$-d tree (k-dimensional tree) to solve this problem. A $k$-d tree is a binary tree that recursively partitions the points based on their coordinates.
The $k$-d tree is built recursively as follows:
\begin{itemize}[leftmargin=1.2em]
    \item Start with the set of points \( S \) and partition it based on one of the coordinates (either $x$ or $y$).
\end{itemize}

    & 
\textcolor{red}{[\textbf{Misleading Claims}: Wrong approach.]} 

We propose using a balanced \textbf{k-d tree} (k-dimensional tree), which is a binary search tree optimized for organizing points in a multidimensional space—in this case, two dimensions.

\vspace{1em}

We build the k-d tree by recursively partitioning the set \( S \) as follows:
At each level of the tree, we choose a splitting dimension.

\\
\begin{itemize}[leftmargin=1.2em]
    \item At the root level, partition the points by the $x$-coordinate: choose the median of the $x$-coordinates as the splitting value, and divide the points into two sets: those with $x$-coordinates smaller than or equal to the median, and those with larger $x$-coordinates.
    \item Recursively repeat this process for the left and right subtrees. Alternate between splitting by $x$-coordinate and $y$-coordinate at successive levels of the tree.
\end{itemize}
\textbf{Time complexity for building:} 
Building the $k$-d tree takes \( O(n \log n) \), because at each level of the recursion, we sort the points and split them, and there are \( \log n \) levels in a balanced tree.
To query for all points within a distance \( d \) from a given point \( q \), we perform a range search on the $k$-d tree:
\begin{itemize}[leftmargin=1.2em]
    \item Start at the root of the tree and compare the query point \( q \) with the splitting coordinate at the current level (either $x$ or $y$).
    \item If the distance from \( q \) to the splitting plane is less than \( d \), recursively search both the left and right subtrees.
    \item If the distance from \( q \) to the splitting plane is greater than \( d \), only search the subtree on the side where \( q \) lies.
    \item At each leaf node, compute the Euclidean distance between the point and \( q \). If this distance is less than or equal to \( d \), report the point.
\end{itemize}
\textbf{Time complexity for querying:} The query time is \( O(\log n + k) \), where \( \log n \) comes from traversing the tree and \( k \) is the number of points reported.
\textcolor{red}{[\textbf{Unjustified Claims}: Not obvious argument.]}

&
\vspace{.2em}
For a two-dimensional plane, we alternate between the \( x \)-coordinate and the \( y \)-coordinate.
We select the median point along the chosen dimension to split the set. This ensures the tree remains balanced, resulting in a depth of \( O(\log n) \).

The points are divided into two subsets: those with coordinates less than or equal to the median and those greater than the median, along the splitting dimension.
This process continues recursively until each leaf node contains a single point or a small number of points.

The preprocessing time to build the k-d tree is \( O(n \log n) \), as sorting is required at each level to find medians.

\vspace{.4em}
\textbf{Query Algorithm}

To find all points within distance \( d \) from a query point \( q \), we perform a range search on the k-d tree:
We start at the root of the tree and compare \( q \)'s coordinate along the splitting dimension with the node's splitting value. 
If the absolute difference between \( q \)'s coordinate and the splitting value is less than or equal to \( d \), both subtrees might contain points within distance \( d \) of \( q \). In this case, we recursively search both subtrees.
If the difference is greater than \( d \), only one subtree needs to be searched—the one on the same side as \( q \).
At each node, if the point stored there is within distance \( d \) from \( q \), we add it to the list of points to report.
The search continues recursively until all relevant subtrees have been explored.

The traversal time through the tree, which is \( O(\log n) \) for a balanced k-d tree. \textcolor{red}{[\textbf{Misleading Claims}: Not true if we go through more branches.]} The time to report the \( k \) points within distance \( d \) of \( q \), which is \( O(k) \), since we check each candidate point individually.
Therefore, the total expected query time is \( O(\log n + k) \).
\end{longtable}

\subsection{Exercise 4: Dynamic Counting}
\label{sect:appendix_1_ex4}
(\emph{Warm-up: Dynamic List Ranking})\\
Suppose you have access to a \emph{static} data structure which does the following: given a collection of $n$ integer-valued elements, it takes $O(n \log n)$ preprocessing time; and after that, given a query element $q$, it returns the number of elements smaller than $q$ in the initial input in $O(\log n)$ time.\footnote{This can be achieved by sorting the elements and then performing binary search} The data structure is static in the sense that if we want to add a new element to the sequence, then we have to spend $O(n \log n)$ time and rebuild the data structure.

We want to use this data structure to solve the following $\emph{dynamic}$ problem. You are given a sequence of $n$ elements $e_1,e_2, \dots, e_n$ one at a time and each time, given a new element $e_k$, you have to tell what is the rank of this element, i.e. how many elements smaller than $e$ are there in the sequence $e_1, e_2, \dots, e_k$. 

\begin{longtable}{|p{0.4\textwidth}|p{0.55\textwidth}|}
\multicolumn{2}{p{0.96\textwidth}}{\textbf{(a)}  Find an algorithm that solves the problem above using the data structure as a black box. After receiving $n$ elements as input, the total amount of time required by the algorithm should be $O(n \log^2 n)$.

Note that you do not have access to an oracle that compares two elements without using the static data structure.

\textit{Hint: Use the data structures of geometrically increasing size.}} \\ 

\multicolumn{2}{p{0.96\textwidth}}{
\vspace{.1em}
\textbf{Master Solution}

When we get a new element, we create a data structure that contains only the new element. Afterwards, if there are 2 data structures containing the same number $2^i$ of elements, we build a new data structure that contains the $2^{i+1}$ elements and discard the two old data structures. Since the size of the data structures increases geometrically, at any moment, we can have at most $\log n$ data structures. Therefore, when a new element comes, we can query all the data structures and report the sum of the rank of the new element in each data structure; this requires $O(\log^2 n)$ time. For $j=1,2,\dots,\log n$, a data structure of size $2^j$ is built after every $2^j$ insertions. So during the insertion of $n$ elements, we build $n/2^j$ times a data structure of size $2^j$. The total time needed by the algorithm is $$ \sum_{j =1}^{\log n} \frac{n}{2^j} 2^{j}j + \sum_{i=1}^n\log^{2} i = O(n \log^2 n) $$.
}
\end{longtable}

\begin{longtable}{|p{0.4\textwidth}|p{0.55\textwidth}|}
\hline
\textbf{GPT-4o} & \textbf{{\o1}} \\ \hline
\endfirsthead
\hline
\textbf{GPT-4o} & \textbf{{\o1}} \\ \hline
\endhead
\hline \multicolumn{2}{r}{\textit{Continued on the next page...}} \\ \hline
\endfoot
\hline
\endlastfoot
We can use the \textbf{doubling trick} to maintain multiple static data structures of geometrically increasing sizes.
    Let \( D_0, D_1, D_2, \dots \) be a sequence of static data structures, where each \( D_i \) can store up to \( 2^i \) elements. Each data structure \( D_i \) allows rank queries in \( O(\log n) \) time, and if a data structure is full, we merge it with a new element to form the next larger data structure.
    For each new element \( e_k \):
    \begin{itemize}[leftmargin=1.2em]
        \item \textbf{Query for rank:} Query all data structures \( D_0, D_1, \dots \) to find how many elements smaller than \( e_k \) are already stored. The query time for each structure is \( O(\log n) \), and there are at most \( \log n \) structures, so the total query time is \( O(\log^2 n) \).
    
        \item \textbf{Insert the element:} If \( D_0 \) is empty, insert \( e_k \) into \( D_0 \). If \( D_0 \) is full, merge it with \( e_k \) and move it to \( D_1 \), and so on. The merging process may propagate up to higher levels of the data structures.
    \end{itemize}
    When two data structures of the same size are full, we merge them and rebuild the larger data structure. Rebuilding a data structure of size \( 2^i \) takes \( O(2^i \log n) \) time. The total time for all merges across the sequence of \( n \) elements is \( O(n \log n) \), as the sizes of the data structures increase geometrically.

    \begin{itemize}[leftmargin=1.2em]
        \item \textbf{Preprocessing:} The total time for merging and rebuilding across all elements is \( O(n \log n) \).
    \end{itemize}

    & 
To solve this problem efficiently, we utilize a strategy that involves maintaining multiple instances of the static data structure over subsets of the input elements. The key idea is to organize the elements into blocks of sizes that are powers of two, enabling us to balance the cost of rebuilding the static data structures with the need to process each new element promptly.

\vspace{.5em}
\textbf{Algorithm Description:}

We maintain \( \lceil \log n \rceil \) static data structures, each corresponding to a block of elements whose sizes are powers of two. Specifically, we organize the sequence of received elements into \textit{runs} of lengths \( 1, 2, 4, 8, \dots \). Each run is associated with a static data structure built over its elements.

When a new element \( e_k \) arrives, we proceed as follows:

1. We create a new run containing \( e_k \) and build a static data structure over this single element. The preprocessing time is \( O(\log 1) = O(1) \).

2. If there is already a run of size \( 1 \), we merge the two runs of size \( 1 \) to form a run of size \( 2 \). We rebuild the static data structure for this new run, which takes \( O(2 \log 2) = O(2) \) time.

3. If there is already a run of size \( 2 \), we merge the new run of size \( 2 \) with the existing one to form a run of size \( 4 \), rebuilding the static data structure in \( O(4 \log 4) = O(8) \) time.

4. We continue this process, merging runs of the same size and rebuilding the corresponding static data structures, until no two runs have the same size.

At each step, we ensure that the runs maintain sizes that are powers of two and that there is at most one run of any given size.

\vspace{.5em}
\textbf{Calculating the Rank of \( e_k \):}

To determine the rank of \( e_k \), we query all the static data structures corresponding to the runs that have been formed so far. Since the runs are disjoint and cover all previously received elements (excluding \( e_k \) if it has been merged into a larger run), we can compute the total number of elements smaller than \( e_k \) by summing the results from each data structure.

\\
\begin{itemize}[leftmargin=1.2em]
        \item \textbf{Query:} Each element query takes \( O(\log^2 n) \), and there are \( n \) elements, so the total query time is \( O(n \log^2 n) \).
        \textcolor{red}{[\textbf{Unjustified Claims}: Correct query time, but why?]}
    \end{itemize}
    Thus, the total time complexity is \( O(n \log^2 n) \), as required.
&

Each query to a static data structure takes \( O(\log n) \) time. Since there are at most \( \lceil \log n \rceil \) runs at any time, the total time to compute the rank of \( e_k \) is \( O(\log n \cdot \log n) = O(\log^2 n) \).

\vspace{.5em}
\textbf{Analysis of Total Time:}

We need to analyze the total time spent on rebuilding the static data structures and processing the queries over all \( n \) elements.
\begin{itemize}[leftmargin=1.2em]
    \item \textbf{Rebuilding Time:} Each element can cause at most \( \lceil \log n \rceil \) merges, and for each merge at level \( i \), the time spent is \( O(2^i \log 2^i) = O(2^i \cdot i) \).

  The total rebuilding time \( T \) can be bounded as:
  \begin{align*}    
  T = \sum_{i=0}^{\lceil \log n \rceil} \left( \frac{n}{2^i} \cdot O(2^i \cdot i) \right) = O\left( n \sum_{i=0}^{\lceil \log n \rceil} i \right)\\ = O(n \log^2 n).
  \end{align*}

  \item \textbf{Query Time:} For each of the \( n \) elements, we perform \( O(\log n) \) queries, each taking \( O(\log n) \) time, resulting in a total query time of \( O(n \log^2 n) \).
\end{itemize}
 
\end{longtable}

\noindent(\emph{Count points below a line dynamically})\\
In this exercise, we devise a \emph{dynamic} version of the data structure for reporting the number of points below a line with nearly the same total update time as the static version. We suppose that we start with an empty data structure. Then, at most $n$ point insertions, at most $n$ point removals, and an arbitrary amount of queries happen in some arbitrary order. 

Our data structure should take $O(\log^2 n)$ time to process each query. Furthermore, the total time used for processing insertions and removals should be $O(n^2 \log n)$, and the total space occupied by the data structure should be $O(n^2)$.

\begin{longtable}{|p{0.4\textwidth}|p{0.55\textwidth}|}
\multicolumn{2}{p{0.96\textwidth}}{\textbf{(b)} Reduce the problem above to the problem of counting the number of lines below a query point when only insertions and queries occur. 
Formally, you are required to demonstrate that, given a data structure capable of counting the number of lines below a query point—supporting both queries and insertions—it is possible to construct a data structure that addresses the problem described above while supporting insertions, deletions, and queries. Additionally, for any insertion of $n$ points into both data structures, the total construction time of the two data structures must be asymptotically equivalent, as should the query and total insertion/deletion times. For a full score, your algorithm must work correctly even when asked to remove points that were not previously inserted. 
} \\ 
\multicolumn{2}{p{0.96\textwidth}}{
\vspace{.1em}
\textbf{Master Solution} 

We can reduce to the problem of counting the number of lines above a point using duality, as seen in the lecture note, it is clear that duality holds also in the dynamic setting. 

    Then, build two identical data structures which don't handle deletions. When you get an insertion, insert the point in the first, when you get a deletion, first check that the point has already been inserted and, if it has, insert the point in the second data structure.\footnote{To check existing insertions, we can use an auxiliary data structure that supports efficient insertion and lookup to record the inserted points. Examples include an AVL tree or a hash table.} When you get a query, query both data structures and report the difference. The time needed for an insertion is $T(\text{\# insertion})$, the time needed for a deletion is $O(\log(k)+T(\text{\# deletion}))$, while the time for a query is $T_{\text{query}}(\text{\# insertion})+ T_{\text{query}}(\text{\# deletion})$, where $T(\cdot)$ and $T_{\text{query}}(\cdot)$ are the insertion and query time of the two data structures, and $k$ is the number of inserted points. 
}
\end{longtable}

\begin{longtable}{|p{0.4\textwidth}|p{0.58\textwidth}|}
\hline
\textbf{GPT-4o} & \textbf{{\o1}} \\ \hline
\endfirsthead
\hline
\textbf{GPT-4o} & \textbf{{\o1}} \\ \hline
\endhead
\hline \multicolumn{2}{r}{\textit{Continued on the next page...}} \\ \hline
\endfoot
\hline
\endlastfoot
We use two data structures:

1. \textbf{Primary Data Structure} \( D \): This is the given black-box data structure, which supports insertions and queries but does not directly support deletions.

2. \textbf{Deletion Marker Set \( M \)}: A secondary data structure (such as a hash set) that keeps track of elements marked for deletion.

\vspace{1em}
We implement the operations as follows.
\begin{itemize}[leftmargin=1.2em]
    \item \textbf{Insertion}: To insert an element \( x \), insert it directly into the primary data structure \( D \).
    \item \textbf{Deletion}: To delete an element \( x \), add \( x \) to the deletion marker set \( M \). The element is not removed from \( D \); instead, \( M \) serves as a filter during queries.
    \item \textbf{Query}: To query the number of elements below a certain value \( q \):
\end{itemize}

&
To achieve this, we utilize a technique that allows us to simulate deletions using the given insertion-only data structure. The key idea is to maintain separate data structures for insertions and deletions and to adjust our query algorithm accordingly.
We maintain two instances of the black-box data structure:
\begin{itemize}[leftmargin=1.2em]
    \item \textbf{Insertion Data Structure} (\( D_{\text{ins}} \)): This data structure handles all insertion operations. It processes all the points that have been inserted into the system.
    
    \item \textbf{Deletion Data Structure} (\( D_{\text{del}} \)): This data structure handles all deletion operations. We treat deletions as insertions into \( D_{\text{del}} \), effectively maintaining the set of points that have been deleted.

When a point is inserted, we insert it into \( D_{\text{ins}} \) in \( O(1) \) time (assuming the insertions are batched or the data structure allows for \( O(1) \) amortized insertion time). 
\textcolor{red}{[\textbf{Unjustified Claims}: Not obvious why \( O(1) \) amortized insertion time.]}
When a point is deleted, we insert it into \( D_{\text{del}} \) in \( O(1) \) time, treating the deletion as an insertion into the deletion data structure.

To answer a query about the number of points below a given line \( \ell \), we proceed as follows: 
We query \( D_{\text{ins}} \) to obtain the total number of inserted points below \( \ell \), denoted \( N_{\text{ins}} \). This takes \( O(Q(n + d)) \) time, as \( D_{\text{ins}} \) contains \( n \) points. 
We query \( D_{\text{del}} \) to obtain the total number of deleted points below \( \ell \), denoted \( N_{\text{del}} \). This also takes \( O(Q(n + d)) \) time, as \( D_{\text{del}} \) contains \( d \) points.
\end{itemize}

\\
\begin{enumerate}[leftmargin=2em]
        \item Perform the query on \( D \) to get the count of all elements below \( q \) (including elements marked for deletion).
        \item Count the number of elements in \( M \) that are below \( q \) and subtract this count from the result of the query on \( D \). This adjustment ensures that deleted elements do not contribute to the final count.
    \end{enumerate}
\begin{itemize}[leftmargin=1.2em]
    \item \textbf{No Requirement for Repeated Insertion/Deletion Handling}: Since we do not need to handle repeated insertions or deletions (such as sequences like (insert \( x \), delete \( x \), insert \( x \))), \( M \) can remain a simple set without additional complexity.
\end{itemize}
\textcolor{red}{[\textbf{Mathematical Errors}: Missing handling the case where we delete an element which has not been inserted yet.]}

We finally analyze the time complexity.
\begin{itemize}[leftmargin=1.2em]
    \item \textbf{Insertion Time}: Each insertion into \( D \) takes \( T(n) \).
    \item \textbf{Deletion Time}: Adding an element to \( M \) takes \( O(1) \), as \( M \) is a hash set.
    \item \textbf{Query Time}: Each query on \( D \) takes \( Q(n) \), plus an additional \( O(\log n) \) time to count deletions in \( M \). This results in an overall query time of \( O(Q(n + d)) \), where \( d \) is the number of deletions.
\end{itemize}
Therefore, given \( n \) insertions and \( d \) deletions, the total time complexity for handling all operations is \( O(T(n + d)) \), with each query taking \( O(Q(n + d)) \), as required.

    & 

\begin{itemize}[leftmargin=1.2em]
\item The actual number of points below \( \ell \) is then calculated as:
   \[
   N = N_{\text{ins}} - N_{\text{del}}.
   \]
\end{itemize}
This approach effectively subtracts the deleted points from the inserted points, giving us the correct count of points below \( \ell \) at the time of the query.
Our algorithm correctly handles deletions of points that were not previously inserted. Since deletions are recorded in \( D_{\text{del}} \) independently of \( D_{\text{ins}} \), the subtraction in the query processing naturally accounts for such cases. If a point was not previously inserted, its inclusion in \( D_{\text{del}} \) does not affect the count of actual points below \( \ell \), as it subtracts zero for that point from the total.
\textcolor{red}{[\textbf{Unjustified Claims}: Too vague reasoning.]}

\vspace{.5em}
\textbf{Time Complexity Analysis:}
\begin{itemize}[leftmargin=1.2em]
    \item \textbf{Insertion and Deletion Time:} Each insertion or deletion operation involves inserting a point into one of the data structures, taking \( O(1) \) amortized time per operation. The total time for \( n \) insertions and \( d \) deletions is \( O(n + d) \). However, since the black-box data structures require rebuilding to accommodate the new points, the total time is \( O(T(n + d)) \).

\item \textbf{Query Time:} Each query involves querying both \( D_{\text{ins}} \) and \( D_{\text{del}} \), each taking \( O(Q(n + d)) \) time. Thus, the total query time is \( O(Q(n + d)) \).
\end{itemize}

\textbf{Space Complexity:}

We maintain two data structures, each storing up to \( n + d \) points. The total space used is \( O(n + d) \).

\vspace{.5em}
\textbf{Edge Cases:}
\begin{itemize}[leftmargin=1.2em]
    \item \textbf{Deletion of Non-Inserted Points:} The algorithm handles deletions of non-inserted points gracefully, as the subtraction during query processing accounts for this without error.

\item \textbf{Repeated Insertions and Deletions:} We are not required to handle repeated insertions and deletions of the same point. Our algorithm assumes that once a point is inserted or deleted, it is not subject to the same operation again.
\end{itemize}

\end{longtable}

\begin{longtable}{|p{0.4\textwidth}|p{0.54\textwidth}|}
\multicolumn{2}{p{0.96\textwidth}}{\textbf{(c)} Devise an algorithm that solves the problem of counting the number of lines below a query point when only insertions and queries occur. In order to get a full score, your algorithm should take $O(\log^2 n)$ time to process each query, the total time used for processing insertions should be $O(n^2 \log n)$, and the total space occupied by the data structure should be $O(n^2)$.

\textit{Hint: Use the strategy from part }(a)\textit{.}} \\ 
\multicolumn{2}{p{0.96\textwidth}}{
\vspace{.1em}
\textbf{Master Solution}

 We use Theorem 3.7 from the lectures as a black-box data structure and we apply the same strategy as in the warm-up.
 
    When we get a new line, we build a data structure with only that line. When we have two data structures with the same size, let's say they contain $2^i$ lines, we build a new data structure that contains $2^{i+1}$ lines and discard the two old data structures.
    When we get a query, we query all the data structures and report the sum of the values we got.
    
    The time needed to process a query is bounded by the number of data structures that we have at each moment times the time to query the biggest data structure, which is $O(\log^2 n)$. The total time for the update after the insertions is $$ \sum_{j=1}^{\log n} \frac{n}{2^j} 2^{2j} \leq O(n^2 \log n).$$
    And the total space occupied is $$\sum_{j=1}^{\log n} 2^{2j} = O(n^2).$$
}
\end{longtable}

\begin{longtable}{|p{0.4\textwidth}|p{0.55\textwidth}|}
\hline
\textbf{GPT-4o} & \textbf{{\o1}} \\ \hline
\endfirsthead
\hline
\textbf{GPT-4o} & \textbf{{\o1}} \\ \hline
\endhead
\hline \multicolumn{2}{r}{\textit{Continued on the next page...}} \\ \hline
\endfoot
\hline
\endlastfoot
We devise the following algorithm for counting lines below a query point using a dynamic range tree:
    \begin{itemize}[leftmargin=1.2em]
        \item The dynamic range tree maintains points in two dimensions (x and y), allowing for efficient insertions and queries.
        \item \textbf{Insertions}: Each point is inserted into the primary x-tree and the corresponding secondary y-tree, taking \( O(\log^2 n) \) time.
        \item \textbf{Queries}: For a query point \( (q_x, q_y) \), we query the range tree to count how many points have \( y < q_y \). This takes \( O(\log^2 n) \) time.
    \end{itemize}
    
    & 
To solve the problem of counting the number of lines below a query point when only insertions and queries occur, we employ a strategy similar to that used in part (a). We organize the lines into groups of geometrically increasing sizes and maintain appropriate data structures over these groups.

We proceed by organizing the inserted lines into batches of sizes that are powers of two. For each batch, we construct a data structure that allows us to count the number of lines below a query point efficiently. We use a \textbf{segment tree} or a \textbf{range tree} to organize the lines in each batch. These trees can answer range counting queries in \( O(\log n) \) time per query.
As lines are inserted, we group them into batches of sizes \( 1, 2, 4, 8, \dots \), similar to the runs in part (a).

\\
\begin{itemize}[leftmargin=1.2em]
        \item \textbf{Query time}: \( O(\log^2 n) \).
        \item \textbf{Total insertion time}: With \( n \) insertions, each taking \( O(\log^2 n) \), the total insertion time is \( O(n^2 \log n) \).
        \item \textbf{Space complexity}: The range tree uses \( O(n^2) \) space, as required.
    \end{itemize}

\textcolor{red}{[\textbf{Misleading Claims}: Doesn't make any sense. No points for this question.]}
&
Each time a batch reaches its capacity (a power of two), we merge it with another batch of the same size to form a larger batch, rebuilding the data structure over the combined set of lines.

\vspace{.5em}
\textbf{Query Processing:}
To count the number of lines below a query point \( q \):
\begin{itemize}[leftmargin=1.2em]
    \item We query each of the \( O(\log n) \) data structures corresponding to the batches.
    \item Since each query to a data structure takes \( O(\log n) \) time, the total time per query is \( O(\log n \cdot \log n) = O(\log^2 n) \).
\end{itemize}
\textbf{Insertion Processing:}
The total time for processing insertions is analyzed similarly to part (a):
\begin{itemize}[leftmargin=1.2em]
\item Each line insertion may cause \( O(\log n) \) merges.
\item Rebuilding the data structure over a batch of size \( 2^i \) takes \( O(2^i \log 2^i) \) time. \textcolor{red}{[\textbf{Mathematical Errors}: Incorrect runtime.]}
\item The total insertion time over all \( n \) lines is:
  \[
  T_{\text{insert}} = O\left( n \log n \cdot \log n \right) = O(n \log^2 n).
  \]
\end{itemize}

However, since we need to support \( O(n) \) insertions and maintain data structures over potentially \( O(n) \) batches (due to the merging of batches), the total space required becomes \( O(n^2) \).

\vspace{.5em}
\textbf{Space Complexity:}
We store \( O(n) \) data structures, each of which may contain up to \( O(n) \) lines (after merges). Therefore, the total space used is \( O(n^2) \). \textcolor{red}{[\textbf{Mathematical Errors}: Incorrect proof for space complexity.]}

\end{longtable}

\newcommand{\xx}{\mathbf{x}}
\newcommand{\yy}{\mathbf{y}}

\newcommand\veczero{\boldsymbol{0}}
\newcommand\vecone{\boldsymbol{1}}

\newcommand\matzero{\boldsymbol{0}}
\newcommand\matone{\boldsymbol{1}}

\section{Second Exam: Exercises and AI-generated Solutions}
\label{sect:appendix_2}

\subsection{Exercise 1: Page Ranking and Farkas Lemma}
\label{sect:appendix_2_ex1}
Consider the following process: a user is browsing a website that contains $n$ pages numbered from 1 to $n$ and when she is on page $i$, she randomly clicks on a link that brings her to page $j$ with probability $p_{ij}$. 

Suppose that when the user enters the website for the first time, she lands on page $i$ with probability $x_i$. After her first click (on one of the links), she will end up on page $j$ with probability $$y_j = \sum_{i =1}^{n} x_i p_{ij}.$$

Assume that $x_i\geq 0$ for all $i=1,\dots,n$, $\sum_{i=1}^n x_i =1$, $p_{ij} \geq 0$ for all $i,j= 1,\dots,n$, and $\sum_j p_{ij} = 1$, for all $i=1,\dots, n$.

\begin{longtable}{|p{0.54\textwidth}|p{0.42\textwidth}|}
\multicolumn{2}{p{0.96\textwidth}}{\textbf{(a)} Prove that the probabilities $y_i$ are well-defined, i.e. show that $y_j \geq 0$, $\forall j\in [n]$ and $\sum_j y_j =1$.} \\ 
\multicolumn{2}{p{0.96\textwidth}}{
\vspace{.1em}
\textbf{Master Solution} 

Since $x_i\geq 0, \forall i$ and $p_{ij} \geq 0, \forall i,j$ it is evident that $y_j \geq 0$ (as we only have non-negative values). Furthermore, $\sum_j y_j = \sum_j \sum_i x_i p_{ij} = \sum_i x_i \sum_j p_{ij} = \sum_i x_i = 1$.
}\\
\hline
\textbf{GPT-4o} & \textbf{{\o1}} \\ \hline
\endfirsthead
\hline
\textbf{GPT-4o} & \textbf{{\o1}} \\ \hline
\endhead
\hline \multicolumn{2}{r}{\textit{Continued on the next page...}} \\ \hline
\endfoot
\hline
\endlastfoot
We are given:
\[
y_j = \sum_{i=1}^n x_i p_{ij}, \quad  \]
\(\text{where} \quad x_i \geq 0,  \sum_{i=1}^n x_i = 1, \quad p_{ij} \geq 0,  \text{ and}  \sum_{j=1}^n p_{ij} = 1 , \forall i.
\)

\vspace{1em}
\emph{Step 1: Prove that $y_j \geq 0 \, \forall j$.}
Since $x_i \geq 0$ and $p_{ij} \geq 0$ for all $i, j$, we have $x_i p_{ij} \geq 0$ for all $i, j$. Therefore, the sum $y_j = \sum_{i=1}^n x_i p_{ij} \geq 0$.

\emph{Step 2: Prove that $\sum_{j=1}^n y_j = 1$.}
Substitute the expression for $y_j$:
\[
\sum_{j=1}^n y_j = \sum_{j=1}^n \sum_{i=1}^n x_i p_{ij}.
\]
Reorder the summation (justified by Fubini's theorem):
\[
\sum_{j=1}^n y_j = \sum_{i=1}^n x_i \left( \sum_{j=1}^n p_{ij} \right).
\]
Since $\sum_{j=1}^n p_{ij} = 1$ for all $i$, we have:
\[
\sum_{j=1}^n y_j = \sum_{i=1}^n x_i \cdot 1 = \sum_{i=1}^n x_i.
\]
Finally, since $\sum_{i=1}^n x_i = 1$, it follows that $\sum_{j=1}^n y_j = 1$.

Thus, the probabilities $y_j$ are well-defined.

    & 
Since \( x_i \geq 0 \) and \( p_{ij} \geq 0 \) for all \( i, j \), it follows that each term \( x_i p_{ij} \geq 0 \). Therefore,
\[
y_j = \sum_{i=1}^n x_i p_{ij} \geq 0.
\]
We then compute the sum of all \( y_j \):
\[
\sum_{j=1}^n y_j = \sum_{j=1}^n \sum_{i=1}^n x_i p_{ij} = \sum_{i=1}^n x_i \left( \sum_{j=1}^n p_{ij} \right).
\]
Since \( \sum_{j=1}^n p_{ij} = 1 \) for all \( i \), we have:
\[
\sum_{j=1}^n y_j = \sum_{i=1}^n x_i (1) = \sum_{i=1}^n x_i = 1.
\]
Therefore, the probabilities \( y_j \) are well-defined.
\end{longtable}

Let $\xx = (x_1, x_2, \dots,x_n)^\top$ and $\yy = (y_1, y_2, \dots, y_n)^\top$, then we can write
$$ \yy = P^\top \xx, \quad \text{where } P  =
  \left[ {\begin{array}{cccc}
    p_{11} & p_{12} & \cdots & p_{1n}\\
    p_{21} & p_{22} & \cdots & p_{2n}\\
    \vdots & \vdots & \ddots & \vdots\\
    p_{n1} & p_{n2} & \cdots & p_{nn}\\
  \end{array} } \right]  
 $$
 
We say that $\xx$ is a \emph{steady state} if 
$$ \xx = P^\top \xx, \quad \text{where } P, x_i \geq 0, \forall i, $$
and $\sum_i x_i =1$.

The original Page Rank algorithm used by Google was actually ranking the web pages based on the score given by the steady state. In the remaining part of the exercise, we want to show that such a state always exists using Farkas lemma.

In order to do so, consider a process like the one described above.

\begin{longtable}{|p{0.4\textwidth}|p{0.55\textwidth}|}
\multicolumn{2}{p{0.96\textwidth}}{\textbf{(b)} Write an LP that has a solution if and only if the steady state exists.} \\ 
\multicolumn{2}{p{0.96\textwidth}}{
\vspace{.1em}
\textbf{Master Solution}

Consider a Markov Chain (MC) with transition matrix $P$. Let 
$$ A = \left[ {\begin{array}{c}
    P^\top - \matone\\
    \vecone^\top
  \end{array} } \right] , \quad \text{ and } b = \left[ {\begin{array}{c}
    \veczero \\
    1
  \end{array} } \right]
$$
The MC has a steady state if and only if the following LP has a solution:
$$
Ax = b, x\geq 0.
$$
This follows directly by construction, i.e. just reflects the 2 necessary and sufficient conditions $x=P^Tx$ and $\sum_i x_i = 1$ in one system of equations.
}
\end{longtable}

\begin{longtable}{|p{0.4\textwidth}|p{0.55\textwidth}|}
\hline
\textbf{GPT-4o} & \textbf{{\o1}} \\ \hline
\endfirsthead
\hline
\textbf{GPT-4o} & \textbf{{\o1}} \\ \hline
\endhead
\hline \multicolumn{2}{r}{\textit{Continued on the next page...}} \\ \hline
\endfoot
\hline
\endlastfoot
To check if the steady state exists, we consider the steady state condition:
\[
x = P^\top x, \quad \text{where} \quad x_i \geq 0, \quad \sum_{i=1}^n x_i = 1.
\]

Rewriting this condition:

\parbox{0.4\textwidth}{
    \begin{align*}   
    &(P^\top - I)x = 0, \\
    \text{where}& \quad x_i \geq 0 \;\; \forall i, \quad \sum_{i=1}^n x_i = 1.
    \end{align*}
    }

    & 
We are required to formulate a linear program that has a solution if and only if the steady state \( \mathbf{x} \) exists, satisfying:
\[
\mathbf{x} = P^\top \mathbf{x}, \quad x_i \geq 0, \quad \sum_{i=1}^n x_i = 1.
\]
Consider the following LP:
\begin{itemize}[leftmargin=1.2em]
    \item Variables: \( \mathbf{x} = (x_1, x_2, \ldots, x_n)^\top \).
    \item Objective Function: Minimize \( 0 \) (since we are only interested in feasibility).
        
\end{itemize}

\\

Introduce an LP to verify feasibility. The LP can be written as:
\[
\text{Find } x \text{ such that:}
\]
\[
(P^\top - I)x \leq 0, \quad x_i \geq 0 \, \forall i, \quad \sum_{i=1}^n x_i = 1.
\]
\textcolor{red}{[\textbf{Mathematical Errors}: Should be $(P^\top - I)x = 0$.]}
This LP has a feasible solution if and only if the steady state $x$ exists.
\textcolor{red}{[\textbf{Unjustified Claims}: No explanation.]}
&
\begin{itemize}[leftmargin=1.2em]
    \item Constraints:
\begin{enumerate}[leftmargin=1.5em]
        \item Steady State Condition:
        $P^\top \mathbf{x} - \mathbf{x} = \mathbf{0}.$
        
        This can be rewritten as:
        $(P^\top - I) \mathbf{x} = \mathbf{0}$,
        where \( I \) is the identity matrix.
        \item Non-negativity:
        $\mathbf{x} \geq \mathbf{0}.$
        \item Probability Sum Constraint:
        $\mathbf{1}^\top \mathbf{x} = \sum_{i=1}^n x_i = 1$,
        where \( \mathbf{1} \) is the vector of ones.
    \end{enumerate}
\end{itemize}
Therefore, the LP is:
\[
\begin{aligned}
& \text{Minimize} && 0 \\
& \text{Subject to} && (P^\top - I) \mathbf{x} = \mathbf{0}, \\
& && \mathbf{x} \geq \mathbf{0}, \\
& && \mathbf{1}^\top \mathbf{x} = 1.
\end{aligned}
\]
This LP has a solution if and only if a steady state \( \mathbf{x} \) exists. 
\textcolor{red}{[\textbf{Unjustified Claims}: No explanation.]}
\end{longtable}

\begin{longtable}{|p{0.4\textwidth}|p{0.55\textwidth}|}
\multicolumn{2}{p{0.96\textwidth}}{\textbf{(c)} Suppose that the LP is impossible, apply Farkas lemma to the LP and arrive at a contradiction. 

\textit{Hint: Prove that the vector guaranteed by Farkas lemma does not exist.}} \\ 
\multicolumn{2}{p{0.96\textwidth}}{
\vspace{.1em}
\textbf{Master Solution}

If the Markov matrix P has no steady state, Farkas theorem implies that there exists a vector $z$ such that
$$ A^\top z \geq 0, \quad b^\top z < 0. $$
Let  $ z = \left[ {\begin{array}{c}
    y \\
   \lambda
  \end{array} } \right].$
We have that
$\lambda <0$ from $b^\top z < 0$ and for all rows $i : \sum_j p_{ij} y_j- y_i +\lambda \ge 0$, thus 
$$ \sum_j p_{ij} y_j- y_i \geq -\lambda >0,\quad \forall i.$$
Let $y_m = \max_i y_i$, look at row $m$ of the inequality above,
$$ \sum_j p_{mj} y_j- y_m \geq -\lambda >0$$
but 
$$ \sum_j p_{mj} y_j- y_m \leq y_m (\sum_j p_{mj} - 1) = 0$$
}
\end{longtable}

\begin{longtable}{|p{0.4\textwidth}|p{0.55\textwidth}|}
\hline
\textbf{GPT-4o} & \textbf{{\o1}} \\ \hline
\endfirsthead
\hline
\textbf{GPT-4o} & \textbf{{\o1}} \\ \hline
\endhead
\hline \multicolumn{2}{r}{\textit{Continued on the next page...}} \\ \hline
\endfoot
\hline
\endlastfoot
The LP from part (b) can be expressed in matrix form as:
\[
\text{Find } x \text{ such that: } Ax \leq b, \quad x \geq 0,
\]
where:
\[
A = P^\top - I, \quad b = 0.
\]
\textcolor{red}{[\textbf{Mathematical Errors}: Missing $\sum_i x_i$.]}
If the LP is infeasible, by Farkas' Lemma, there exists a vector $y \in \mathbb{R}^n$ such that:
\[
y^\top A \geq 0, \quad y^\top b < 0, \quad \text{and } y \geq 0.
\]

\textcolor{red}{[\textbf{Misleading Claims}: Below is wrong.]}
Substituting $A = P^\top - I$ and $b = 0$, this means:
\[
y^\top (P^\top - I) \geq 0, \quad y^\top 0 < 0, \quad \text{and } y \geq 0.
\]

The condition $y^\top b < 0$ simplifies to:
\[
0 < 0,
\]
which is a contradiction. Therefore, the assumption that the LP is infeasible is false.

Hence, the steady state $x$ must exist.

    & 

Suppose that the LP formulated in part (b) is infeasible, i.e., there is no \( \mathbf{x} \geq \mathbf{0} \) satisfying:
\[
(P^\top - I) \mathbf{x} = \mathbf{0}, \quad \mathbf{1}^\top \mathbf{x} = 1.
\]
Farkas' Lemma states that for a matrix \( A \) and vector \( \mathbf{b} \), exactly one of the following holds:
\begin{enumerate}
    \item There exists \( \mathbf{x} \geq \mathbf{0} \) such that \( A \mathbf{x} = \mathbf{b} \).
    \item There exists \( \mathbf{y} \) such that \( A^\top \mathbf{y} \geq \mathbf{0} \) and \( \mathbf{y}^\top \mathbf{b} < 0 \).
\end{enumerate}
Since the first condition is assumed false (the LP is infeasible), there must exist a vector \( \mathbf{y} \) satisfying the second condition.
Let’s define:
\[
A = \begin{bmatrix}
P^\top - I \\
\mathbf{1}^\top
\end{bmatrix}, \quad \mathbf{b} = \begin{bmatrix}
\mathbf{0} \\
1
\end{bmatrix}.
\]
Here, \( A \) is an \( (n+1) \times n \) matrix, and \( \mathbf{b} \) is an \( (n+1) \)-dimensional vector.
By Farkas' Lemma, there exists \( \mathbf{y} \in \mathbb{R}^{n+1} \) such that:
\[
A^\top \mathbf{y} \geq \mathbf{0}, \quad \mathbf{y}^\top \mathbf{b} < 0.
\]
We partition \( \mathbf{y} \) as:
\[
\mathbf{y} = \begin{bmatrix}
\tilde{\mathbf{y}} \\
\lambda
\end{bmatrix}, \quad \tilde{\mathbf{y}} \in \mathbb{R}^n, \quad \lambda \in \mathbb{R}.
\]
We have
\[
A^\top \mathbf{y} = \left( P^\top - I \right)^\top \tilde{\mathbf{y}} + \mathbf{1} \lambda = (P - I) \tilde{\mathbf{y}} + \lambda \mathbf{1} \geq \mathbf{0},
\]
and
\[
\mathbf{y}^\top \mathbf{b} = \tilde{\mathbf{y}}^\top \mathbf{0} + \lambda \cdot 1 = \lambda < 0.
\]
Thus, \( \lambda < 0 \).
Letting \( s = -\lambda > 0 \), our inequality becomes:
\[
(P - I) \tilde{\mathbf{y}} - s \mathbf{1} \geq \mathbf{0},
\]
which can be rewritten as:
\[
P \tilde{\mathbf{y}} \geq \tilde{\mathbf{y}} + s \mathbf{1}.
\]
For each component \( i \):
\[
\sum_{j=1}^n p_{ij} \tilde{y}_j \geq \tilde{y}_i + s.
\]

\\

&

However, since \( p_{ij} \geq 0 \) and \( \sum_{j=1}^n p_{ij} = 1 \), the left-hand side is a convex combination of \( \tilde{y}_j \).
Let \( y_{\max} = \max_{j} \tilde{y}_j \).
Then:
\[
\sum_{j=1}^n p_{ij} \tilde{y}_j \leq y_{\max} \sum_{j=1}^n p_{ij} = y_{\max}.
\]
So we have:
\[
y_{\max} \geq \tilde{y}_i + s, \quad \text{for all } i.
\]
This implies:
\[
\tilde{y}_i \leq y_{\max} - s, \quad \text{for all } i.
\]
Thus, all components \( \tilde{y}_i \) are bounded above by \( y_{\max} - s \).
Now, consider the inequality for \( i \) such that \( \tilde{y}_i = y_{\max} \):
\[
\sum_{j=1}^n p_{ij} \tilde{y}_j \geq y_{\max} + s.
\]
But since \( \sum_{j=1}^n p_{ij} = 1 \) and \( p_{ij} \geq 0 \), we have:
\[
\sum_{j=1}^n p_{ij} \tilde{y}_j \leq y_{\max}.
\]
Therefore:
\[
y_{\max} \geq y_{\max} + s,
\]
which simplifies to:
\[
0 \geq s.
\]
This contradicts \( s > 0 \).
\end{longtable}

\subsection{Exercise 2: NOT-OR Circuits Using Linear Programming}
\label{sect:appendix_2_ex2}

A digital circuit consists of a network of wires connected through logic gates. Each wire carries a signal, which can be either 0 (representing \emph{false}) or 1 (representing \emph{true}). In this exercise, we consider a class of circuits called \emph{NOT-OR circuits}, which consist of three sequential layers.
\begin{itemize}
	\item \textbf{Input Layer}: The circuit has \( n \) input wires, where each wire \( i \) (for \( i = 1, 2, \dots, n \)) carries either a 1 or a 0. The sequence of values on these input wires is called the \emph{input configuration}.
	\item \textbf{Negation Layer}: The \( n \) input wires are passed through a layer that outputs \( 2n \) wires. For each input \( x_i \), this layer produces both \( x_i \) and its negation, denoted as \(\operatorname{not}(x_i)\). These values are referred to as \emph{intermediate wires}.
	\item \textbf{OR Layer}: The intermediate wires are then fed into a layer that outputs \( m \) wires, each representing a \emph{OR-gate} value \( C_j \in \{0,1\} \) computed from a subset of the intermediate wires (for \( j = 1, 2, \dots, m \)). Each OR-gate value is 1 if at least one of its intermediate wires has value 1.
    Multiple OR-gates may receive the same intermediate wire as input.
\end{itemize}

\begin{figure}[h]
	\centering
	\includegraphics[scale=0.4]{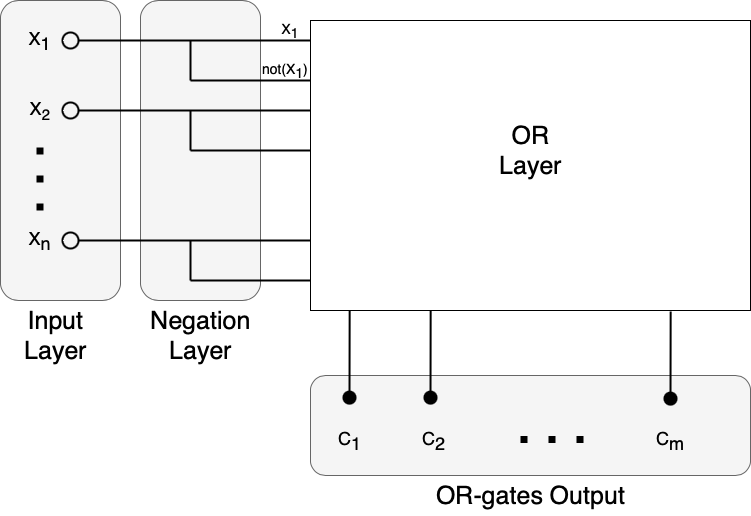}
	\caption{A scheme of a NOT-OR circuit}
\end{figure}

In the first part of the exercise, we consider circuits where each OR-gate takes as input exactly 5 intermediate wires. It can be shown that it is NP-hard to determine whether there exists an input configuration such that all OR-gates output the value 1. Instead, we aim to find an input configuration that maximizes the number of OR-gates with output value 1. Let OPT represent the maximum number of OR-gates that can have value 1 for any input configuration.

\begin{longtable}{|p{0.4\textwidth}|p{0.55\textwidth}|}
\multicolumn{2}{p{0.96\textwidth}}{\textbf{(a)} Assign the value 1 or 0 to each input wire independently with probability \( \frac{1}{2} \). Prove that, in expectation, this assignment results in at least \( \frac{31}{32} \ \text{OPT} \) OR-gates having value 1.} \\ 
\multicolumn{2}{p{0.96\textwidth}}{
\vspace{.1em}
\textbf{Master Solution}

Assign the value 0 or 1 with probability 1/2 independently to each input wire. For each OR-gate $j$, let $Y_j$ be a variable representing its value and let $Y= \sum_j Y_j$ denote the number of OR-gates with value 1. Fix an OR-gate $j\in S$, since it is the conjunction of exactly 5 intermediate wires, then it is true with a probability at least $1-(1/2)^5$, i.e. $\expec{}{Y_j} \geq 31/32$. By linearity of expectation, we have that $$ \expec{}{Y} = \sum_{j=1}^m \expec{}{Y_j} \geq \frac{31}{32} m .$$
This is clearly at least 31/32 of the optimal value.
}
\end{longtable}

\begin{longtable}{|p{0.45\textwidth}|p{0.5\textwidth}|}
\hline
\textbf{GPT-4o} & \textbf{{\o1}} \\ \hline
\endfirsthead
\hline
\textbf{GPT-4o} & \textbf{{\o1}} \\ \hline
\endhead
\hline \multicolumn{2}{r}{\textit{Continued on the next page...}} \\ \hline
\endfoot
\hline
\endlastfoot
We are tasked with assigning values to the input wires independently with probability $\frac{1}{2}$, and proving that this random assignment achieves at least $\frac{31}{32} \cdot \text{OPT}$ OR-gates having value $1$ in expectation.

\textbf{Analysis}: Each OR-gate $C_j$ takes exactly $5$ intermediate wires as input. For $C_j$ to output $1$, at least one of its $5$ inputs must have a value of $1$. Let $y_i$ be the value of the $i$-th input wire, which is assigned independently with probability $\frac{1}{2}$. The negated wire $\text{not}(y_i)$ is assigned $1 - y_i$.

For each OR-gate $C_j$, the probability that all of its $5$ input wires are $0$ is:
\begin{align*}  
\Pr(\text{all inputs to } C_j \text{ are } 0) 
\\= \left(1 - \frac{1}{2}\right)^5 = \frac{1}{32}.
\end{align*}
Thus, the probability that $C_j$ outputs $1$ is:

  \parbox{0.4\textwidth}{
    \begin{align*}   
        \Pr(C_j = 1) &= 1 - \Pr(\text{all inputs to } C_j \text{ are } 0) \\
        &= 1 - \frac{1}{32} = \frac{31}{32}.
    \end{align*}
    }
    
In expectation, the number of OR-gates that output $1$ is:
\begin{align*}
\mathbb{E}\left[\text{Number of OR-gates with output } 1\right] \\= \frac{31}{32} \cdot \text{OPT}.
\end{align*}

\textbf{Conclusion}: The expected number of OR-gates with output $1$ is at least $\frac{31}{32} \cdot \text{OPT}$, as required.

    & 

Consider any OR-gate $C_j$. It takes exactly 5 intermediate wires as input. Each intermediate wire is either $x_i$ or $\text{not}(x_i)$ for some $i$.
Since each input wire $x_i$ is assigned 1 or 0 independently with probability $\frac{1}{2}$, we have:
\begin{itemize}[leftmargin=1.2em]
    \item For $x_i$:
  \[
  P(x_i = 1) = \frac{1}{2}, \quad P(x_i = 0) = \frac{1}{2}.
  \]
  \item For $\text{not}(x_i)$:

  \parbox{0.4\textwidth}{
    \begin{align*}   
     &P(\text{not}(x_i) = 1) = P(x_i = 0) = \frac{1}{2}, \\
     &P(\text{not}(x_i) = 0) = P(x_i = 1) = \frac{1}{2}.
    \end{align*}
    }
    
\end{itemize}
Thus, for any intermediate wire (whether $x_i$ or $\text{not}(x_i)$), the probability that it is 0 is $\frac{1}{2}$.
Since an OR-gate outputs 0 only if all its inputs are 0, the probability that $C_j$ outputs 0 is:
\[
P(C_j = 0) = \left( \frac{1}{2} \right)^5 = \frac{1}{32}.
\]
And the probability that $C_j$ outputs 0 is:
\[
P(C_j = 1) = 1 - P(C_j = 0) = 1 - \frac{1}{32} = \frac{31}{32}.
\]
Let $E[\text{OR}_1]$ denote the expected number of OR-gates outputting 1. Then:
\[
E[\text{OR}_1] = \sum_{j=1}^m P(C_j = 1) = m \cdot \frac{31}{32}.
\]
Since $\text{OPT} \leq m$, we have:
\[
E[\text{OR}_1] = m \cdot \frac{31}{32} \geq \frac{31}{32} \cdot \text{OPT}.
\]
\end{longtable}

In the second part of the exercise, we consider a NOT-OR circuit where each OR-gate may take as input an arbitrary number of intermediate wires. Each output OR-gate \( C_j \) controls a switch that regulates an output current \( f_j \geq 0 \). The \emph{total output} of the circuit is defined as \( \sum_{j=1}^m C_j f_j \). Given a circuit and outputs $f_1,\ldots,f_m$, our goal is to find an input configuration that maximizes this total output.

\begin{longtable}{|p{0.45\textwidth}|p{0.5\textwidth}|}
\multicolumn{2}{p{0.96\textwidth}}{\textbf{(b)} Design a polynomial-time deterministic algorithm that returns an input configuration achieving at least half of the maximum possible output current. 

	\textit{Hint: Consider an input configuration and its negation.}} \\ 
\multicolumn{2}{p{0.96\textwidth}}{
\vspace{.1em}
\textbf{Master Solution}

The algorithm is simple. Consider the following two input configurations. Config. A: all the wires carry the value 0; Config. B: all the wires carry the value 1. Compute the output of Config. A and Config. B and take the one that satisfies the most OR-gates. 

In order to prove that we get at least half of the maximum possible output current, we show that at least half of the OR-gates have output 1. In fact, an OR-gate $C_i$ cannot have an output value of 0 for both Config. A and B. This means that if we denote with $c_A$ and $c_B$, the number of OR-gates satisfied respectively for Config. A and Config. B, we have that $c_A +c_B \geq m$. We conclude that $\max\setof{c_A, c_B} \geq m/2$.
}\\
\hline
\textbf{GPT-4o} & \textbf{{\o1}} \\ \hline
\endfirsthead
\hline
\textbf{GPT-4o} & \textbf{{\o1}} \\ \hline
\endhead
\hline \multicolumn{2}{r}{\textit{Continued on the next page...}} \\ \hline
\endfoot
\hline
\endlastfoot
We aim to design a polynomial-time deterministic algorithm that achieves at least half of the maximum possible output current.

\textbf{Algorithm:}

Given a NOT-OR circuit with $n$ input wires, proceed as follows:
1. Compute the output current for the given input configuration $x$.
2. Compute the output current for the negation of the input configuration $\text{not}(x)$.
3. Select the configuration (either $x$ or $\text{not}(x)$) that results in the higher output current.
\textcolor{red}{[\textbf{Unjustified Claims}: Needs more argumentation, you are basically already assuming the thing you want to show here.]}

\textbf{Justification:} 

Each OR-gate $C_j$ is connected to the values of the intermediate wires $x_i$ and $\text{not}(x_i)$. Therefore, either $x$ or $\text{not}(x)$ must contribute at least half of the maximum output current. Since the algorithm evaluates both configurations, it achieves at least half of the maximum possible output current.

\textbf{Conclusion:} 

This deterministic algorithm runs in polynomial time and guarantees at least half of the maximum output current.
\textcolor{red}{[\textbf{Misleading Claims}: Wrong claim.]}

    & 
\textbf{Algorithm:}
\begin{enumerate}[leftmargin=1.2em]
    \item Compute the total output when all input wires are assigned 0. Let this output be $T_0$.
    \item  Compute the total output when all input wires are assigned 1. Let this output be $T_1$.
    \item Choose the input configuration (all zeros or all ones) that yields the higher total output.
\end{enumerate}

\textbf{Proof of correctness:}

Consider any input configuration $S$ and its negation $\bar{S}$ (where each input is flipped). For each OR-gate $C_j$, at least one of $C_j$ and $\bar{C}_j$ outputs 1 because the inputs to $C_j$ in $S$ are the negation of its inputs in $\bar{S}$. Therefore:
\[
C_j + \bar{C}_j \geq 1.
\]
Let $T_{\text{max}}$ be the maximum possible total output. Then:
\[
T(S) + T(\bar{S}) = \sum_{j=1}^m (C_j + \bar{C}_j) f_j \geq \sum_{j=1}^m f_j \geq T_{\text{max}}.
\]
Therefore, at least one of $T(S)$ or $T(\bar{S})$ is at least $\frac{1}{2} T_{\text{max}}$.
\end{longtable}

\begin{longtable}{|p{0.45\textwidth}|p{0.5\textwidth}|}
\multicolumn{2}{p{0.96\textwidth}}{\textbf{(c)} Show that the optimal solution to the following linear program provides an upper bound on the maximum possible output current for the circuit.
   \[
   \max \sum_{j=1}^m z_j f_j
   \]
   subject to:
   \[
   \sum_{i: x_i \in C_j} y_i + \sum_{i: (\operatorname{not}(x_i))\in C_j} (1 - y_i) \geq z_j, \quad \forall j=1,2,\dots, m
   \]
   \[
   0 \leq y_i \leq 1, \quad 0 \leq z_j \leq 1
   \]} \\ 
\multicolumn{2}{p{0.96\textwidth}}{\textbf{Master Solution}

We show that given a solution to the maximum output current problem, we build a solution for the LP whose cost equals the output current from the solution. Given a solution to the maximum output current problem, assign $y_i = x_i$. Then, assign $z_j=1$ if and only if the OR-gate $C_i$ has value 1. It is easy to see that the just-built solution of the LP is feasible and that its cost is exactly the output current of the maximum output current solution.
}
\end{longtable}

\begin{longtable}{|p{0.48\textwidth}|p{0.5\textwidth}|}
\hline
\textbf{GPT-4o} & \textbf{{\o1}} \\ \hline
\endfirsthead
\hline
\textbf{GPT-4o} & \textbf{{\o1}} \\ \hline
\endhead
\hline \multicolumn{2}{r}{\textit{Continued on the next page...}} \\ \hline
\endfoot
\hline
\endlastfoot
We are asked to show that the optimal solution to the following linear program (LP) provides an upper bound on the maximum possible output current:
\[
\max \sum_{j=1}^m z_j f_j,
\]
subject to:

  \parbox{0.4\textwidth}{
    \begin{align*}   
        &\sum_{i : x_i \in C_j} y_i + \sum_{i : \text{not}(x_i) \in C_j} (1 - y_i) \geq z_j,  \\
        &\forall j = 1, 2, \dots, m,\quad0 \leq y_i \leq 1, \quad 0 \leq z_j \leq 1.
    \end{align*}
    }

\textbf{Interpretation:} 
\begin{itemize}[leftmargin=1.2em]
    \item The variable $y_i$ represents the probability of setting input wire $x_i$ to $1$.
    \item The variable $z_j$ represents the output value of OR-gate $C_j$, which depends on its inputs.
\end{itemize}
The constraints ensure that $z_j$ is at least the logical OR of its inputs, and the objective function $\sum_{j=1}^m z_j f_j$ seeks to maximize the total output current weighted by $f_j$.

    & 
Consider any input configuration, where each input wire $x_i$ is assigned a value $y_i \in \{0,1\}$. For each OR-gate $C_j$, let $z_j \in \{0,1\}$ represent its output.
The assignments satisfy:
\[
\sum_{i: x_i \in C_j} y_i + \sum_{i: \text{not}(x_i) \in C_j} (1 - y_i) \geq z_j.
\]
This is because if at least one input to $C_j$ is 1, then $z_j = 1$.
The total output current is:
\[
\sum_{j=1}^m z_j f_j.
\]
By relaxing $y_i, z_j$ to lie in $[0,1]$, we obtain the LP:
\[
\max \sum_{j=1}^m z_j f_j
\]
subject to:
\[
\sum_{i: x_i \in C_j} y_i + \sum_{i: \text{not}(x_i) \in C_j} (1 - y_i) \geq z_j, \quad \forall j,
\]
\[
0 \leq y_i \leq 1, \quad 0 \leq z_j \leq 1.
\]
\\
\textbf{Optimality:} 

Any feasible solution to the LP represents a fractional relaxation of the circuit problem, where inputs are allowed to take fractional values. Thus, the LP provides an upper bound on the maximum achievable output current for the circuit.
&
Since any feasible integer solution corresponds to an input configuration, and the LP relaxes this to continuous variables, the optimal LP value $\text{LP}^*$ satisfies:
\[
\text{LP}^* \geq T_{\text{max}}.
\]
\end{longtable}

\begin{longtable}{|p{0.42\textwidth}|p{0.56\textwidth}|}
\multicolumn{2}{p{0.96\textwidth}}{\textbf{(d)} Let \( y^*, z^* \) be an optimal solution to the LP in (c). Construct an input configuration by assigning each input wire \( i \) the value 1 with probability \( y^*_i \) and the value 0 with probability \( 1 - y^*_i \). Prove that, in expectation, the total output current of this solution is at least \( \left( 1 - \frac{1}{e} \right) \) times the optimal output.

   	\textit{Hint (AM-GM Inequality): For non-negative values \( x_1, x_2, \dots, x_\ell \), the inequality \[ \left( \prod_{i=1}^\ell x_i \right)^{1/\ell} \leq \frac{\sum_{i=1}^\ell x_i}{\ell} \] holds.}
    
   	\textit{Hint: The function \( f(x) = 1 - \left( 1 - \frac{x}{\ell} \right)^\ell \) is concave for \( x \in [0, 1] \) and \( \ell \in \mathbb{N}^+ \).}} \\ 

\multicolumn{2}{p{0.96\textwidth}}{
\vspace{.1em}
\textbf{Master Solution}

Let $\ell_j$ be the numbers of intermediate wires that appear in the OR-gate $C_j$.
\begin{eqnarray*}
&& \sum_j w_j \left(1-\prod_{i:x_i \in C_j} (1- y^*_i)\prod_{i: \overline{x}_i \in C_j} y^*_i\right)\\
& \geq & \sum_j w_j \left(1- \left(\frac{\sum_{i:x_i \in C_j} (1- y^*_i)+ \sum_{i: \overline{x}_i \in C_j} y^*_i }{l_j}\right)^{l_j}\right) \\
& = & \sum_j w_j \left(1- \left(\frac{l_j - \sum_{i:x_i \in C_j} y^*_i- \sum_{i: \overline{x}_i \in C_j} (1-y^*_i) }{l_j}\right)^{l_j}\right) \\
& \geq & \sum_j w_j \left(1- \left(1-\frac{z^*_j}{l_j}\right)^{l_j}\right) \\
& \geq & \sum_j w_j \left(1 - \left(1- \frac{1}{l_j}\right)^{l_j}\right) z^*_j \\
& \geq & \min_j \left(1-\left(1- \frac{1}{l_j}\right)^{l_j}\right) \sum_j w_j z^*_j \\
& \geq & \left(1-\frac{1}{e}\right) opt
\end{eqnarray*}
}
\end{longtable}

\begin{longtable}{|p{0.4\textwidth}|p{0.55\textwidth}|}
\hline
\textbf{GPT-4o} & \textbf{{\o1}} \\ \hline
\endfirsthead
\hline
\textbf{GPT-4o} & \textbf{{\o1}} \\ \hline
\endhead
\hline \multicolumn{2}{r}{\textit{Continued on the next page...}} \\ \hline
\endfoot
\hline
\endlastfoot
Let $(y^*, z^*)$ be an optimal solution to the LP in part (c). Construct an input configuration by assigning each input wire $x_i$ the value $1$ with probability $y_i^*$ and $0$ with probability $1 - y_i^*$. We aim to show that, in expectation, the total output current of this solution is at least $\left(1 - \frac{1}{e}\right)$ times the optimal output.

\textbf{Analysis:}

The probability that OR-gate $C_j$ outputs $1$ is:
\begin{multline*}
    \Pr(C_j = 1) = \\
    1 - \prod_{i : x_i \in C_j}(1 - y_i^*) \prod_{i : \text{not}(x_i) \in C_j} y_i^*.
\end{multline*}

Using the concavity of the function $f(x) = 1 - (1 - x)^\ell$ (as hinted), we conclude that the expected value of $z_j$ is at least:
\[
\mathbb{E}[z_j] \geq \left(1 - \frac{1}{e}\right) z_j^*.
\]

Thus, the expected total output current is:
\[
\mathbb{E}\left[\sum_{j=1}^m z_j f_j\right] \geq \left(1 - \frac{1}{e}\right) \sum_{j=1}^m z_j^* f_j.
\]

\textbf{Conclusion:} 

In expectation, the total output current achieved by the probabilistic assignment is at least $\left(1 - \frac{1}{e}\right)$ times the optimal output. 
\textcolor{red}{[\textbf{Unjustified Claims}: You are missing way too many steps, this is not a sufficient derivation.]}
    & 
Let $(\mathbf{y}^*, \mathbf{z}^*)$ be an optimal solution to the LP in part (c). We construct an input configuration by assigning each input wire $x_i$ to 1 with probability $y_i^*$ and to 0 with probability $1 - y_i^*$, independently.

For each OR-gate $C_j$, we compute the expected output.
The probability that $C_j$ outputs 0 is:
\begin{itemize}[leftmargin=1.2em]
    \item For $x_i \in C_j$:
  \[
  P(x_i = 0) = 1 - y_i^*.
  \]
  \item For $\text{not}(x_i) \in C_j$:
  \[
  P(\text{not}(x_i) = 0) = P(x_i = 1) = y_i^*.
  \]
\end{itemize}

Therefore, the probability that an input wire to $C_j$ is 0 is:
\[
p_s = \begin{cases}
1 - y_i^*, & \text{if } s = x_i, \\
y_i^*, & \text{if } s = \text{not}(x_i).
\end{cases}
\]
Since the inputs are independent:
\[
P(C_j = 0) = \prod_{s \in S_j} p_s.
\]
The expected output of $C_j$ is:
\[
E[C_j] = 1 - P(C_j = 0) = 1 - \prod_{s \in S_j} p_s.
\]
From the LP constraint:
\[
\sum_{i: x_i \in C_j} y_i^* + \sum_{i: \text{not}(x_i) \in C_j} (1 - y_i^*) \geq z_j^*.
\]
Define $L_j$ as the left-hand side of the constraint:
\[
L_j = \sum_{s \in S_j} q_s,
\]
where:
\[
q_s = \begin{cases}
y_i^*, & \text{if } s = x_i, \\
1 - y_i^*, & \text{if } s = \text{not}(x_i).
\end{cases}
\]
\\

&

Since $p_s = 1 - q_s$, we have:
\[
\ln P(C_j = 0) = \sum_{s \in S_j} \ln p_s = \sum_{s \in S_j} \ln (1 - q_s).
\]
Using the inequality $\ln(1 - x) \leq -x$ for $x \in [0,1]$, we get:
\[
\ln P(C_j = 0) \leq -\sum_{s \in S_j} q_s = -L_j.
\]
Exponentiating both sides:
\[
P(C_j = 0) \leq e^{-L_j}.
\]
Therefore:
\[
E[C_j] = 1 - P(C_j = 0) \geq 1 - e^{-L_j}.
\]

Since $L_j \geq z_j^*$, we have:
\[
E[C_j] \geq 1 - e^{-z_j^*}.
\]
For $x \in [0,1]$, the function $f(x) = 1 - e^{-x}$ satisfies:
\[
1 - e^{-x} \geq \left( 1 - \frac{1}{e} \right) x.
\]
At $x = 0$, both sides are 0. The derivative of $1 - e^{-x}$ at $x = 0$ is $1$. The linear function $(1 - \frac{1}{e}) x$ is below $1 - e^{-x}$ for $x \in [0,1]$.
So
\[
E[C_j] \geq \left( 1 - \frac{1}{e} \right) z_j^*.
\]
And the expected total output current is:

\parbox{0.45\textwidth}{
\begin{align*}
    E\left[ \sum_{j=1}^m C_j f_j \right] &= \sum_{j=1}^m E[C_j] f_j \geq \left( 1 - \frac{1}{e} \right) \sum_{j=1}^m z_j^* f_j\\ 
&= \left( 1 - \frac{1}{e} \right) \text{LP}^*.
\end{align*}
}

From part (c), we have $\text{LP}^* \geq T_{\text{max}}$.
Therefore, the randomized algorithm achieves, in expectation, at least $\left( 1 - \dfrac{1}{e} \right)$ times the optimal total output current.
\end{longtable}

\subsection{Exercise 3: Pizza Slicing}
\label{sect:appendix_2_ex3}
Suppose that we have a rectangular pizza of size $n\times m$. The pizza can be seen as a collection of $nm$ squares, half topped with mushrooms and half topped with peppers. We want to slice the pizza in slices of size $2\times 1$ so that each slice contains a square topped with mushrooms and one with peppers. Devise a polynomial time algorithm that given the topping for each square of the pizza finds the number of ways in which it is possible to slice the pizza into slices that satisfy the above conditions.

\begin{figure}[h]
	\centering
	\includegraphics[scale=0.4]{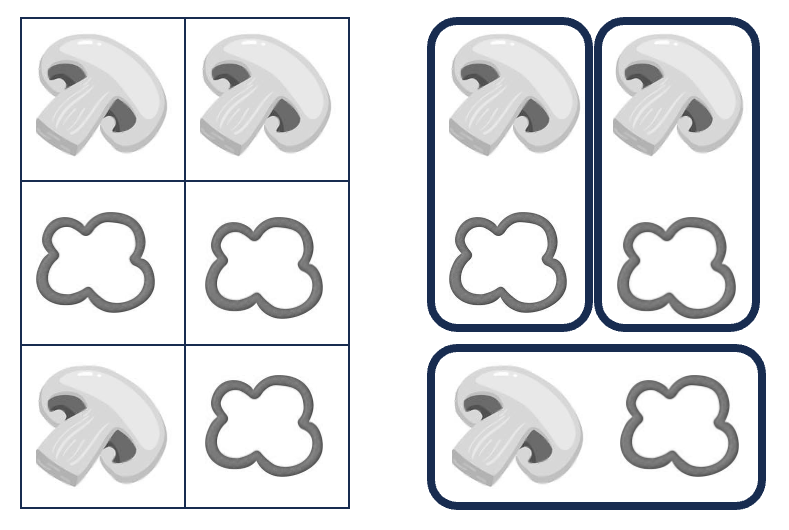}
	\caption{A $ 3 \times 2$ pizza and slicing of the pizza into three slices each containing one mushroom square and one pepper square.}
\end{figure}
\noindent\textbf{Master Solution}

Build a graph where each node represents either a square topped with mushrooms or a square topped with peppers. Connect the nodes that are adjacent to the pizza and contain different toppings. There is a bijection between each perfect matching in the graph and the way to slice the pizza. Furthermore, the graph must be planar by construction hence we can compute the number of perfect matchings using the algorithm seen in the lectures.

\begin{longtable}{|p{0.44\textwidth}|p{0.55\textwidth}|}
\hline
\textbf{GPT-4o} & \textbf{{\o1}} \\ \hline
\endfirsthead
\hline
\textbf{GPT-4o} & \textbf{{\o1}} \\ \hline
\endhead
\hline \multicolumn{2}{r}{\textit{Continued on the next page...}} \\ \hline
\endfoot
\hline
\endlastfoot
\textcolor{red}{[\textbf{Misleading Claims}: Wrong approach. The way the recurrence relation is set up is wrong. As a proof of this consider that in your algorithm dp(i,j) can be at most dp(i, j-2) + dp(i-2, j), but it's easy to think of cases where dp(i,j) is much larger than that.]}

We can model this problem as finding a perfect matching in a bipartite graph, where:
- Each mushroom square and pepper square corresponds to a vertex in the graph.
- An edge connects a mushroom vertex to a pepper vertex if they are adjacent in the pizza (i.e., they form a valid \( 2 \times 1 \) slice).
The solution involves the following steps:
\begin{enumerate}[leftmargin=1.2em]
    \item Construct a bipartite graph \( G = (M, P, E) \), where:  \( M \) represents all mushroom squares;
      \( P \) represents all pepper squares.
       An edge \( (m, p) \in E \) exists if square \( m \) (a mushroom) and square \( p \) (a pepper) are adjacent in the pizza.

    \item Find all perfect matchings in \( G \). Each perfect matching corresponds to a valid way to slice the pizza.
\end{enumerate}

The total number of perfect matchings in \( G \) can be computed using dynamic programming, as described below.
Define a state \( dp(i, j) \) as the number of valid ways to slice the first \( i \times j \) sub-pizza of the \( n \times m \) pizza.

\vspace{.5em}
\textbf{Base Case:}

\( dp(0, j) = dp(i, 0) = 1 \), since an empty sub-pizza has only one way (do nothing).

\vspace{.5em}
\textbf{Transition:}

To compute \( dp(i, j) \), consider the possible ways to slice the square \( (i, j) \):

&

\textbf{Algorithm Steps:}
\begin{enumerate}[leftmargin=1.2em]
    \item Construct the Bipartite Graph $G$:
    
        Let $M$ be the set of squares topped with mushrooms.
        Let $P$ be the set of squares topped with peppers.
        The vertex set is $V = M \cup P$.
        For each square, connect it via an edge to its adjacent squares (up, down, left, right) that have the opposite topping.
        This results in a bipartite graph $G = (V, E)$, where edges only connect vertices from $M$ to $P$.

    \item Count the Number of Perfect Matchings in $G$:

        Since $G$ is a bipartite planar graph, we can compute the number of perfect matchings in polynomial time.
        We can use the \textbf{Kasteleyn's algorithm} or \textbf{Pfaffian orientation} to compute the number of perfect matchings.

    \item Implement the Algorithm Efficiently:

Use appropriate data structures to represent the graph.
 Use existing algorithms for counting matchings in planar graphs.
   Ensure that the implementation runs in polynomial time.

\end{enumerate}

\textbf{Details:}

1. Constructing the Bipartite Graph

Each square of the pizza corresponds to a vertex. The toppings partition the vertices into two disjoint sets:
\begin{itemize}[leftmargin=1.2em]
    \item $M = \{ \text{Squares topped with mushrooms} \}$
    \item $P = \{ \text{Squares topped with peppers} \}$
\end{itemize}

 2. Counting Perfect Matchings in a Bipartite Planar Graph

Counting the number of perfect matchings in a general graph is a \#P-complete problem. However, for planar graphs, and specifically for bipartite planar graphs, there exist polynomial-time algorithms.

Kasteleyn showed that for planar graphs, by assigning directions and weights (called a Pfaffian orientation), the number of perfect matchings can be computed efficiently.
\\

1. Slice horizontally: If squares \( (i, j) \) and \( (i, j-1) \) form a valid \( 2 \times 1 \) slice (i.e., one mushroom and one pepper), then:
   \[
   dp(i, j) \mathrel{+}= dp(i, j-2).
   \]
   
2. Slice vertically: If squares \( (i, j) \) and \( (i-1, j) \) form a valid \( 2 \times 1 \) slice, then:
   \[
   dp(i, j) \mathrel{+}= dp(i-2, j).
   \]

\textbf{Complexity:}
\begin{itemize}[leftmargin=1.2em]
    \item Constructing the graph \( G \): \( O(nm) \) (each cell checks adjacency).
    \item Computing \( dp(i, j) \): \( O(nm) \), as each state depends on at most two prior states.
\end{itemize}
Thus, the total complexity is \( O(nm) \), which is polynomial.


&

\begin{itemize}[leftmargin=1.2em]
    \item Pfaffian Orientation: An orientation of the edges such that every face (except possibly the outer face) has an odd number of edges oriented clockwise.
    \item Construction: Such an orientation can be found in linear time for planar graphs.
    \item Adjacency Matrix: Construct a skew-symmetric matrix $A$, where $A_{uv} = 1$ if there is an edge from $u$ to $v$, $A_{uv} = -1$ if there is an edge from $v$ to $u$, and $A_{uv} = 0$ otherwise.
    \item Pfaffian: The square of the Pfaffian of $A$ equals the determinant of $A$, and this determinant gives the square of the number of perfect matchings.
    \item Computation: Compute $\det A$ using standard determinant algorithms (e.g., LU decomposition).
\end{itemize}

\textbf{Algorithm Outline:}
\begin{enumerate}[leftmargin=1.2em]
    \item Find a Pfaffian Orientation of $G$:

 Use a planar embedding of $G$.
 Orient the edges to satisfy Kasteleyn's condition.

    \item Construct the Skew-Symmetric Adjacency Matrix $A$:

 For vertices $u$ and $v$ connected by an edge:
If the edge is directed from $u$ to $v$, set $A_{uv} = 1$ and $A_{vu} = -1$.
If the edge is directed from $v$ to $u$, set $A_{uv} = -1$ and $A_{vu} = 1$.

    \item Compute the Pfaffian or Determinant of $A$:
    
 Compute $\det A$.
 The number of perfect matchings is $\sqrt{|\det A|}$.

\end{enumerate}

\\

&

\textbf{Complexity analysis:} The above steps can be carried out in polynomial time:
\begin{itemize}[leftmargin=1.2em]
    \item Constructing $G$: Takes $O(nm)$ time.
    \item Planar Embedding and Pfaffian Orientation: Planar embedding can be found in linear time for grid graphs. The Pfaffian orientation can be computed in linear time.
    \item Constructing $A$: Takes $O(N^2)$ time, where $N = nm$.
    \item Computing $\det A$: Determinant of a skew-symmetric matrix of size $N \times N$ can be computed in $O(N^\omega)$ time, where $\omega$ is the matrix multiplication exponent (currently around 2.373). For practical purposes, standard algorithms suffice.
\end{itemize}

\vspace{1em}
\textbf{Additional Notes:}
\begin{itemize}[leftmargin=1.2em]
    \item The constraint that half of the squares are mushrooms and half are peppers ensures that a perfect matching is possible (since the bipartite graph has an equal number of vertices on both sides).
    \item The planarity of the graph is crucial for the polynomial-time computation.
    \item This method leverages advanced concepts from combinatorial enumeration and graph theory but can be implemented efficiently for practical sizes of $n$ and $m$.
\end{itemize}
\end{longtable}

\subsection{Exercise 4: Shortest Cycle in a Graph}
\label{sect:appendix_2_ex4}
This exercise aims to develop a fast algorithm that finds the weight of the cycle with the minimum weight in a weighted  directed graph. Consider a simple directed graph $G=(V,E)$ with a weight function $w:E \mapsto \setof{1,2,\dots, W}$.
The weight of a cycle is defined as the sum of the weights of the edges in the cycle.
Let $n=|V|$ and assume that $W=O(n^{100})$ and that the cycle with minimum weight in $G$ is unique. Our algorithm should run in  $O(WM(n)n^{o(1)})$ time\footnote{$M(n)$ is the time required to multiply two $n\times n$ matrices.}. To do this,  introduce a variable $x_{ij}$ for each edge $(i,j) \in E$ and another extra variable $y$ and define an $n\times n$ matrix $C$ where 
$$
C_{ij} = \begin{cases}
x_{ij} y^{w((i,j))}, & \text{if } (i,j) \in E;\\
0, & \text{otherwise.}
\end{cases}
$$
Furthermore, let  $$ p(\xx,y) :=\det(C+I)-1 ,$$
where $I$ is the identity matrix and $$ q(y) := p(\vecone, y)$$ be the polynomial in $y$ obtained by setting $x_{ij}=1$ for $i,j =1,2,\dots n$.
Finally, let $d_{\min}$ and $c_{\min}$ be the degree and the coefficient of the monomial of smallest degree that appears in $q$.

\begin{longtable}{|p{0.4\textwidth}|p{0.55\textwidth}|}
\multicolumn{2}{p{0.96\textwidth}}{\textbf{(a)} Prove that $d_{\min}$ equals the weight of the shortest cycle in $G$.} \\ 
\multicolumn{2}{p{0.96\textwidth}}{
\vspace{.1em}
\textbf{Master Solution} 

 $\det(C+I) = \sum_\sigma sign(\sigma) \prod_i (C+I)_{i,\sigma(i)}$ The term $\prod_i (C+I)_{i,\sigma(i)}$ is non-zero if and only if the set $(i, \sigma(i))$ when $\sigma(i)\neq i$ is a collection of cycles in $G$. Furthermore, each term contains the variables $x_{i, \sigma(i)}$ for all the edges of the cycles and the exponent of $y$ is the total weight of the cycles. We conclude that the terms of the smallest degree in $\det(C+I)$ are 1 and then a term that corresponds to the shortest cycle.
}
\end{longtable}

\begin{longtable}{|p{0.4\textwidth}|p{0.58\textwidth}|}
\hline
\textbf{GPT-4o} & \textbf{{\o1}} \\ \hline
\endfirsthead
\hline
\textbf{GPT-4o} & \textbf{{\o1}} \\ \hline
\endhead
\hline \multicolumn{2}{r}{\textit{Continued on the next page...}} \\ \hline
\endfoot
\hline
\endlastfoot
The determinant $\det(C + I)$ captures all walks in the graph, including cycles. Each cycle in $G$ contributes a monomial term to $q(y)$ of the form $c y^w$, where $w$ is the total weight of the cycle, and $c$ is a coefficient related to the cycle.
The smallest degree $d_{\min}$ of $q(y)$ corresponds to the smallest weight $w$ among all cycles in $G$, which is the weight of the shortest cycle.
Thus, $d_{\min}$ equals the weight of the shortest cycle in $G$.
\textcolor{red}{[\textbf{Unjustified Claims}: Didn't explain why is there no cancellation in the determinant sum, and didn't explain the term $-1$ in the formula.]}
    & 

We are given the polynomial
\[
q(y) = \det(C + I) - 1,
\]
where \( C \) is an \( n \times n \) matrix defined by
\[
C_{ij} =
\begin{cases}
y^{w((i, j))}, & \text{if } (i, j) \in E; \\
0, & \text{otherwise}.
\end{cases}
\]
Our goal is to show that the minimal degree \( d_{\min} \) of any monomial in \( q(y) \) is equal to the weight of the shortest (minimum weight) cycle in \( G \).
Recall that the determinant of an \( n \times n \) matrix \( M \) is given by:
\[
\det(M) = \sum_{\sigma \in S_n} \operatorname{sgn}(\sigma) \prod_{i=1}^n M_{i, \sigma(i)},
\]
where \( S_n \) is the set of all permutations of \( \{1, 2, \ldots, n\} \) and \( \operatorname{sgn}(\sigma) \) is the sign of permutation \( \sigma \).
The matrix \( C + I \) has entries:
\[
(C + I)_{ij} =
\begin{cases}
1 + y^{w((i, i))}, & \text{if } i = j; \\
y^{w((i, j))}, & \text{if } (i, j) \in E \text{ and } i \neq j; \\
0, & \text{otherwise}.
\end{cases}
\]
Since there are no loops, \( w((i, i)) \) is undefined, so \( (C + I)_{ii} = 1 \).

\\

&

Therefore, \( (C + I)_{ii} = 1 \) and \( (C + I)_{ij} = y^{w((i, j))} \) for \( i \neq j \) if \( (i, j) \in E \), and zero otherwise.
The determinant expansion involves terms corresponding to permutations \( \sigma \) in \( S_n \). For each permutation \( \sigma \), the term is:
\[
\operatorname{sgn}(\sigma) \prod_{i=1}^n (C + I)_{i, \sigma(i)}.
\]
The identity permutation \( \sigma = \text{id} \) contributes:
\[
\operatorname{sgn}(\text{id}) \prod_{i=1}^n (C + I)_{i, i} = 1.
\]
Therefore, \( \det(C + I) \) includes the constant term 1.
For \( \sigma \neq \text{id} \), the term \( \prod_{i=1}^n (C + I)_{i, \sigma(i)} \) is non-zero only if \( (i, \sigma(i)) \in E \) for all \( i \) where \( \sigma(i) \neq i \).
The degree of the monomial contributed by \( \sigma \) is:
\[
\deg(\sigma) = \sum_{i: \sigma(i) \neq i} w((i, \sigma(i))).
\]
A cycle in \( G \) corresponds to a cyclic permutation \( \sigma \) where \( \sigma \) permutes a subset of indices in a cycle and fixes the others. The degree of the monomial corresponding to this cycle is equal to the total weight of the edges in the cycle.
Therefore, the minimal degree \( d_{\min} \) among all monomials in \( q(y) \) corresponds to the minimal total weight among all cycles in \( G \).
Thus, \( d_{\min} \) equals the weight of the shortest cycle in \( G \).
\textcolor{red}{[\textbf{Unjustified Claims}: Not clear why is there no cancellation in the sum happening.]}
\end{longtable}

\noindent\textbf{(b)} Devise an algorithm that finds the weight of the shortest cycle in $G$ with probability at least $1-1/n$. The algorithm should run in $O(WM(n)\log^{50} (n) )$ time.

\noindent\textit{Hint: Use Theorem~\ref{thm:hint} stated below.}

\begin{theorem}\label{thm:hint}
Let $A(y)$ be an $n\times n$ matrix where the entries $A_{i,j}$ are polynomials over $y$ with coefficients in a finite field $F$ and have degree at most $d$. Then we can compute $\det(A(y))$ in $O(dM(n)(\log (n) + \log(d))^{50})$ operations.
\end{theorem} 

\vspace{.1em}
\noindent\textbf{Master Solution}

Let $p$ be a prime number greater that $n^2$. Substitute the values of $x_{ij}$ with uniformly at random values in $\setof{1,2,\dots,p}$ in $p(x,y)$. Compute the determinant using the Theorem. By Schwartz-Zippel lemma if $p$ is big enough with probability at least $1-1/n$, the monomial of smallest degree in $y$ will appear. Return the degree of the monomial of the smallest degree.

\begin{longtable}{|p{0.48\textwidth}|p{0.5\textwidth}|}
\hline
\textbf{GPT-4o} & \textbf{{\o1}} \\ \hline
\endfirsthead
\hline
\textbf{GPT-4o} & \textbf{{\o1}} \\ \hline
\endhead
\hline \multicolumn{2}{r}{\textit{Continued on the next page...}} \\ \hline
\endfoot
\hline
\endlastfoot
\textbf{Algorithm:}
\begin{enumerate}[leftmargin=1.2em]
    \item Matrix Construction: Construct the $n \times n$ matrix $C$, where:
     \[
     C_{ij} =
     \begin{cases}
     x_{ij} y^{w((i,j))}, & \text{if } (i,j) \in E, \\
     0, & \text{otherwise}.
     \end{cases}
     \]
\item Polynomial Computation:
Compute the determinant:
     \[
     p(x, y) = \det(C + I) - 1.
     \]
Substitute $x_{ij} = 1$ for all edges $(i, j) \in E$ to obtain the polynomial:
     \[
     q(y) = p(1, y).
     \]

\item Finding the Minimum Degree:
Use a randomized approach to evaluate $q(y)$ at several points $y$ from a finite field of size $O(n^2)$.

Perform polynomial interpolation to recover the coefficients of $q(y)$ and determine the smallest degree $d_{\min}$.

\item Output: Return $d_{\min}$ as the weight of the shortest cycle.
\end{enumerate}

\textbf{Correctness:}

 The determinant $\det(C + I)$ sums over all walks in the graph. Subtracting 1 removes the contribution from the empty walk, leaving only cycles. 

 The polynomial $q(y)$ encodes all cycles in $G$, and the smallest degree of $q(y)$ corresponds to the shortest cycle.
&

We will design an algorithm that efficiently computes \( d_{\min} \) using Theorem 1. The key idea is to perform a binary search on the possible degrees and use determinant computations modulo \( y^k \) to identify the minimal degree.

\vspace{.5em}
\textbf{Algorithm:}
\begin{enumerate}[leftmargin=1.2em]
    \item Initialize: Let \( D_{\max} = nW \) (since the maximum possible total weight of a cycle is \( nW \)). Set \( k_{\text{low}} = 1 \), \( k_{\text{high}} = D_{\max} \).
    
    \item Binary Search:
        While \( k_{\text{low}} < k_{\text{high}} \):
        \begin{enumerate}
            \item Set \( k = \left\lfloor \dfrac{k_{\text{low}} + k_{\text{high}}}{2} \right\rfloor \).
            \item Compute \( q_k(y) = q(y) \mod y^{k} \).
            \item If \( q_k(y) \equiv 0 \) (all coefficients up to degree \( k-1 \) are zero):
                 Set \( k_{\text{low}} = k + 1 \).
            \item Else:
                Set \( k_{\text{high}} = k \).
       \end{enumerate}

    \item Output:
      After the loop, \( d_{\min} = k_{\text{low}} \).

\end{enumerate}
Since the minimal cycle is unique, \( d_{\min} \) correctly represents the minimal cycle weight.

\vspace{.5em}
\textbf{Complexity analysis:}
\begin{itemize}[leftmargin=1.2em]
    \item Computing \( q_k(y) \):
         We need to compute \( q(y) \mod y^{k} = \det(C + I) \mod y^{k} - 1 \).
        \textcolor{red}{[\textbf{Mathematical Errors}: What does this mean?]}
         The entries of \( C + I \) are polynomials in \( y \) of degree at most \( W \).
         Using Theorem 1, we can compute \( \det(C + I) \mod y^{k} \) in \( O(k M(n) (\log n + \log k)^{50}) \) time.
   
    \item Number of Iterations:
         The binary search runs for \( O(\log D_{\max}) = O(\log nW) = O(\log n) \) iterations (since \( W = O(n^{100}) \)).

    \item Total Time Complexity:
         Each iteration takes \( O(k M(n) (\log n + \log k)^{50}) \) time.
         Since \( k \leq D_{\max} = O(n^{101}) \), \( \log k = O(\log n) \).
         Therefore, each iteration takes \( O(W M(n) \log^{50}(n)) \) time.
         Total time complexity is \( O(W M(n) \log^{51}(n)) \).
\end{itemize}
\\

\textbf{Complexity:}
\begin{enumerate}[leftmargin=1.2em]
    \item \emph{Matrix Construction:} Constructing $C$ requires $O(n^2)$ operations.
\item \emph{Determinant Computation:} Using Theorem 1, $\det(C + I)$ can be computed in:
     \[
     O(M(n)(\log(n) + \log(d))^{5/2}),
     \]
     where $d$ is the degree of $q(y)$.
     \textcolor{red}{[\textbf{Mathematical Errors}: You need a finite field to use the theorem.]}
\item \emph{Randomized Evaluation:} Evaluating $q(y)$ at random points from a finite field requires $O(M(n))$ operations.
\item \emph{Polynomial Interpolation:} Interpolating a polynomial with degree $O(W)$ takes $O(W)$ evaluations and $O(W^2)$ time.
\end{enumerate}

\textbf{Overall Complexity:}

The dominant term is $O(WM(n)n^{o(1)})$, where $W$ is the maximum edge weight, $M(n)$ is the time complexity for multiplying two $n \times n$ matrices, and $n^{o(1)}$ accounts for logarithmic factors.

\vspace{1em}
\textbf{Probability of Success:}

By choosing random points $y$ from a sufficiently large finite field, the algorithm succeeds with probability at least $1 - 1/n$.

    &

\end{longtable}

\newpage
\section{Supplemental Experiments}
\label{sect:extended_data}
\subsection{Random Orientation}
Let $G=(V,E)$ be a graph and let $\overrightarrow{G}$ be a graph obtained by orienting the edges of $G$ independently with probability $1/2$ in either direction. Given $\overrightarrow{G}$, define the skew symmetric matrix $A_s$ such that
$$ A_s(i,j) = \begin{cases}
    +1, &(i,j) \in E \text{ with orientation from $i$ to $j$},\\
    -1, &(i,j) \in E \text{ with orientation from $j$ to $i$},\\
    0, & otherwise.
\end{cases}$$

\begin{longtable}{|p{0.45\textwidth}|p{0.5\textwidth}|}
\multicolumn{2}{p{0.96\textwidth}}{\textbf{(a)}  Let $\text{pm}(G)$ denotes the number of \emph{perfect matchings} in $G$, prove that $$\expec{}{\det{A_s(\overrightarrow{G})}} = \text{pm}(G).$$ 

\textit{Hint: Use the definition of determinant and the linearity of expectation.}} \\ 
\multicolumn{2}{p{0.96\textwidth}}{
\vspace{.1em}
\textbf{Master Solution} 

$$\expec{}{\det{A_s(\overrightarrow{G})}} = \expec{}{ \sum_\pi sign(\pi) \prod_i a_{i, \pi(i)}} = \sum_\pi sign(\pi) \expec{}{ \prod_i a_{i, \pi(i)}}.$$
If there exists $j$ such that $\pi(j) =j$ or $(j,\pi(j)) \notin E(G)$, then $ \prod_i a_{i, \pi(i)} = 0$. Further, if there exists $j$ such that $\pi(\pi(j)) \neq j$, then we have that $ \expec{}{ \prod_i a_{i, \pi(i)}} = \expec{}{ \prod_{i\neq j} a_{i, \pi(i)}}\expec{}{a_{j,\pi(j)}} = 0$ because $a_{j,\pi(j)}$ is independent from the other terms and has expectation $0$. We are left with only the permutations $\pi$ such that $\pi(\pi(i)) =i$, $\forall i$; for these, the term $ \prod_i a_{i, \pi(i)} = 1 = sign(\pi)$ because each variable appears exactly 2 times in the sum and they consist of even cycles. Finally it is not hard to see that there exists a bijection between these permutations and the perfect matchings of $G$. This concludes the argument.
}
\end{longtable}

\begin{figure}[h]
    \centering
    \includegraphics[width=16cm]{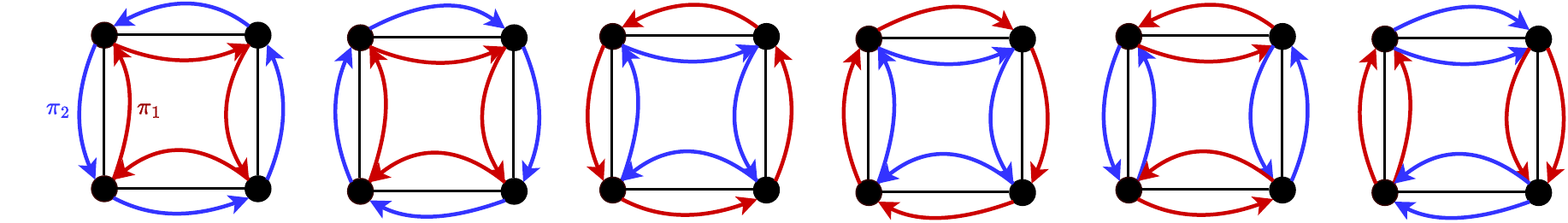}
    \caption{The six options of permutations on an (even) cycle. Notice that the same permutation types exist on a longer (even) cycle. The red arrows go from vertex $i$ to vertex $\pi_1(i)$ while the blue arrows go from vertex $i$ to vertex $\pi_2(i)$.}
    \label{fig:matching_cycles}
\end{figure}

\begin{longtable}{|p{0.45\textwidth}|p{0.5\textwidth}|}
\hline
\textbf{GPT-4o} & \textbf{{\o1}} \\ \hline
\endfirsthead
\hline
\textbf{GPT-4o} & \textbf{{\o1}} \\ \hline
\endhead
\hline \multicolumn{2}{r}{\textit{Continued on the next page...}} \\ \hline
\endfoot
\hline
\endlastfoot
The matrix \( A_s(G) \) is a skew-symmetric matrix associated with the random orientation \( \vec{G} \) of the edges of \( G \). Each edge \( (i, j) \in E \) is directed from \( i \) to \( j \) with probability \( \frac{1}{2} \) and from \( j \) to \( i \) with probability \( \frac{1}{2} \), making \( A_s(i, j) = +1 \) or \( -1 \) with equal probability.

The determinant of a skew-symmetric matrix can be expressed as a sum over permutations:

\[
\det(A_s) = \sum_{\sigma \in S_n} \text{sgn}(\sigma) \prod_{i=1}^{n} A_s(i, \sigma(i)),
\]

where \( S_n \) is the set of all permutations of \( n \) elements. Since \( A_s(i, j) \) is nonzero only for edges in \( E \), only permutations corresponding to perfect matchings contribute to the determinant. 
\textcolor{red}{[\textbf{Unjustified Claims}: Why only perfect matchings? What about $\text{sgn}(\sigma)$?]} \textcolor{red}{[\textbf{Misleading Claims}: Also, all the perfect matchings correspond to a permutation.]}

Given that the random orientation assigns equal probabilities to \( +1 \) and \( -1 \), every term in the determinant corresponding to a perfect matching contributes with equal probability. The linearity of expectation simplifies the determinant’s expectation to the count of perfect matchings:

\[
\expec{}{\det(A_s(\vec{G}))} = \text{pm}(G).
\]

    & 
Recall the standard permutation expansion of the determinant:

\[
\det(A_s(\vec{G})) = \sum_{\sigma \in S_n} \text{sgn}(\sigma) \prod_{i=1}^{n} A_s(\vec{G})( i, \sigma(i) ).
\]

Since \( A_s(\vec{G}) \) is skew-symmetric with \( \pm1 \) on edges and 0 on non-edges, any nonzero product

\[
\prod_{i=1}^{n} A_s(\vec{G})(i, \sigma(i))
\]

must correspond to a permutation \( \sigma \) that is a disjoint union of 2-cycles, matching vertices in pairs. In other words, \( \sigma \) must correspond exactly to a perfect matching in \( G \). (Cycles of length \( \geq 3 \) would yield zero or canceling contributions in a skew-symmetric matrix.)

If \( M \) is a perfect matching, there are exactly \( 2^{n/2} \) ways to orient each matched edge so that \( \sigma \) realizes those transpositions with nonzero contribution (one orientation for each pair, times two per edge). Each such orientation occurs with probability \( \frac{1}{2^{n/2}} \). The sign of each transposition in a skew-symmetric matrix contributes consistently \( +1 \) (having two factors of \( \pm1 \) for each edge in the matched cycle). Therefore, each perfect matching contributes exactly 1 in expectation. Summing over all \( \text{pm}(G) \) matchings, we conclude:

\[
\expec{}{ \det(A_s(\vec{G})) } = \text{pm}(G).
\]

\end{longtable}

\noindent\textbf{(b)}  Given two perfect matchings $M_1,M_2$ of $G$, denote with $a(M_1,M_2)$ the number of cycles in $G$ made of (alternating) edges contained in $M_1$ and $M_2$  (or equivalently the cycles in $M_1 \text{ xor } M_2$), and denote with $\mathcal{M}$ the set of all perfect matchings in $G$. Prove that $$\expec{}{\left(\det{A_s(\overrightarrow{G})}\right)^2} = \sum_{M_1 \in \mathcal{M}} \sum_{M_2 \in \mathcal{M}} 3^{a(M_1,M_2)}.$$

\noindent \textit{Hint: 
You may find inspiration in the proof of Theorem 5.3. First deal with permutations whose directed graph uses edges that do not appear in $G$.
Next argue about the contribution from permutations where this graph contains at least one cycle of odd length, and finally the rest.}

\vspace{1em}
\noindent\textbf{Master Solution} 

We first plug in the definition of the determinant $$ \expec{}{\left(\det{A_s(\overrightarrow{G})}\right)^2} =  \expec{}{\left( \sum_\pi sign(\pi) \prod_i a_{i, \pi(i)}\right)^2}.$$
We partition the set of permutations into a set
$$ A := \{ \pi: \exists j: \pi(j) = j \lor (j, \pi(j)) \notin E(G)\}$$ and its complement $$B := \{ \pi: \pi \notin A \}.$$
Every permutation in $A$ contributes a zero term to the product, and therefore we have 
\parbox{0.96\textwidth}{
\begin{align*}
    \expec{}{\left( \sum_\pi sign(\pi) \prod_i a_{i, \pi(i)}\right)^2} &= \expec{}{\left( \sum_{\pi \in B} sign(\pi) \prod_i a_{i, \pi(i)}\right)^2} \\
    &= \expec{}{\sum_{\pi_1 \in B} \sum_{\pi_2 \in B} sign(\pi_1) \cdot sign(\pi_2) \prod_i a_{i, \pi_1(i)} \prod_i a_{i, \pi_2(i)}}\\
    &= \expec{}{\sum_{\pi_1 \in B} \sum_{\pi_2 \in B} val(\pi_1) \cdot val(\pi_2)}
\end{align*}
}
where $val(\pi) := sign(\pi) \prod_i a_{i, \pi(i)}$. We then define another set of permutation $C \subset B$:
$$C := \{\pi \in B: \text{all cycles in $\pi$ are of even length}\}.$$

Next, we show that 
\begin{align}
    \label{eq:odd_out}
    \expec{}{\sum_{\pi_1 \in B} \sum_{\pi_2 \in B} val(\pi_1) \cdot val(\pi_2)} = \expec{}{\sum_{\pi_1 \in C} \sum_{\pi_2 \in C} val(\pi_1) \cdot val(\pi_2)}.
\end{align}
To do so, we fix some permutation $\pi_1 \in B$ that contains an odd cycle and an arbitrary permutation $\pi_2 \in B$. Then consider the odd cycle $c$ incident to the lowest indexed vertex among the odd cycles in $\pi_1$. If this odd cycle does not appear in $\pi_2$ in some direction, there is a term $a_{ij}$ that only appears once in $val(\pi_1) \cdot val(\pi_2)$. Therefore $\expec{}{val(\pi_1) \cdot val(\pi_2)} = 0$ because the edge directions are chosen independently. Similarly, this holds for every (odd) cycle in $\pi_1$. Therefore the cycle $c$ is also incident to the lowest indexed vertex on an odd cycle in $\pi_2$. We now pair permutations 1: ($\pi_1, \pi_2$), 2: ($\pi'_1 = \pi_1$ with $c$ reversed, $\pi_2$), 3: ($\pi_1$, $\pi'_2 = \pi_2$ with $c$ reversed) and 4: ($\pi'_1$, $\pi'_2$). Since $$val(\pi_1) \cdot val(\pi_2) = - val(\pi'_1) \cdot val(\pi_2) = - val(\pi_1) \cdot val(\pi'_2) = val(\pi'_1) \cdot val(\pi'_2)$$ these 4 cancel out and this establishes (\ref{eq:odd_out}).

Finally, we show that
\begin{align}
    \label{eq:final_result}
    \expec{}{\sum_{\pi_1 \in C} \sum_{\pi_2 \in C} val(\pi_1) \cdot val(\pi_2)} = \sum_{M_1 \in \mathcal{M}} \sum_{M_2 \in \mathcal{M}} 3^{a(M_1,M_2)}.
\end{align}

We fix two matchings $M_1$ and $M_2$ and we let the support $E_s$ denote the set of edges that are either in $M_1$, in $M_2$ or in both. This support consists of even cycles and isolated edges. Given a fixed support $E_s$, we are interested in the amount of (ordered) matching pairs that yield $E_s$. For every isolated edge there is exactly one option: Both $M_1$ and $M_2$ have to contain it. But for each cycle there are exactly two possible options. Therefore, the amount of matchings is given by the $2^{\#\text{even cycles}}$. Now, we consider the amount of pairs of permutations $\pi_1, \pi_2$ that live on this support. For each isolated edge $(i,j)$, there is exactly one choice for both $\pi_1$ and $\pi_2$. For each even cycle however, there are $6$ choices. These are displayed in Figure \ref{fig:matching_cycles}. Therefore, the total number of permutations supported is $6^{\#\text{even cycles}}$, and there is a mapping from each pair of matchings yielding the support to $3^{\#\text{even cycles}}$ permutations. In such a way, each permutation is allocated to a pair of matchings. This establishes (\ref{eq:final_result}) and concludes the solution to this exercise.

\begin{longtable}{|p{0.44\textwidth}|p{0.55\textwidth}|}
\hline
\textbf{GPT-4o} & \textbf{{\o1}} \\ \hline
\endfirsthead
\hline
\textbf{GPT-4o} & \textbf{{\o1}} \\ \hline
\endhead
\hline \multicolumn{2}{r}{\textit{Continued on the next page...}} \\ \hline
\endfoot
\hline
\endlastfoot

The square of the determinant expands as:

\[
\det(A_s)^2 = \]
\[ \sum_{\sigma, \tau \in S_n} \text{sgn}(\sigma) \text{sgn}(\tau) \prod_{i=1}^{n} A_s(i, \sigma(i)) A_s(i, \tau(i)).
\]

A term contributes to the expectation only if every edge in the product appears an even number of times (due to the random orientation). This condition is satisfied when \( \sigma \) and \( \tau \) correspond to matchings \( M_1 \) and \( M_2 \) that form alternating cycles. \textcolor{red}{[\textbf{Unjustified Claims}: Why? This needs a proof.]} Each cycle of length \( \ell \) contributes a factor of \( 3^{\ell-1} \), as every edge in the cycle has three configurations (same direction, reversed direction, or alternating). \textcolor{red}{[\textbf{Misleading Claims}: This is not the correct reason.]}

Therefore:

\[
\expec{}{(\det(A_s(\vec{G})))^2} = \sum_{M_1, M_2 \in M} 3^{a(M_1, M_2)},
\]

where \( a(M_1, M_2) \) is the number of alternating cycles in \( M_1 \oplus M_2 \).

    & 
We use 

\[
(\det(A_s))^2 = \det(A_s) \cdot \det(A_s)
\]

and expand each determinant via permutations. A pair \( \sigma, \tau \in S_n \) yields a nonzero-oriented product only if \( \sigma \) and \( \tau \) together correspond to unions of disjoint cycles using edges of \( G \). In particular, if \( \sigma \leftrightarrow M_1 \) and \( \tau \leftrightarrow M_2 \) are two perfect matchings, their union is a collection of cycles, each of which must be even (length \( 2 \) or more, since each vertex has degree 2 in the union). \textcolor{red}{[\textbf{Unjustified Claims}: Why all the other cancel out?]}

When averaged over random orientations, each even cycle of length \( 2k \) contributes a factor of 3, since among its \( 2^{2k} \) possible orientations, exactly 2 orientations give a nonzero skew-symmetric product (clockwise or counterclockwise), but each orientation has probability \( \frac{1}{2^{2k}} \). Careful sign tracking shows the net effect is a factor 3 per cycle (rather than just 2) because each even cycle can appear in exactly three types of contributing orientations (two single-cycle orientations plus one subtle additional contribution),
or equivalently one can see it as $\sum_{c=0}^{2^2k} \dots$ leading to a factor of 3 per cycle. \textcolor{red}{[\textbf{Misleading Claims}: This is not the correct reason.]}

Summing over all pairs \( M_1, M_2 \) and all cycles in \( (M_1 \cup M_2) \), we obtain:

\[
\expec{}{ (\det(A_s(\vec{G})))^2 } = \sum_{M_1 \in M} \sum_{M_2 \in M} 3^{\alpha(M_1,M_2)},
\]

where \( \alpha(M_1, M_2) \) is the number of cycles in the union of the two matchings.

\end{longtable}

\begin{longtable}{|p{0.4\textwidth}|p{0.55\textwidth}|}
\multicolumn{2}{p{0.96\textwidth}}{\textbf{(c)}  Let $$\var{\det{A_s(\overrightarrow{G}_i)}} = \expec{}{\left(\det{A_s(\overrightarrow{G})}\right)^2} - \left(\expec{}{\det{A_s(\overrightarrow{G})}}\right)^2. $$  Design a randomized algorithm that computes the determinant of at most  $N=2  \var{\det{A_s(\overrightarrow{G}_i)}}$ matrices, runs in $O(N\ \text{poly}(|V|))$ time, and returns $x$ such that  
  $$\prob{}{ |\text{pm}(G)-x|  \geq 1} \leq \frac{1}{2} .$$} \\

\multicolumn{2}{p{0.96\textwidth}}{\textbf{Master Solution}

Sample $N$ independent random orientations for the edges of $G$ and let $\overrightarrow{G}_1, \dots, \overrightarrow{G}_N$ be these graph. The algorithm returns $ \frac{1}{N} \sum_{i=1}^{N}\det{A_s(\overrightarrow{G}_i})$. The variance of this quantity is given by $1/N$ times the variance of an individual run by independence of the runs and $var[X + Y] = \var[X] + \var[Y]$ for $X$ and $Y$ independent, and the fact that $var[c \cdot X] = c^2 \cdot var[X]$. That is 
$$\sqrt{ \var{\frac{1}{N} \sum_{i=1}^{N}\det{A_s(\overrightarrow{G}_i})}} = \frac{1}{\sqrt{N}} \sqrt{ \var{\det{A_s(\overrightarrow{G}_i})}}.$$
By Chebyshev's inequality:
  $$ \prob{}{\bigg | \expec{}{\det{A_s(\overrightarrow{G})}} - \frac{1}{N} \sum_{i=1}^{N}\det{A_s(\overrightarrow{G}_i}) \bigg |  \geq k \sqrt{ \var{\frac{1}{N} \sum_{i=1}^{N}\det{A_s(\overrightarrow{G}_i})}} } \leq \frac{1}{k^2} $$
  For $k = \sqrt{2}$, we obtain the desired bound. 
Each iteration of the algorithm computes a determinant. Using gaussian elimination, this can be implemented in polynomial time. 
}
\end{longtable}

\begin{longtable}{|p{0.45\textwidth}|p{0.5\textwidth}|}
\hline
\textbf{GPT-4o} & \textbf{{\o1}} \\ \hline
\endfirsthead
\hline
\textbf{GPT-4o} & \textbf{{\o1}} \\ \hline
\endhead
\hline \multicolumn{2}{r}{\textit{Continued on the next page...}} \\ \hline
\endfoot
\hline
\endlastfoot

The variance of \( \det(A_s(\vec{G})) \) is given by:

\[
\text{var}(\det(A_s(\vec{G}))) =
\]
\[\expec{}{(\det(A_s(\vec{G})))^2} - (\expec{}{\det(A_s(\vec{G})})^2.
\]

Using the results of (a) and (b), we obtain:

\[
\text{var}(\det(A_s(\vec{G})))=
\]
\[ \sum_{M_1, M_2 \in M} 3^{a(M_1, M_2)} - (\text{pm}(G))^2.
\]

\textbf{Randomized algorithm:}  
To estimate \( \text{pm}(G) \), we compute \( \det(A_s(\vec{G})) \) for multiple random orientations of \( G \). Using Chebyshev’s inequality:

\[
P\left(\left| \det(A_s(\vec{G})) - \text{pm}(G) \right| \geq 1 \right)
\leq
\]
\[\frac{\text{var}(\det(A_s(\vec{G})))}{12}.
\]

    & 

From (a) and (b), we derive:
\begin{align*}
\text{var}[\det(A_s(\vec{G}))] = \expec{}{ (\det(A_s(\vec{G})))^2 } \\- (\expec{}{ \det(A_s(\vec{G})) })^2.
\end{align*}

Define the random variable \( X = \det(A_s(\vec{G})) \). Note that \( \expec{}{X} = \text{pm}(G) \), so \( X \) is an unbiased estimator for the number of perfect matchings. We approximate \( \text{pm}(G) \) by taking N independent orientations \( \vec{G}_1, \dots, \vec{G}_N \), computing 

\[
X_k = \det(A_s(\vec{G}_k))
\]

for each \( k \), and letting 

\[
X = \frac{1}{N} \sum_{k=1}^{N} X_k.
\]

Then \( \expec{}{X} = \text{pm}(G) \) as well, and 

\[
\text{var}(X) = \frac{\text{var}(X)}{N}.
\]
\\

Run the algorithm \( N = 2 \cdot \text{var}(\det(A_s(\vec{G}))) \) times, compute the average of \( \det(A_s(\vec{G})) \), and return this average as an estimate of \( \text{pm}(G) \). By Chebyshev’s inequality, the probability of error is at most \( 1/2 \).\textcolor{red}{[\textbf{Unjustified Claims}: How are you using Chebyschev's inequality?]}

&

By Chebyshev’s inequality, for \( \kappa > 0 \),

\[
P[|X - \text{pm}(G)| \geq \kappa] \leq \frac{\text{var}(X)}{N \kappa^2}.
\]

Choosing \( N = 2\text{var}(X) \) and \( \kappa = 1 \), we get 

\[
P[ |X - \text{pm}(G)| \geq 1 ] \leq \frac{1}{2}.
\]

Hence, with probability at least \( \frac{1}{2} \), \( X \) approximates \( \text{pm}(G) \) to within 1. Computing each determinant can be done in \( O(|V|^\omega) \), so the total runtime is

\[
O(N \cdot |V|^\omega).
\]

This demonstrates how \( \det(A_s(\vec{G})) \) can be used as a randomized estimator.
\end{longtable}

\begin{longtable}{|p{0.4\textwidth}|p{0.55\textwidth}|}
\multicolumn{2}{p{0.96\textwidth}}{\textbf{(d)}  Find a collection of graphs $G$ such that $\var{\det{A_s(\overrightarrow{G})}}$, where $\overrightarrow{G}$ is a random orientation of $G$, is \emph{not} polynomially bounded in the size of $G$, i.e. there is no polynomial $p$ such that $\var{\det{A_s(\overrightarrow{G})}} \leq p(|V(G)|+|E(G)|)$ . (This means that the algorithm we just developed does not run in polynomial time for all the input graphs.) 

\textit{Hint: Consider multiple copies of a graph containing only one cycle.}} \\ 
\multicolumn{2}{p{0.96\textwidth}}{
\vspace{.1em}
\textbf{Master Solution} 

Fix $n \in \N$, divisible by 4 and consider the graph $G$ made by the union of $n/4$ copies of the cycle graph on 4 edges $C_4$. 

Since $C_4$ has two perfect matchings, we have that $$pm(G) = 2^{n/4}.$$ Consider the pairs of matchings $M_1, M_2 \in \mathcal{M}$ that are disjoint (i.e. they do not share any edge). There are exactly $2^{n/4}$ such pairs because for each copy of $C_4$ we have 2 possible combinations. Furthermore, if $M_1$ and $M_2$ are disjoint, the graph obtained by the \textit{xor} of the edges is $G$ and $a(M_1,M_2) = n/4$. Using the results from the previous part of the exercise,
$$\var{\det{A_s(\overrightarrow{G}_i)}} = \expec{}{\left(\det{A_s(\overrightarrow{G})}\right)^2} - \left(\expec{}{\det{A_s(\overrightarrow{G})}}\right)^2 $$
$$ = \sum_{M_1 \in \mathcal{M}} \sum_{M_2 \in \mathcal{M}} 3^{a(M_1,M_2)} - \text{pm}(G)^2$$
$$ \geq \sum_{M_1, M_2 \text{ disjoint}} 3^{a(M_1,M_2)} - \text{pm}(G)^2 $$
$$ = 2^{n/4} \ 3^{n/4} - 2^{ n/2}$$
$$ \geq \frac{3}{2} 2^{ n/2} - 2^{ n/2} = \frac{1}{2} 2^{ n/2}$$
}
\end{longtable}

\begin{longtable}{|p{0.45\textwidth}|p{0.5\textwidth}|}
\hline
\textbf{GPT-4o} & \textbf{{\o1}} \\ \hline
\endfirsthead
\hline
\textbf{GPT-4o} & \textbf{{\o1}} \\ \hline
\endhead
\hline \multicolumn{2}{r}{\textit{Continued on the next page...}} \\ \hline
\endfoot
\hline
\endlastfoot

Consider a graph \( G \) consisting of \( k \) disjoint cycles of length \( n \). Each cycle independently contributes a factor to \( \det(A_s(\vec{G})) \). The number of alternating cycles \( a(M_1, M_2) \) grows exponentially with the size of the graph, causing \( \expec{}{(\det(A_s(\vec{G})))^2} \) and \( \text{var}(\det(A_s(\vec{G}))) \) to grow exponentially. 

Thus, no polynomial bound exists for \( \text{var}(\det(A_s(\vec{G}))) \)  in terms of \( |V| \)  or \( \text{pm}(G) \).

    & 

The above approach yields a randomized approximation only if \( \text{var}[\det(A_s(\vec{G}))] \) is not too large. However, it is possible to construct graphs whose variance grows super-polynomially in \( |V| \). The key idea is to build a graph \( G \) from multiple copies of a small base graph that contains exactly one cycle.

For instance, consider a single cycle \( C_4 \) with a chord or two to ensure it has a perfect matching. Then, replicate it \( k \) times disjointly. Each copy contributes substantial independent randomness to the determinant, causing the combined variance to grow exponentially in \( k \). Since \( |V| \) is linear in \( k \), this implies that
$\text{var}[\det(A_s(\vec{G}))]$ cannot be bounded by any fixed polynomial in $|V|$.

Therefore, there is no universal polynomial \( p(|V(G)|) \) such that 

\[
\text{var}[\det(A_s(\vec{G}))] \leq p(|V(G)|) + \text{pm}(G),
\]

for all graphs \( G \). This means that the randomized method sketched in part (c) may require a super-polynomial number of samples \( N \), making it infeasible for polynomial-time execution on all inputs.
\end{longtable}

\end{document}